\documentclass[11pt]{article}
\usepackage[margin=1in]{geometry}
\usepackage[utf8]{inputenc}
\usepackage{amsmath,amssymb}
\usepackage{bm}
\usepackage{graphicx}
\usepackage{hyperref}
\usepackage{natbib}
\usepackage{parskip}

\usepackage{listings}%

\usepackage{caption}
\usepackage{float}
\usepackage{xcolor}

\usepackage{wrapfig}
\usepackage{graphicx}

\usepackage{pifont}% http://ctan.org/pkg/pifont
\newcommand{\cmark}{\textcolor{green!60!black}{\ding{51}}}%
\newcommand{\xmark}{\textcolor{red!70!black}{\ding{55}}}%
\usepackage{comment}
\newcommand{\vect}[1]{{\mathbf {#1} }}

\newcommand{\timeop}[0]{\op{T}}
\newcommand{\linop}[0]{L}
\DeclareMathOperator{\nonlinop}{\op{N}}

\newcommand{\bra}[1]{\langle#1\rvert} % Bra
\newcommand{\ket}[1]{\lvert#1\rangle} % Ket
\newcommand{\qprod}[2]{ \langle #1 | #2 \rangle} %Inner Product
\newcommand{\braopket}[3]{\langle #1 | #2 | #3\rangle} % Matrix Element
\newcommand{\op}[1]{\mathbf{ #1 }} % 5x5 matrices and diff operators
\newcommand{\conj}[1]{ {#1}^\ast }

\newcommand{\wmode}{{\Psi}^\pm_\textrm{w}}
\newcommand{\iomode}{{\Psi}_\textrm{io}}
\newcommand{\gmode}{{{\Psi}}_\textrm{g}}

\newcommand{\mdamode}{{{\Psi}}_{\textrm{mda}}}

\newcommand{\etat}{\eta}
\newcommand{\etae}{\eta_\textrm{e}}
\newenvironment{abstractlike}[1]
{
\begin{small}
    \begin{center}
        {\bfseries #1}
    \end{center}
    \begin{quote}
}
{
    \end{quote}
    \end{small}
}

\begin{document}

\title{Measuring fluxes between wave and geostrophic features in rotating non-hydrostatic flows with variable stratification}

\author{%
  Jeffrey J.~Early$^{1}$ \and
  Gerardo Hern\'andez-Due\~nas$^{2}$ \and
  Leslie M.~Smith$^{3,4}$ \and
  Cimarron Wortham$^{1}$ \and
  M.-Pascale Lelong$^{1}$%
}

\date{}

\maketitle

\begin{abstractlike}{Key Points}
\begin{itemize}
\item A decomposition of wave and geostrophic total energy applicable to non-hydrostatic flows with arbitrary stratification
\item Application to realistic mid-ocean simulations with geostrophic-only and geostrophic-plus-wave forcing
\item Identification of inverse geostrophic cascade, forward wave cascade, and strong geostrophic-to-wave transfer
\end{itemize}
\end{abstractlike}

\begin{abstract}

A challenge in physical oceanography is quantifying the energy content of waves and balanced flows and the fluxes that connect these reservoirs with their sources and sinks. Methodological limitations have prevented decompositions for realistic flows with non-hydrostatic motions and variable stratification.

We present a framework that separates the flow into wave and geostrophic components using the principle that waves have no Eulerian available potential vorticity signature. Starting from new expressions for available energy and potential vorticity conservation, we construct a basis of wave and geostrophic modes, complete and orthogonal with respect to quadratic approximations of the conserved quantities. Using the resulting non-hydrostatic projection operators, the nonlinear equations of motion are expressed as coupled wave and geostrophic equations, quantifying cascade and transfer fluxes of wave and geostrophic energy.

We apply the method to non-hydrostatic mid-ocean simulations with geostrophic mean-flow, near-inertial, and tidal forcing. From these experiments, we construct source-sink-reservoir diagrams for exact and quadratic fluxes, quantifying the fluxes between geostrophic and wave components. Because the cascade fluxes obey total energy conservation, we construct energy flow diagrams within the wave and geostrophic reservoirs and diagnose nonlocal transfers. The simulations show a geostrophic inverse cascade, a forward wave cascade, and a direct transfer of geostrophic to wave energy, with no indication of a forward geostrophic cascade. The mean-flow-only simulation shows weak spontaneous wave emission during spin-up, which diminishes to zero. Finally, we evaluate the decomposition by comparing linearized and fully conserved available potential vorticity, finding that errors become significant at scales below 15\,km.
\end{abstract}

\begin{abstractlike}{}
$^{1}$NorthWest Research Associates, Seattle, WA, USA \\
$^{2}$National Autonomous University of Mexico Campus Juriquilla, Quer\'etaro, Mexico \\
$^{3}$Department of Atmospheric and Oceanic Sciences, University of Wisconsin, Madison, WI, USA \\
$^{4}$Department of Mathematics, University of Wisconsin, Madison, WI, USA
\end{abstractlike}

\newpage

\section{Introduction}

% Context
Identifying energy pathways, from sources to sinks through the different reservoirs, remains one of the central challenges of oceanography. \citet{wunsch2004-arfm} constructed a `strawman energy budget' from numerous observational sources with large uncertainties. In numerical models, on the other hand, producing a detailed energy budget hinges on (1) partitioning the flow into distinct constituents, i.e., the wave and balanced flow, and (2) precisely measuring their respective energies and associated energy fluxes. While it is relatively easy to compute  {\em total} energies and {\em total} energy fluxes from numerical simulations, finding a decomposition for waves and vortex motions that is grounded in meaningful and measurable dynamics, e.g.~a timescale separation, and that satisfies energy orthogonality between the flow constituents is far more challenging. Here, the term `vortex' refers to the geostrophic or balanced part of the flow with dynamics linked to potential vorticity conservation. 
Not all linear flow decompositions satisfy energy orthogonality, yet this property is essential for uniquely interpreting energy transfers between reservoirs.

% Present state-of-the-art
Attempts at understanding and quantifying wave-vortex interactions have a long history and use a variety of methodologies. One approach is to use simplified geometry, typically either shallow-water or triply-periodic models. In both cases the linearized equations of motion admit both slow vortical modes and fast internal gravity wave modes proportional to Fourier modes \citep{lien1992-jpo,salmon1998-book,riley2000-arfm}. The separation is dynamical in nature, based on potential vorticity inversion. This approach allows an energetically orthogonal decomposition where all modes have a unique signature of total energy \citep[e.g.][]{lelong1991-jfm, bartello1995-jas, smith1999-pof, smith2002-jfm,waite2006-jfm,waite2006b-jfm, remmel2009-jfm, hernandez2014-jfm, eden2020-jpo, hernandez2021-jpo, thomas2021-jfm}. An advantage of complete orthogonal decompositions is that interactions can be intentionally restricted to isolate important interactions \citep[e.g.][]{hernandez2014-jfm,eden2019-jpo,hernandez2021-jpo} while the full nonlinear equations of motion remain unapproximated in general, thereby eliminating at least one source of uncertainty when interpreting results. The disadvantage of this approach is the constrained model geometry, which severely limits realism. 
Another approach is to use asymptotic models \citep[e.g.][]{warn1995-qjrms,young1997-jmr,xie2015-jfm, wagner2016-jfm, rocha2016-jpo, xie2020-jfm}. This approach offers the advantage of isolating the physical process of interest, but the disadvantage of approximating the equations of motion and thus the conservation laws. 
A final approach is to use a primitive equation model with realistic boundary conditions, and construct a filter to decompose the output \citep[e.g.][]{gertz2009-jpo, taylor2016-jpo, barkan2017-jpo, taylor2020-jpo, barkan2024-jpo, shaham2025-james}. The advantage is that no physics is lost in the model. The disadvantages are that these decompositions typically treat kinetic energy only, precluding analysis of inertial ranges, and rely on linear-theory scale separations to partition the flow.

% The primary challenges in quantifying sources, sinks, and fluxes are methodological in nature, centered around the separation of wave and balanced motions. In numerical models, measuring the {\em total} energy and {\em total} energy flux from the forcing terms is relatively straightforward. Far more challenging is finding a decomposition that (1) separates the flow into wave and vortical motions and (2) has energy orthogonality between these two types of motion.
% The terminology here is that the word `vortex' refers to the geostrophic, or balanced, part of the flow with dynamics linked to potential vorticity conservation. Resolving this ambiguity in nomenclature is the first challenge---any separation must have a definition tied to the dynamics that is meaningful and quantifiable, e.g., a temporal scale separation between wave and vortex motions. The second challenge is that wave and vortex motions must be energetically orthogonal---the total wave energy added to the total vortex energy must equal the total energy of the fluid. In contrast, an arbitrary linear decomposition of the fluid will not satisfy this property as total energy is a nonlinear property and it would be impossible to determine how energy transfers from one reservoir to another. A complete and orthogonal wave-vortex decomposition valid in vertically bounded domains with arbitrary stratification is needed.

% Importance of energy/enstrophy orthogonality and completeness
One of the most important properties of rotating stratified fluids is the existence of inertial cascades \cite{kraichnan1967-pof,charney1971-jas,salmon1980-gafd}. In quasigeostrophic flow, energy exhibits an inverse cascade, from small to large scales, while at the same time potential enstrophy cascades from large to small scales. To observe these dual cascades simultaneously requires a basis of eigenmodes that are orthogonal with respect to both energy and potential enstrophy. Wave energy, by contrast, cascades forward from large to small scales, e.g. \citet{mccomas1977-jgr,zakharov1992-book,polzin2011-rg,wu2023-jfm}.

The concepts of completeness and orthogonality are critical for discussing spectra and fluxes between flow components. To demonstrate this, let the state of a geophysical fluid be given by $(u,v,w,\eta,p)$ for fluid velocity $(u,v,w)$ with vertical displacement $\eta$ and pressure $p$. Any basis capable of representing all physically realizable states of the fluid is said to be \emph{complete}. To lowest order, the total volume-integrated \emph{available energy} of the rotating, non-hydrostatic Boussinesq fluid is
\begin{equation}
\label{eqn:intro-total-energy}
    \mathcal{E}\left[ (u,v,w,\eta) \right] = \frac{1}{2} \int \left( u^2 + v^2 + w^2 + N^2 \eta^2 \right) dV
\end{equation}
where $N^2(z)$ is the squared buoyancy frequency. If the fluid can be linearly decomposed into two distinct parts such that $\left(u,v,w,\eta \right)=\left(u_1+u_2,v_1+v_2,w_1+w_2,\eta_1+\eta_2 \right)$ then
\begin{equation}
\label{eqn:intro-energy-sum}
    \mathcal{E}\left[\left(u,v,w,\eta \right) \right] = \mathcal{E}\left[ \left(u_1,v_1,w_1,\eta_1 \right) \right] + \mathcal{E}\left[ \left(u_2,v_2,w_2,\eta_2 \right) \right] + \epsilon_{12}
\end{equation}
where $\epsilon_{12}$ is a quadratic cross-term. For a sparse-data decomposition, the hope is that the cross-term $\epsilon_{12}$ is small, but for a full-knowledge decomposition it may be possible to find solutions where  $\epsilon_{12}=0$ and the decomposition is \emph{energetically orthogonal}. 
With energy orthogonality as a guiding principle, we can define a set of \emph{reservoirs} as a partition of the full fluid state, each with  distinct energy.
%The primary partition considered below is into wave and vortex components, but other partitions are possible, e.g.~by space or time scale.
Because potential enstrophy is also quadratic at lowest order, a wave–vortex decomposition must satisfy enstrophy orthogonality following rules analogous to \eqref{eqn:intro-energy-sum}.
% With energy orthogonality as a guiding principle, we can define a set of \emph{reservoirs} as a partition of the full fluid state, and each reservoir has distinct energy. The primary partition considered below is into wave and vortex components, but other partitions are possible.

When a basis is both complete and energetically orthogonal then each basis member has a unique energy signature---a feature that allows one to uniquely interpret that energy is moving \emph{from} somewhere \emph{to} somewhere else. Moreover, the lack of cross-terms also allows us to write energy spectra for separate flow components because the total energy is the sum of the squares of the amplitudes of the basis functions. This is exactly analogous to a Fourier series which partitions the total variance of a periodic function in terms of orthogonal Fourier modes. Importantly, energy orthogonality is not just a choice, but a requirement for constructing an energy spectrum.

% A consistent picture has emerged from the variety of methodologies, although, as noted by \citet{taylor2020-jpo}, the terminology used is quite varied and the relationship between the proposed physical phenomena is not always clear. A few of the key ideas are as follows,
% \begin{itemize}
%     \item Spontaneous generation, the excitation of internal gravity waves from balanced motions, has been shown to be very small. See \citet{vanneste2013-arfm} for a historical review.
%     \item The introduction of wave forcing reduces the amount of geostrophic energy.
%     \item Wave forcing may introduce some form of `stimulated loss of balance', either causing a direct transfer of geostrophic energy to the wave field, or causing a forward cascade of energy within the geostrophic reservoir toward smaller scales.
% \end{itemize}

% Development of mode decompositions
Finding a complete and orthogonal basis that represents wave and geostrophic flows in a realistic domain is not trivial. 
While the formulation of the internal gravity wave energy spectrum from orthogonal wave modes by \citet{garrett1972-gfd} is sufficient for describing the energy content of interior flow, eigenmodes with an explicit free-surface are necessary when surface (or barotropic) waves are present. \citet{kelly2016-jpo} constructed an energetically orthogonal basis for hydrostatic waves with an explicit free-surface using vertical modes first described by \citet{olbers1986-igw}. This solved a decades-long `spurious energy conversion' problem which, when viewed through the lens of energy orthogonality, was caused by attempts to use a non-orthogonal basis.
\citet{smith2013-jpo} proposed an orthogonal basis for quasigeostrophic flows in three dimensions with variable stratification. \citet{yassin2021-jmp} generalizes the orthogonality condition for a variety of boundary conditions.
Recent work has extended the normal mode decomposition to the shallow-water equations on the sphere including theoretical advances to include variable stratification \cite{zagar2020-book,vasy2021-qjrms}. In this case, the geostrophic mode is now interpreted as a Rossby mode, and the method has been applied both numerically and diagnostically. As precursor to the work in this manuscript, \citet{early2021-jfm} extended prior wave-vortex decompositions to include non-hydrostatics with variable stratification, although without explicit derivation of the projection operators necessary to recast the non-hydrostatic equations of motion.
% cite: \cite{early2021-jfm} citeA: \citet{early2021-jfm} author year: \citeauthorNP{early2021-jfm} \citeyear{early2021-jfm}.

% Task
The purpose of this paper is to develop a method for precisely measuring the energy fluxes, including the inertial cascades and transfers, in a non-hydrostatic model with realistic stratification and forcing from a mean flow, winds, and tide. We define waves as the flow component with no Eulerian signature of available potential vorticity and decompose the flow according to the linear available potential vorticity. The result is a complete decomposition of the fluid flow where each mode satisfies quadratic energy and potential enstrophy orthogonality, the lowest order approximation of conserved nonlinear quantities. With this method we map the total energy flow for our semi-realistic simulations, and compare them to other similar results.
%We take the opinionated view that waves have no Eulerian signature of available potential vorticity and construct a method that decomposes the flow according its linear approximation.

\subsection{Outline}
\label{sec:outline}

For this work we consider how wave forcing shapes the geostrophic field. Wave forcing may introduce some form of `stimulated loss of balance', either causing a direct transfer of geostrophic energy to the wave field, or causing a forward cascade of energy within the geostrophic reservoir toward smaller scales. To demonstrate our approach, we contrast simulations for two primary configurations, the first with mean-flow forcing only and the second with additional near-inertial and tidal forcing, resulting in simulations that are similar in design to \citet{taylor2016-jpo} and \citet{barkan2017-jpo}. Although a fair amount of theoretical work is required to analyze the simulations, we begin section~\ref{sec:simulations} by describing the numerical setup.

A key feature of our decomposition is its relationship to available energy and available potential vorticity (APV), two conserved quantities of the nonlinear system. Section~\ref{sec:exact-conservation-laws} presents what we claim is the `best' definition for APV  in variable stratification on an $f$-plane, which matches the intention of \citet{wagner2015-jfm}, but in closed form as described in \citet{early2022-arxiv}. Here we derive, for the first time, the exact closed-form expressions for available energy and available potential enstrophy conservation in a forced, rotating non-hydrostatic fluid with variable stratification.

The advantage of using \emph{available} energy and potential vorticity is that they linearize directly to the corresponding quantities derived from the linearized equations of motion, as shown in section~\ref{sec:linearization}. 
% Because of this construction, it will be possible to directly compare the exact nonlinear quantities with their lower-order counterparts used in the decomposition and flux calculations. 
The complete energetically and enstrophically orthogonal solution set and associated projection operators are summarized in Table~\ref{tab:solutions} for the wave and geostrophic solutions, Table~\ref{tab:vertical-mode-projection} for the vertical mode equations and Table~\ref{tab:solution-projection} for the projection operators. Additional details are provided in appendix~\ref{sec:orthogonal-solutions} and in \citet{early2024-arxiv}.

With this new framework, the equations of motion are rewritten in terms of the wave-vortex modes (section~\ref{sec:energy-fluxes}) and we derive the triad interactions which govern the cascades within the reservoirs and transfers between them. Our results are discussed in section~\ref{sec:results}. Finally, in section~\ref{sec:discussion} we utilize the relationship between APV and the linearized decomposition to discuss the errors that arise. The manuscript contains both the derivation of the decomposition from conserved quantities (partially relegated to Appendix~\ref{sec:orthogonal-solutions}) and its application to simulations. The focus is on establishing the method implied by the theoretical framework, leaving more comprehensive physical analysis to future work.

\begin{figure}
    \centering
    \includegraphics[width=\linewidth]{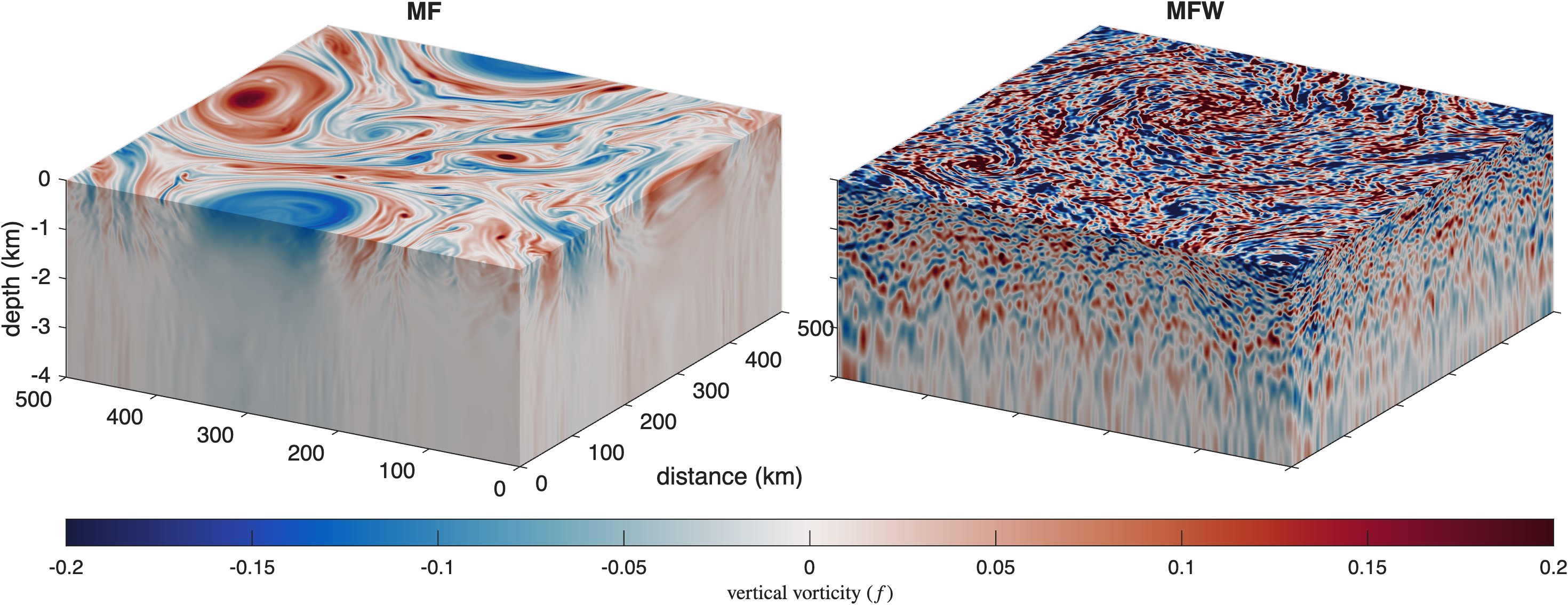}
    \caption{Vertical component of relative vorticity $\zeta$ for mean flow forcing only (MF, left) and mean flow \& wave forcing (MFW, right) simulations under steady-state conditions.}
    \label{fig:two_sim_comparison}
\end{figure}

%%%%%%%%%%%%%%%%%%%%%%%%
%
\section{Numerical experiments}
\label{sec:simulations}
%
%%%%%%%%%%%%%%%%%%%%%%%%

% Equations of motion, friction and damping
Numerical experiments are carried out using a pseudo-spectral code that solves the forced, nonlinear, non-hydrostatic, Boussinesq equations:
\begin{subequations}
\label{eqn:boussinesq}
\begin{align}
\label{x-momentum}
 \frac{du}{dt} - f v   =& - \frac{1}{\rho_0} \partial_x p_\textrm{tot} + \mathcal{S}_u \\ \label{y-momentum}
 \frac{dv}{dt} + f u  =&  - \frac{1}{\rho_0}\partial_y p_\textrm{tot} + \mathcal{S}_v   \\ \label{z-momentum}
 \frac{dw}{dt}  =& - \frac{1}{\rho_0}\partial_z p_\textrm{tot} -\frac{1}{\rho_0}g \rho_\textrm{tot} + \mathcal{S}_w  \\ \label{thermodynamic}
\frac{d \rho_\textrm{tot}}{dt}=& \mathcal{S}_\rho  \\ 
\label{continuity}
\partial_x u + \partial_y v + \partial_z w =& 0, 
\end{align}
\end{subequations}
where $\vect{u} = (u,v,w)$ is the fluid velocity in Cartesian coordinates with position vector $\vect{x} = (x,y,z)$, $p_\textrm{tot}$ is the total pressure, $\rho_\textrm{tot}$ is the total density, $f$ is the constant Coriolis parameter, and $\frac{d}{dt}$ is the material derivative. The momentum and thermodynamic equations are forced by $\mathcal{S}_\vect{u}$ and $\mathcal{S}_\rho$, respectively. The domain is 
$[0,L_x]\times [0,L_y] \times [-D,0]$ with periodic boundaries in $(x,y)$, rigid, free-slip ($w=0$) and flat at $z=0$ and $z=-D$. Reference density $\rho_0 \equiv \rho_\textrm{tot}(z=0)$ is taken to be constant. The boundary conditions do not incorporate topography or surface buoyancy anomalies. 
The pseudo-spectral code uses Fourier modes in the horizontal and the complete, orthogonal vertical basis described in \S~\ref{sec:linearization} and Appendix~\ref{sec:orthogonal-solutions}.
All simulations include quadratic bottom friction with drag coefficient $C_d=10^{-3}$, vertical diffusivity of $\kappa_z=10^{-5}\,\mathrm{m^2\,s^{-1}}$, anti-aliasing, and adaptive spectral vanishing viscosity to maintain numerical stability.  

% Domain and geostrophic forcing
We present two numerical simulations designed to capture the interaction between internal wave and geostrophically balanced motions.
The numerical setup is designed to represent semi-realistic forcing conditions found throughout the world oceans within the limitations of the model geometry. Domain dimensions are 500 km $\times$ 500 km in the horizontal and 4000 m in the vertical. Stratification is exponential $N^2(z) = N_0^2 \exp( 2 z / b)$ with canonical Garrett-Munk parameters $N_0 = 3$ cycles per day and scale depth $b=1300\,\mathrm{m}$. 
The model domain and relative vorticity fields are shown in Figure~\ref{fig:two_sim_comparison}.

% Initial sping-up from rest, MF and MFW simulations
The model is initially spun up from rest with only geostrophic dynamics and mean-flow forcing at a resolution of $256^2 \times 43$ for 30,000 days. This forcing maintains a zonal mean flow $U(y) = u_0 \sin(10 \pi y/L_x)$ with $u_0=0.0065\,\mathrm{m\,s^{-1}}$ at the surface, decaying to zero at $z=-D$. This preliminary spin-up provides the initial condition for the two simulations analyzed. These two simulations restart with fully nonlinear and non-hydrostatic dynamics. The first simulation, called MF, maintains the same mean-flow forcing $U$. The second simulation, called MFW, has mean flow and wave forcing. Wave forcing consists of near-inertial oscillation forcing of waves with frequencies near $f$ and semi-diurnal ($M_2$) tidal forcing for a narrow band of waves around the local $M_2$ tidal frequency. The wave forcing restores the amplitude of waves in the selected frequency bands to levels set by the \citet{garrett1972-gfd} spectrum.
Importantly, the amplitude of the forced modes are held fixed, rather than the flux. This allows the system to draw energy from the forced mode at a rate determined by the nonlinear model dynamics.

\begin{figure}
    \centering
    \includegraphics[width=0.8\linewidth]{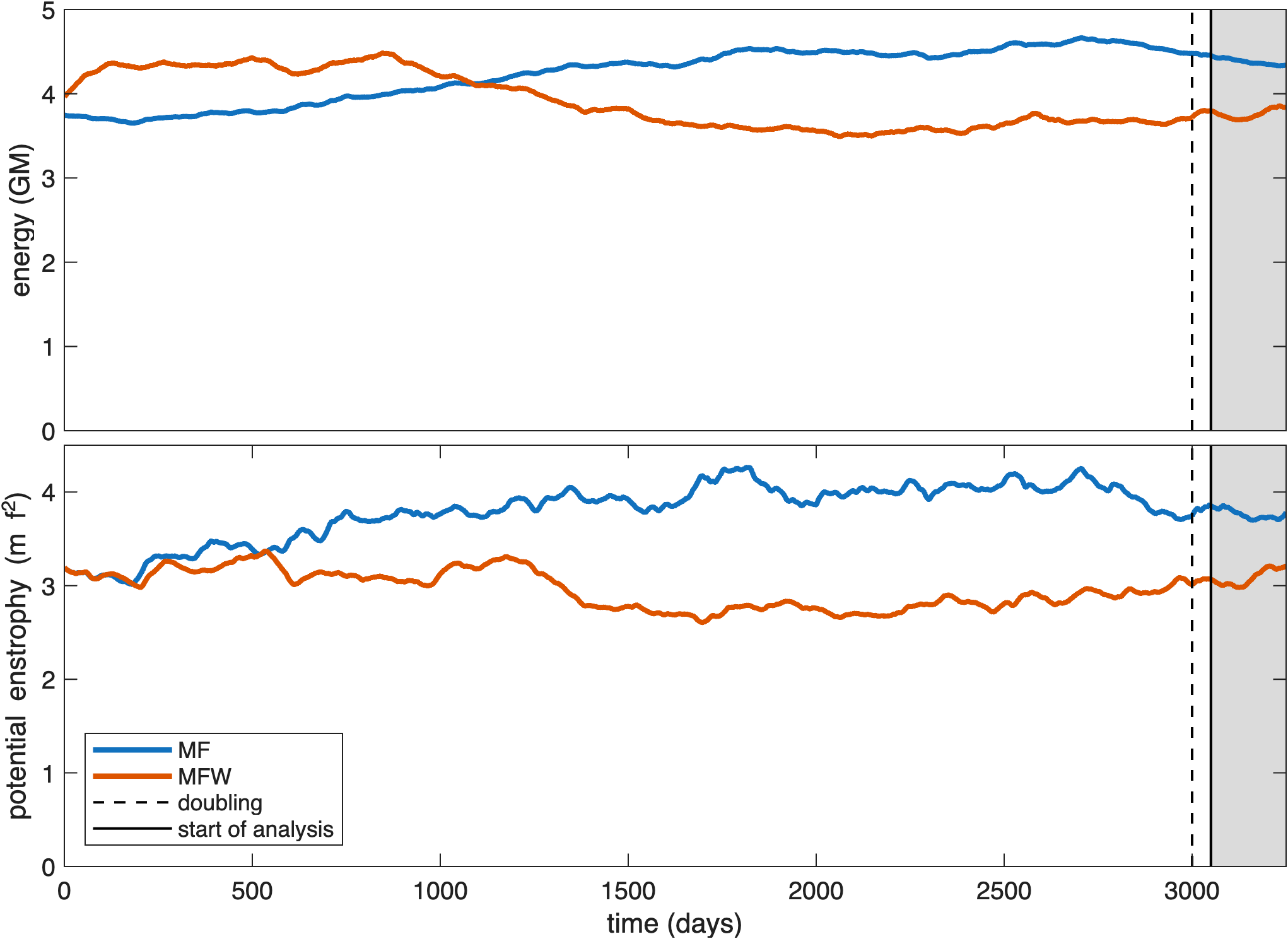}
    \caption{The area-averaged depth-integrated energy (top) and enstrophy (bottom) time series for the two simulations. The initial spin-up period at $256^2 \times 43$ resolution is 3000 days, at which time the resolutions are doubled to $512^2 \times 86$, as indicated by the dashed vertical line. The gray box highlights the period of steady-state analysis, from day 3050--3250.}
    \label{fig:energy_time_series}
\end{figure}

% Resolution
After the simulations reach an approximate steady-state (3000 days, indicated by a dashed line in Figure~\ref{fig:energy_time_series}), the resolution is doubled from $256^2 \times 43$ to $512^2 \times 86$ and the simulations continue for an additional 250 days. This doubling results in a nominal resolution of 975 m in the horizontal and 15 m between the smallest two points on the stretched grid in the vertical. Due to spectral de-aliasing, the effective resolution is somewhat less. The analysis period covers the final 200 days of the high-resolution simulation, as shown in Figure~\ref{fig:energy_time_series}. The units of energy in Figure~\ref{fig:energy_time_series} (and others) are expressed as area-averaged, depth-integrated per unit density, which is m$^3$ s$^{-2}$ in SI units. In the internal wave literature 3.7 m$^3$ s$^{-2}$ is often expressed as `1 GM', as this is the typical energy level of the oceanic internal gravity wave field established by Garrett \& Munk in the 1970s (see also Figure~7 of \citet{leboyer2021-jpo}). Energy fluxes (m$^3$ s$^{-3}$ in SI units) are expressed in GM/yr.

\begin{figure}
    \centering
    \includegraphics[width=1.0\linewidth]{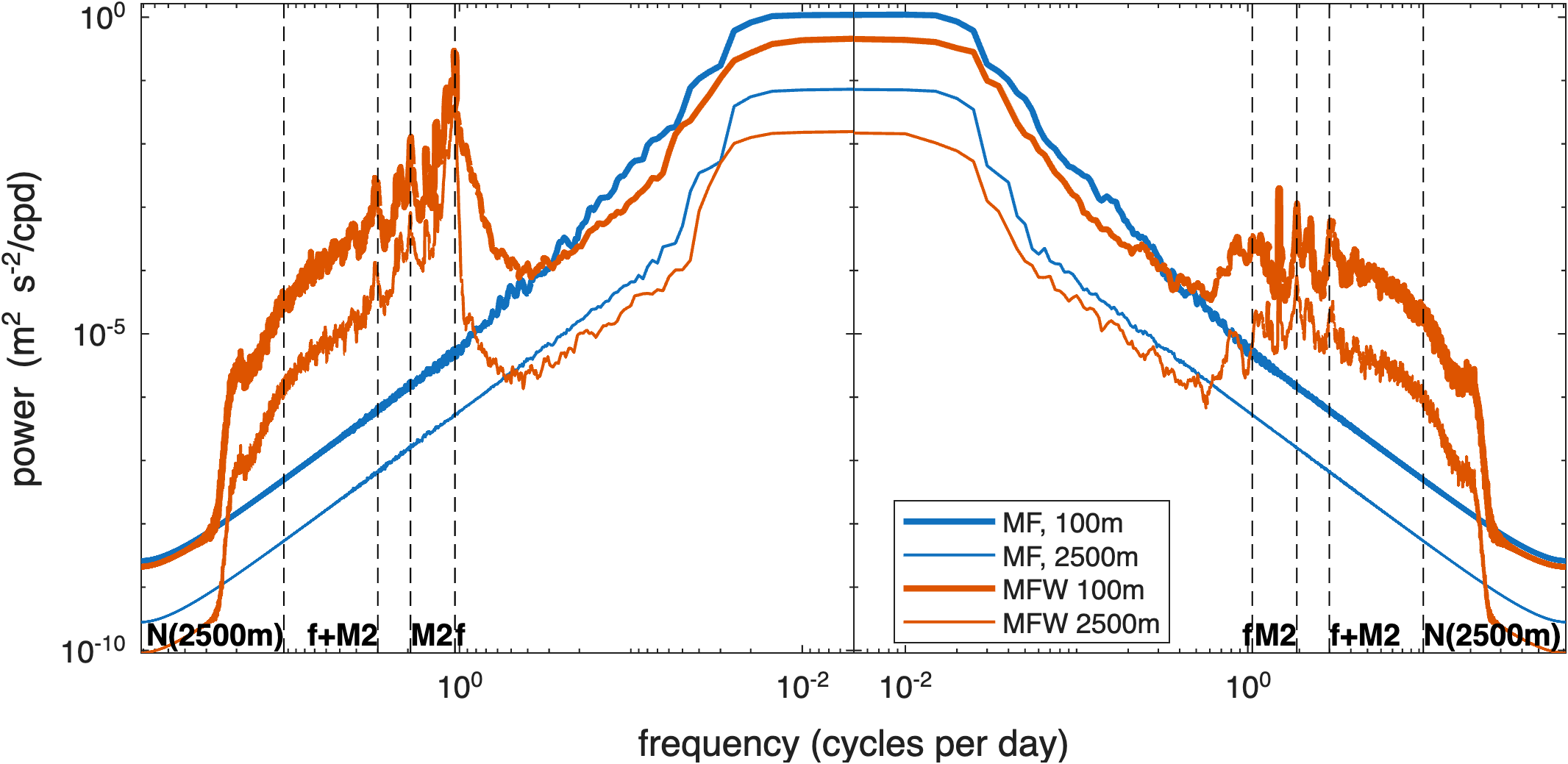}
    \caption{Rotary spectrum of the horizontal velocity field $u + i v$ at a synthetic mooring in the simulations.}
    \label{fig:mooring}
\end{figure}

% Preliminary description
Qualitatively and quantitatively, the MF and MFW simulations show distinct characteristics once steady-state is achieved, as seen in Figures~\ref{fig:two_sim_comparison} and \ref{fig:mooring}.
The MF simulation visually resembles a quasigeostrophic simulation, despite the non-hydrostatic dynamics, while the addition of wave forcing produces much smaller scale features in relative vorticity (Figure~\ref{fig:two_sim_comparison}).
Figure~\ref{fig:mooring} shows the rotary spectra from a synthetic mooring in both simulations, computed from the complex velocity $u+i v$ following \citet{gonella1972-dsr} and \citet{jlab2024}. The rotary spectrum highlights the distinct signatures of internal gravity waves in MFW, with clear differences between cyclonic (positive frequency) and anticyclonic (negative frequency) rotations. Both simulations show the characteristic low-frequency peak of geostrophic motions. The MF simulation shows evidence for the emergence of a (very) weak wave field in the $256^2 \times 43$ simulation (not shown) from spontaneous generation, which disappears over the course of the 3250-day experiment \cite{vanneste2013-arfm}.

 The MF simulation has a root-mean square (rms) Rossby number, $(\partial_x v - \partial_y u)/f$, of 0.032, with minimum and maximum of $-0.24$ and $0.24$, respectively, while the MFW simulation  has rms Rossby number of $0.054$, with minimum and maximum of $-0.41$ and $0.43$ respectively. These values are typical of mid-ocean conditions \cite{chelton2011-pio,wunsch2024-po}. The rotary spectra from the model mooring in Figure~\ref{fig:mooring} can be compared with the mooring spectra in \citet{leboyer2021-jpo} and show that both the low-frequency and high-frequency motions are also consistent with mid-ocean values. 
 % Thus, any conclusions drawn from these simulations are broadly applicable.

%%%%%%%%%%%%%%%%%%%%%%%%
%
\section{Nonlinear energy and potential enstrophy conservation}
\label{sec:exact-conservation-laws}
%
%%%%%%%%%%%%%%%%%%%%%%%%

% Before separating the fluxes into wave and geostrophic components, we  compute the exact energy and potential enstrophy fluxes in the Boussinesq system. The goal is to derive exact, closed form expressions for energy and potential enstrophy conservation in terms of the forcing in \eqref{eqn:boussinesq}. The area-averaged, depth-integrated energy and energy fluxes from this expression are then used to construct \emph{some} of the source-reservoir-sink values in Figure~\ref{fig:sources_sinks}, indicated using square brackets $[ \; ]$ in the diagram. The decomposition into wave and geostrophic components will be explained later.

Before separating the fluxes into wave and geostrophic components, we compute the exact energy and potential enstrophy fluxes for the Boussinesq system. These closed-form conservation statements, expressed in terms of the forcing in  \eqref{eqn:boussinesq}, provide the area-averaged, depth-integrated energy and flux values shown in brackets $[ \; ]$ in Figure~\ref{fig:sources_sinks}. The decomposition itself is presented later.

%%%%%%%%%%%%%%%%%%%%%%%%
\subsection{The no-motion and flattened isopycnal states}
\label{subsec:nomotionsoln}
%%%%%%%%%%%%%%%%%%%%%%%%

The most trivial, but important, solution to the equations of motion \eqref{eqn:boussinesq} is the \textit{no-motion} state obtained by setting $\vect{u}(\vect{x},t) =(0,0,0)$. Achieving this state requires adiabatically rearranging the fluid density $\rho_\textrm{tot}(\vect{x},t)$ to remove horizontal pressure and density gradients by moving parcels to their no-motion heights. The no-motion solution is denoted as $(u,v,w,p,\rho)=\left(0,0,0,p_\textrm{nm}(z),\rho_\textrm{nm}(z) \right)$ where $p_\textrm{nm}$ is defined as
\begin{equation}
\label{nm-sol}
     p_\textrm{nm}(z) 
     = -g \int_0^z  \rho_\textrm{nm}(\xi) d\xi,
\end{equation}
so that $\partial_z p_\textrm{nm}(z) \equiv -g \rho_\textrm{nm}(z)$ and the equations of motion are satisfied.

A fully general energy budget relative to the no-motion density would include time dependence, allowing diapycnal mixing to modify the background state. Although some diapycnal mixing occurs in our simulations, we restrict the analysis to 200 days (Figure~\ref{fig:energy_time_series}) and therefore neglect its small effects on the energy budget. The no-motion density differs from the mean density by a `mean density anomaly' (mda), a function of $z$ that may be small or large (e.g., in an inverted fluid). This quantity is discussed further in section~\ref{sec:mda-solution}.

% A more general formulation of the energy budget relative to the no-motion density would include time dependence, thereby allowing diapycnal mixing to alter the background state. Although the simulations considered here do have some diapycnal mixing, we restrict the analysis to only 200 days (see Figure~\ref{fig:energy_time_series}) and thus ignore the relatively small effects of diapycnal mixing on the energy budget.

% Importantly, the no-motion density is not the same as the mean density. These two quantities differ by a `mean density anomaly' (mda).  The mda is a function of $z$. It may be small or large ({\it e.g.}, for an inverted fluid) and is further discussed in section~\ref{sec:mda-solution}.

% The total density is often partitioned into the the no-motion (or mean)\begin{color}{red} this parenthetical is now confusing?\end{color} and the excess (or perturbation) density $\rho_\textrm{e}$, i.e.,  \begin{color}{red} Replace by ?\end{color}

The total density is partitioned into $\rho_\textrm{nm}$ and an excess density $\rho_\textrm{e}$ (analogous to mean and perturbation), 
\begin{equation}
    \rho_\textrm{tot}(\vect{x},t) \equiv   \rho_\textrm{nm}(z) + \rho_\textrm{e}(\vect{x},t).
\end{equation}
Alternatively, following \citet{holliday1981-jfm}, the total density can be written in terms of the vertical displacement of each parcel $\eta(\vect{x},t)$, 
\begin{equation}
\label{def:eta}
    \rho_\textrm{tot}(\vect{x},t) \equiv  \rho_\textrm{nm}(z - {\etat}(\vect{x},t)).
\end{equation}
Using this definition, the thermodynamic equation \eqref{thermodynamic} becomes
\begin{equation}
\label{thermodynamic-eta1}
    \frac{d}{dt} \left( z - \eta \right) = \frac{1}{\rho^\prime_\textrm{nm}(z-\eta)} \mathcal{S}_\rho.
\end{equation}
In the absence of forcing, $z-\eta$ is therefore conserved, just like total density.

%%%%%%%%%%%%%%%%%%%%%%%%
\subsection{Energy conservation with forcing}
\label{subsec:energy-conservation}
%%%%%%%%%%%%%%%%%%%%%%%%

To derive an energy conservation statement, we multiply the thermodynamic equation \eqref{thermodynamic} by $g \eta$ and use \eqref{nm-sol} together with \eqref{def:eta}-\eqref{thermodynamic-eta1} to show that
\begin{equation}
\label{quadratic-thermo}
    \frac{d}{dt} \left[ g \eta \rho_\textrm{tot} -  p_\textrm{nm}(z-\eta) \right] = g w \rho_\textrm{tot} + g \eta \mathcal{S}_\rho,
\end{equation}
where 
\begin{equation}
    \frac{d }{dt} p_\textrm{nm}(z-\eta) = -g \frac{\rho_\textrm{nm}(z-\eta)}{\rho^\prime_\textrm{nm}(z-\eta)} \mathcal{S}_\rho
\end{equation}
%which 
follows from \eqref{nm-sol} and \eqref{thermodynamic-eta1}. The total energy equation is now formed in the usual way by taking the inner product of the momentum equations \eqref{x-momentum}-\eqref{z-momentum} with $\vect{u}$ and summing, such that
\begin{equation}
\label{total-energy-lorenz}
     \frac{d}{dt} \left( \frac{1}{2} \vect{u}^2 + \frac{g}{\rho_0} \eta \rho_\textrm{nm}(z - \eta) -  \frac{1}{\rho_0} p_\textrm{nm}(z-\eta) \right) = - \frac{1}{\rho_0} \vect{u} \cdot \nabla p_\textrm{tot} +\vect{u} \cdot \vect{\mathcal{S}}_\vect{u} + \frac{1}{\rho_0} g \eta \mathcal{S}_\rho.
\end{equation}
The term $g \eta \rho_\textrm{nm}(z-\eta)/\rho_0$ in \eqref{total-energy-lorenz} is the available potential energy (APE) of \citet{lorenz1955-tellus}, while the $p_\textrm{nm}(z - \eta)$ term arises as a consequence of buoyancy forcing. Thus, Casimir invariance is lost, meaning alternative reference states are no longer permitted. Only when buoyancy forcing is absent does the total derivative of  $p_\textrm{nm}(z-\eta)$ in \eqref{quadratic-thermo} vanish, recovering the familiar case \cite{shepherd1993-ao,roullet2008-jfm}.

 To construct an energy conservation statement in terms of the APE of \citet{holliday1981-jfm}, we define the excess (perturbation) pressure $p_\textrm{e}$ relative to the no-motion pressure \eqref{nm-sol},
\begin{equation}
    p_\textrm{tot}(\mathbf{x},t) = p_\textrm{nm}(z) + p_\textrm{e}(\mathbf{x},t).
    \label{def:p_e}
\end{equation}
Using \eqref{def:p_e}, \eqref{total-energy-lorenz} may be rewritten as
\begin{equation}
\label{total-energy-lorenz-pre}
     \frac{d}{dt} \left( \frac{1}{2} \vect{u}^2 + \frac{g}{\rho_0} \eta \rho_\textrm{nm}(z - \eta) -  \frac{1}{\rho_0} p_\textrm{nm}(z-\eta) + \frac{1}{\rho_0} p_\textrm{nm}(z) \right) = - \frac{1}{\rho_0} \vect{u} \cdot \nabla p_\textrm{e} +\vect{u} \cdot \vect{\mathcal{S}}_\vect{u}+ \frac{1}{\rho_0} g \eta \mathcal{S}_\rho.
\end{equation}
by simply moving the $p_\textrm{nm}(z)$ to the right-hand side. This expression is alternatively written as,
\begin{equation}
\label{total-energy-hm}
     \frac{d}{dt} \left( \frac{1}{2} \vect{u}^2 - \frac{1}{\rho_0}\int_0^\etat g \xi \partial \rho_\textrm{nm}(z- \xi) d \xi \right) = - \frac{1}{\rho_0} \vect{u} \cdot \nabla p_\textrm{e} +\vect{u} \cdot \vect{\mathcal{S}}_\vect{u} + \frac{1}{\rho_0} g \eta \mathcal{S}_\rho.
\end{equation}
using the transformation noted in \citet{holliday1981-jfm}. Using the definitions,
\begin{equation}
    \label{eqn:KE-APE-definition}
    \textrm{KE} \equiv \rho_0 \frac{1}{2} \vect{u}^2,  \quad \textrm{APE} \equiv - \int_0^\etat g \xi \partial \rho_\textrm{nm}(z- \xi) d \xi ,
\end{equation}
the volume integral of \eqref{total-energy-hm} reduces to,
\begin{equation}
\label{eqn:volume-integrated-exact-energy}
    \partial_t \int \left( \textrm{KE} + 
    \textrm{APE} \right) dV =  \int \left( \vect{u} \cdot \vect{\mathcal{S}}_\vect{u} + \frac{1}{\rho_0} g \eta \mathcal{S}_\rho \right) dV, 
\end{equation}
where the right-hand-side defines the energy flux from the forcing.

The expression in \eqref{total-energy-hm} offers an exact, closed-form energy conservation law similar to \citet{tailleux2018-jfm}, but with two key differences: buoyancy forcing $\mathcal{S}_\rho$ removes the reference state invariance, and the result is expressed in terms of vertical displacement $\eta$. The latter is essential to our approach, which will linearize the equations of motion and conserved quantities about $\eta$ in section~\ref{sec:linearization}. This derivation also directly links the point-wise \eqref{total-energy-hm} and volume-integrated \eqref{eqn:volume-integrated-exact-energy} conservation statements sought after in \citet{winter2013-jfm} where they use $z_\ast$ in place of $z-\eta$. Constructing $\rho_\textrm{nm}(z)$ requires sorting densities $\rho_i$ and iteratively determining the corresponding $z_i$ satisfying $\rho_i = \rho_\textrm{nm}(z_i)$, after which $\eta$ follows from \eqref{def:eta} using the bisection method. APE is then efficiently computed to high-precision from the three terms in \eqref{total-energy-lorenz-pre}.

% Using \eqref{eqn:volume-integrated-exact-energy} and Figure~\ref{fig:energy_time_series}, we are now able to explain exact time-averaged total energies and source-sink fluxes for the HS-G and NHS-GW numerical experiments, as presented in Figure~\ref{fig:sources_sinks}.  By comparing figures \ref{fig:energy_time_series} and \ref{fig:sources_sinks}, one can see that average values of the total energy in the analysis window match with the bracketed number underneath the middle-left axis-label `Reservoirs.'  Reservoir energies that contain a parenthetical change in energy indicate the deviation from steady-state over that time period, which must be considered when closing energy budgets.

\begin{figure}[t]
\centering
\includegraphics[height=6cm]{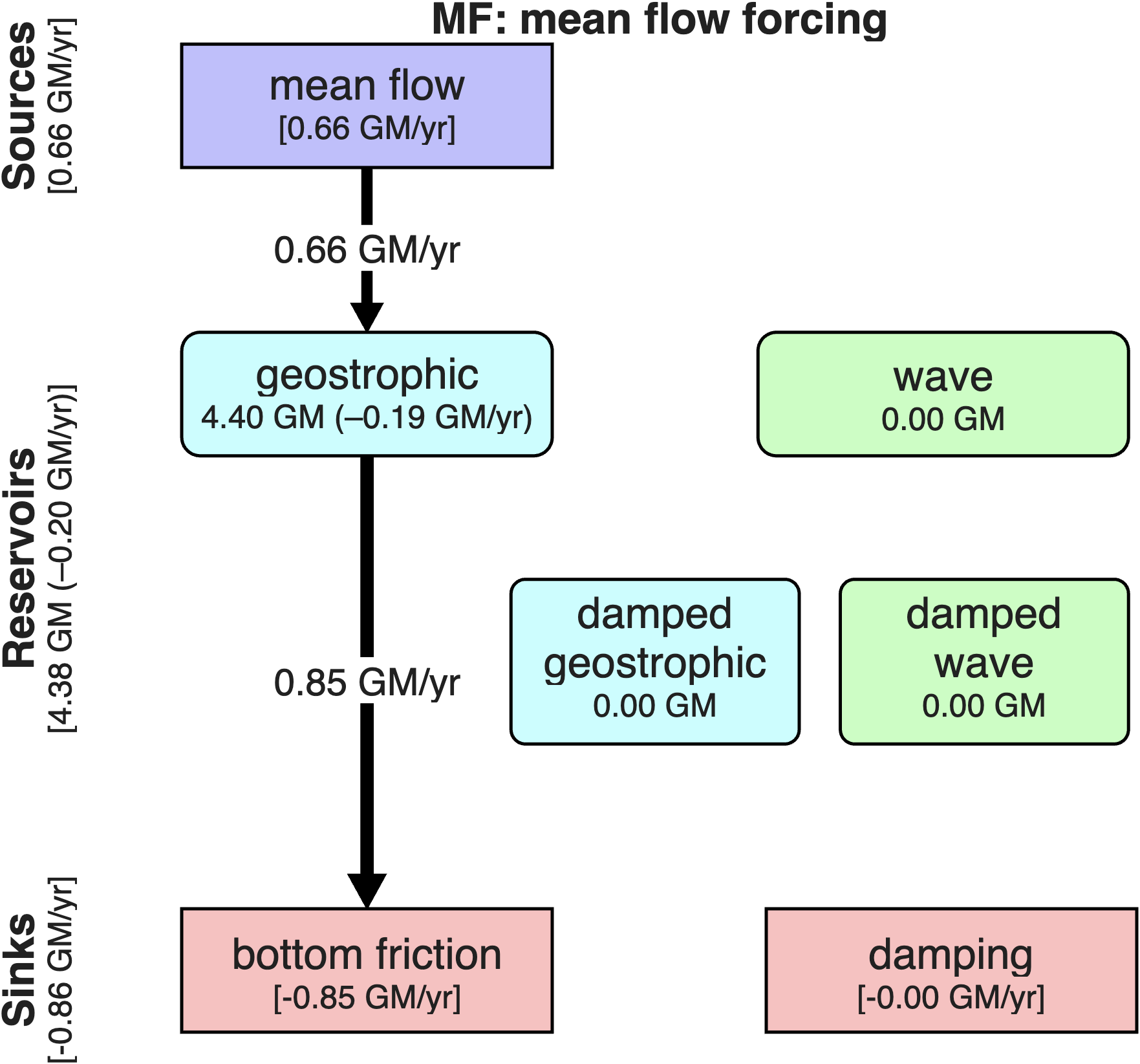}
% \hspace{0.5cm}
% \includegraphics[height=4.5cm]{figures/sources_sinks_run9.png}
% \hspace{0.5cm}
\hfill
\includegraphics[height=6cm]{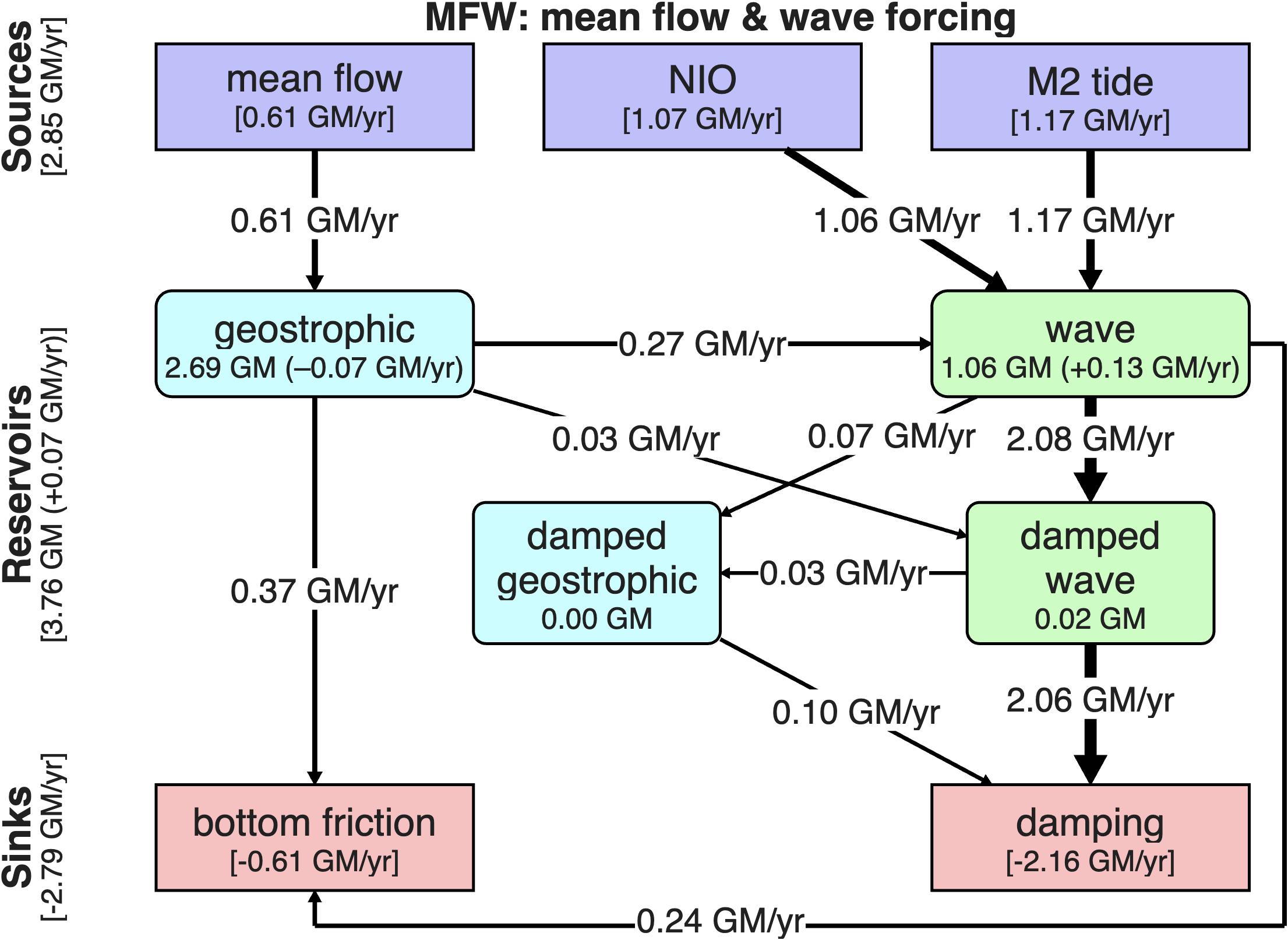}
\caption{Energy sources, sinks and reservoirs for the MF and MFW simulations. Energy and energy fluxes shown in brackets are computed from the exact expression \eqref{eqn:volume-integrated-exact-energy} and shown for each forcing and the total system energy. The energy and energy fluxes in and between the wave and geostrophic reservoirs are computed with the quadratic approximation and are shown without brackets. All values are averages over the analysis period and energy reservoirs include the gain or loss over the period in parentheses, which is required to close the energy flux budgets. Fluxes with less than $0.01$ GM/yr are not shown in the diagram. }
\label{fig:sources_sinks}
\end{figure}

Using \eqref{eqn:volume-integrated-exact-energy}, we are now able to explain exact time-averaged total energies and source-sink fluxes as presented in Figure~\ref{fig:sources_sinks}---the decomposition into wave and geostrophic motions at different scales is presented in section~\ref{subsec:closure}.  Exact total energy and energy flux are shown in brackets, and any time rate-of-change is indicated in parentheses. A \emph{source/sink} has net positive/negative flux. The rate-of-change must be considered when closing energy budgets. The sources, sinks and deviation from steady-state sum to zero up to the precision reported here using daily output of the 200 day analysis period. The total energy flux of the nonlinear advection term (which ideally should sum to zero) has a root-mean square value of 0.0039 GM/yr, and a mean value of 0.00011 GM/yr for the 200 day analysis period. However, the forcing fluxes have significant natural temporal variability such that the budgets close to a precision of 0.01 GM/yr with daily output---increased precision requires averaging over more frequent output. The vertical diffusivity is not shown in the figure as its energy flux is below the reported precision.

Ignoring the separation into wave and geostrophic reservoirs for the moment, the total energy and fluxes in Figure~\ref{fig:sources_sinks} show significant quantitative differences between the two simulations. First, the presence of wave forcing causes the total energy of the fluid to decrease, compared the MF simulation. Second, the energy flux from the mean flow also decreases in the presence of waves. Thirdly, all the energy flux from mean flow to the geostrophic reservoir in the MF simulation is removed through bottom friction, indicating an inverse cascade. In contrast, the MFW simulation shows significant energy removal by small-scale damping.

\begin{figure}
\includegraphics[height=8.0cm]{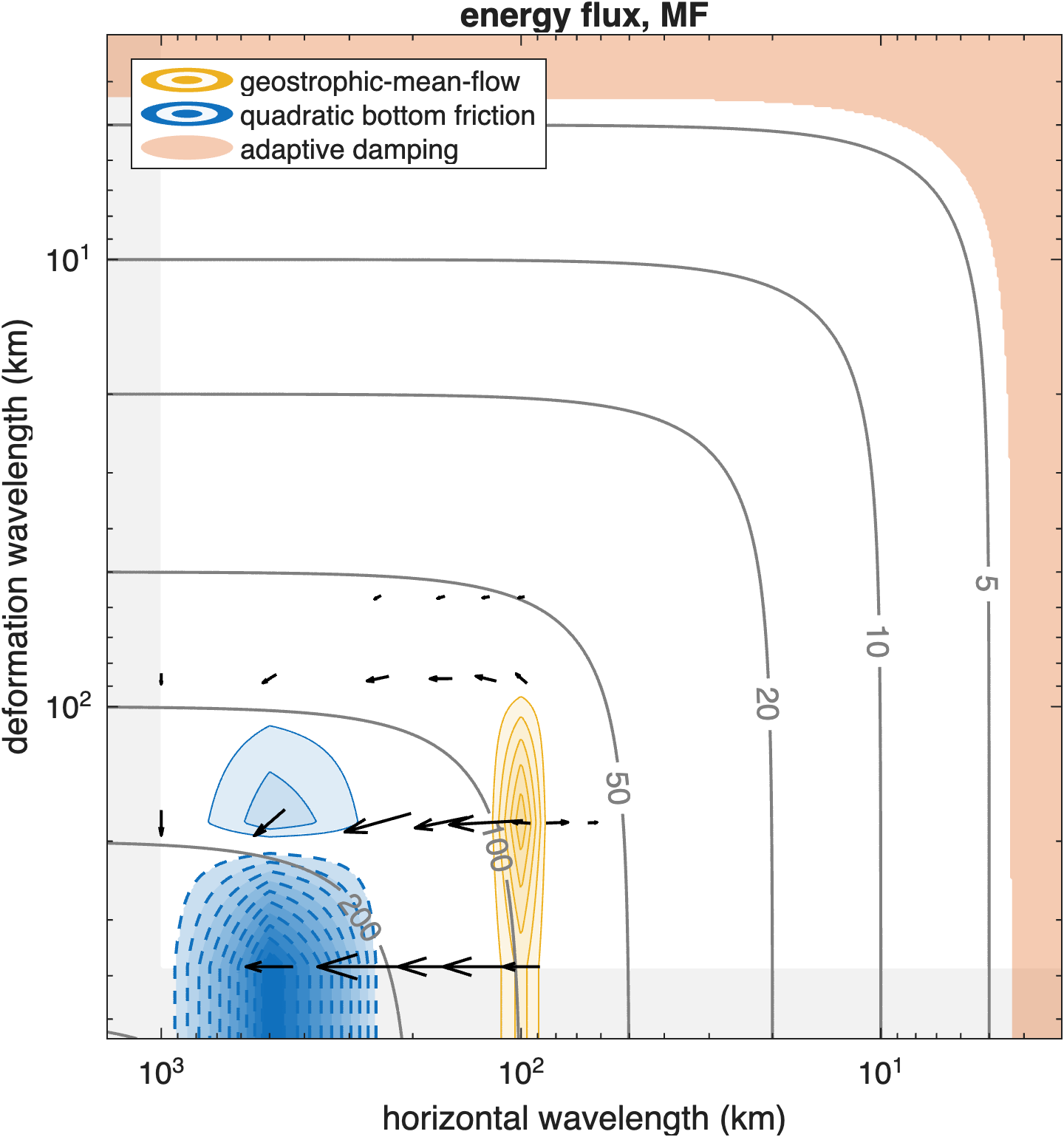}
\hfill
\includegraphics[height=8.0cm]{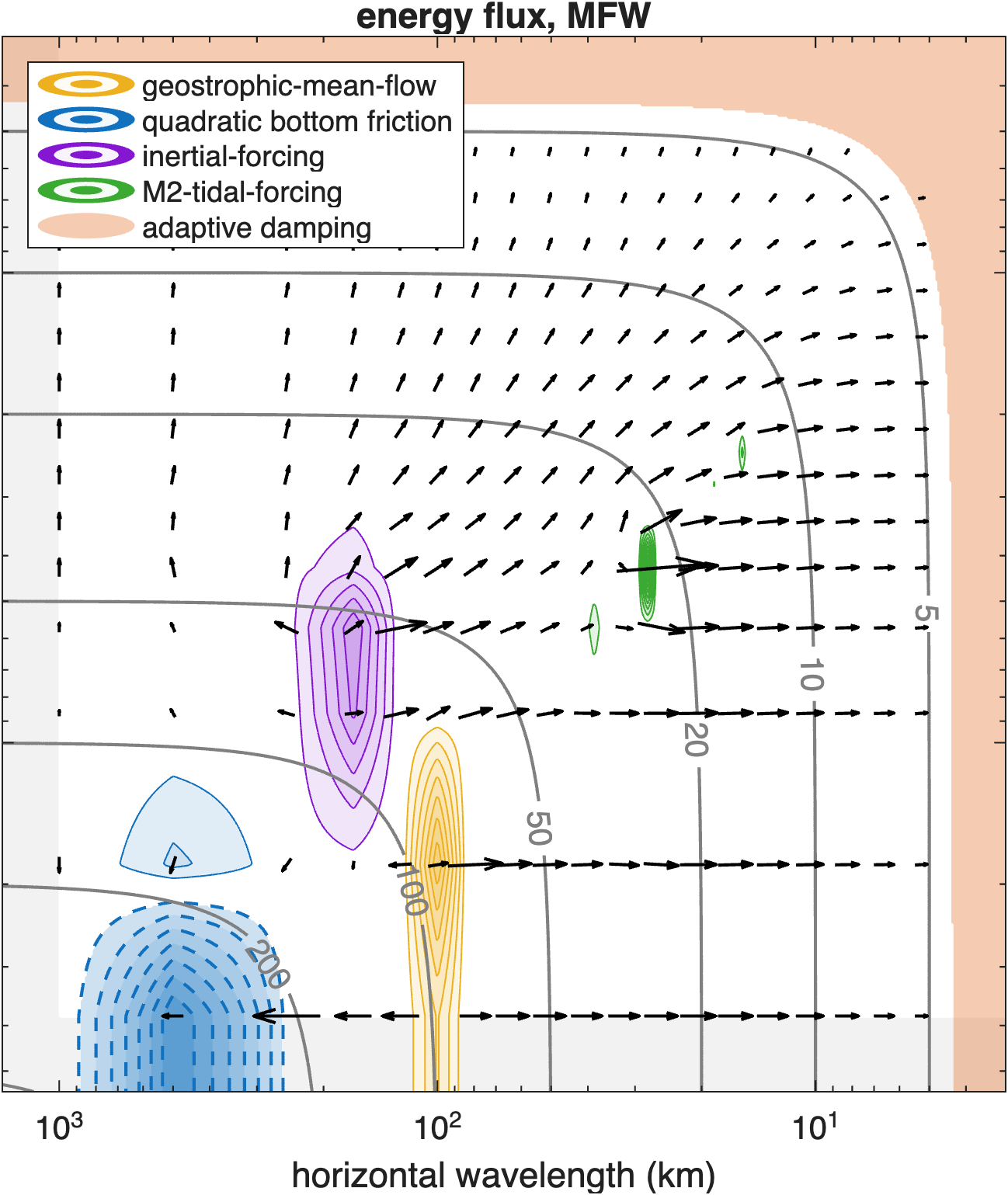}
\caption{Energy sources (solid contours), sinks (dashed contours), and advective flux (arrows) in the MF (left) and MFW (right) simulations.}
\label{fig:energy_flux_2D_flow}
\end{figure}

Figure~\ref{fig:energy_flux_2D_flow} helps to further visualize the flux of energy from the sources to the sinks, by computing the spectral density of the forcing fluxes using the spectrum defined in appendix~\ref{appendix:modal-spectra} and constructing a energy flux vector field with divergence matching the nonlinear flux term. The method will be discussed in detail in section~\ref{subsec:2d-fluxes}. The figures are drawn in the space of deformation wavelength (a measure of vertical scale) vs horizontal wavelength---integration of the flux spectral densities over the entire space results in the total flux values reported in Figure~\ref{fig:sources_sinks}, but some fluxes, most notably small scale damping, are spread out over so much space that the flux density amplitudes are too small to be shown.

The energy flux in the MF simulation in Figure~\ref{fig:energy_flux_2D_flow} matches what one would expected for quasigeostrophic dynamics, despite the non-hydrostatic dynamics considered here. Energy flows from mean flow forcing to larger scales, where it is mostly removed by the bottom friction, although the nonlinear nature of quadratic bottom friction causes a small portion energy flux at higher vertical mode. Adding the wave forcing weakens the flux from the mean-flow, as previously noted, but energy still fluxes to larger scale. The most notable difference is that the inertial and tidal forcing cause a significant flux of energy downscale. This picture will be expanded after the tools to separate the geostrophic and wave motions are developed below.

%%%%%%%%%%%%%%%%%%%%%%%%
\subsection{Available potential enstrophy conservation with forcing}
\label{subsec:enstrophy-conservation}
%%%%%%%%%%%%%%%%%%%%%%%%

Here we derive a novel form of potential enstrophy conservation, based on available potential vorticity (APV). The term APV was first used in \citet{wagner2015-jfm}, but later co-opted by \citet{early2022-arxiv} for a qualitatively similar quantity in closed form used here. In order to compute the potential enstrophy fluxes, we rederive the quantity defined in \citet{early2022-arxiv}, but now include the forcing terms.
%in order to construct an equation for potential enstrophy conservation. 

The derivation of APV conservation begins in the usual way, by taking the curl of the momentum equations \eqref{x-momentum}-\eqref{z-momentum} and forming an evolution equation for absolute vorticity $\vec{\omega}+ f \hat{\vect{z}}$, where $\vec{\omega} \equiv \nabla \times \vect{u}$ is the relative vorticity. APV follows from using $z - \eta$ as the conserved quantity, rather than $\rho_\textrm{tot}$ so we re-express \eqref{thermodynamic-eta1} as
\begin{equation}
\label{thermodynamic-eta2}
    \frac{d \eta}{dt} - w = \mathcal{S}_\eta
\end{equation}
by redefining the forcing as $\mathcal{S}_\eta = - \mathcal{S}_\rho/\rho^\prime_\textrm{nm}(z-\eta)$
for notational convenience. To form an evolution equation for APV then requires taking the dot product of the absolute vorticity equation with $\nabla (z - \eta)$, and adding that to the dot product of the gradient of the thermodynamic equation \eqref{thermodynamic-eta1}  with absolute vorticity $\vec{\omega} + f \hat{\vect{z}}$. The result is the expression,
\begin{equation}
\label{eqn:APVwithforcing}
    \frac{d}{d t} \textrm{APV} = - \left( \vec{\omega} + f \hat{\vect{z}} \right) \cdot \nabla \mathcal{S}_\eta +  \nabla (z-\eta) \cdot \left[ \nabla \times \mathcal{S}_\mathbf{u} \right]
\end{equation}
where
\begin{equation}
\label{eqn:APV-definition}
    \textrm{APV} \equiv \partial_x v - \partial_y u - f \partial_z \eta - \left( \nabla \times \vect{u} \right) \cdot \nabla \eta
\end{equation}
and the constant $f$ has been dropped, following the more rigorous treatment in \citet{early2024-arxiv}.

%Note that the linear forcing terms in \eqref{eqn:APVwithforcing} %are $\partial_x F_y - \partial_y F_x -  f \partial_z G_\eta$ 
% have the same functional form as QGPV forcing, which is why APV is such a useful quantity. \begin{color}{red}The last statement may not be obvious to all readers.  Can we expand a bit?\end{color} [

Available potential enstrophy is defined as,
\begin{equation}
\label{eqn:exact-available-potential-enstrophy-def}
    Z \equiv \frac{1}{2} \textrm{APV}^2,
\end{equation}
so that conservation of available potential enstrophy is given by,
\begin{equation}
\label{potential-enstrophy-conservation}
     \frac{d}{dt} Z = \textrm{APV} \cdot \left( \partial_x \mathcal{S}_v - \partial_y \mathcal{S}_u -  f \partial_z \mathcal{S}_\eta \right) + \textrm{APV} \cdot \left(- \vec{\omega} \cdot \nabla \mathcal{S}_\eta - \nabla \eta \cdot \left[ \nabla \times \mathcal{S}_\mathbf{u} \right] \right).
\end{equation}
The first term on the right-hand-side is approximately the leading order term in quasigeostrophic scaling, and the second term is the nonlinear correction required to make the expression exact.

The volume-averaged potential enstrophy fluxes follow from the right-hand side of the volume integral,
\begin{equation}
\label{volume-integrated-exact-enstrophy}
    \partial_t \int Z \, dV =  \int \left[ \textrm{APV} \cdot \left( \partial_x \mathcal{S}_v - \partial_y \mathcal{S}_u -  f \partial_z \mathcal{S}_\eta \right) + \textrm{APV} \cdot \left(- \vec{\omega} \cdot \nabla \mathcal{S}_\eta - \nabla \eta \cdot \left[ \nabla \times \mathcal{S}_\mathbf{u} \right] \right) \right] dV.
\end{equation}
Just as with the energy budget derived above, we have neglected temporal changes in the no-motion state.

\begin{figure}
\includegraphics[height=8.0cm]{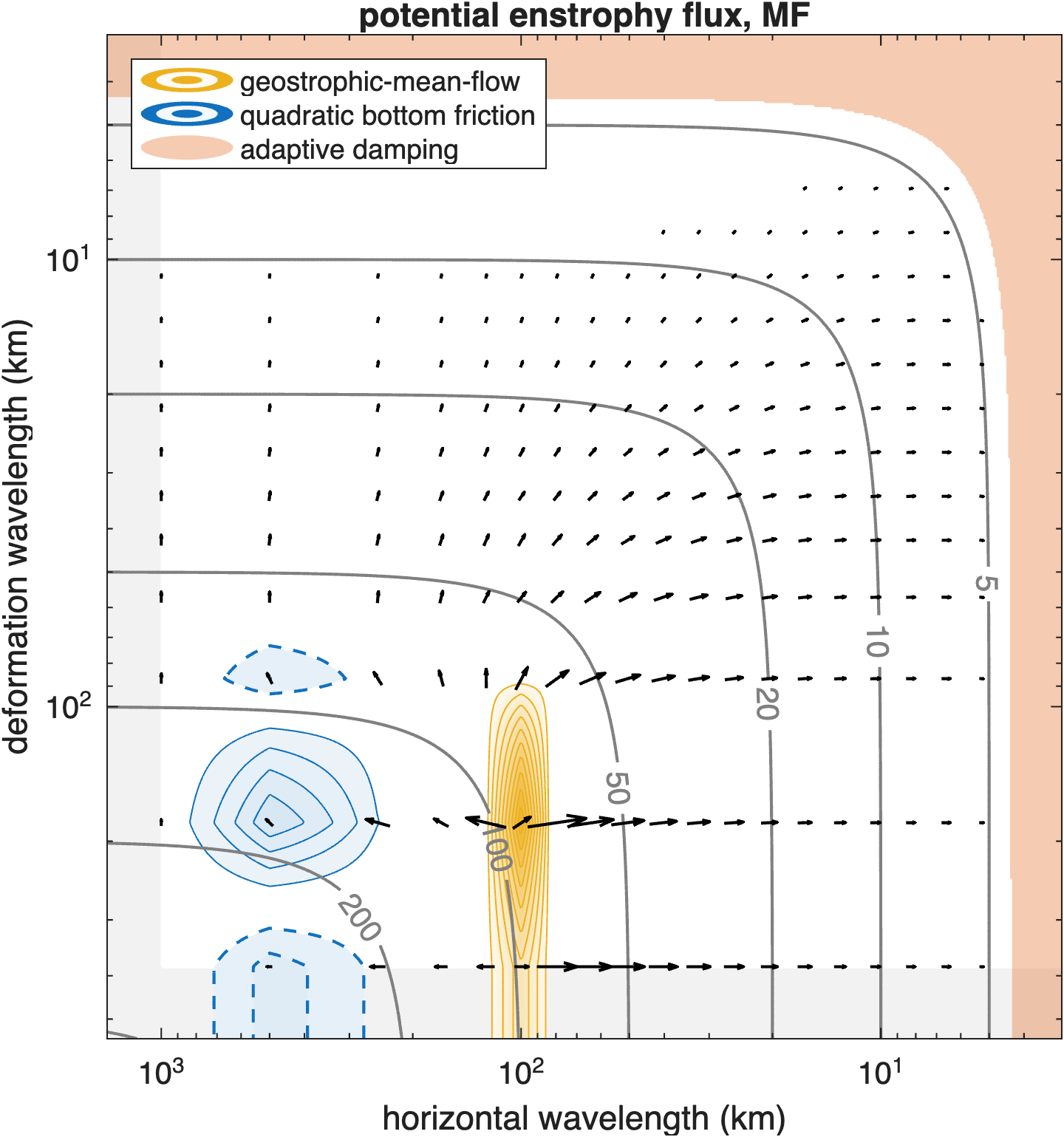}
\includegraphics[height=8.0cm]{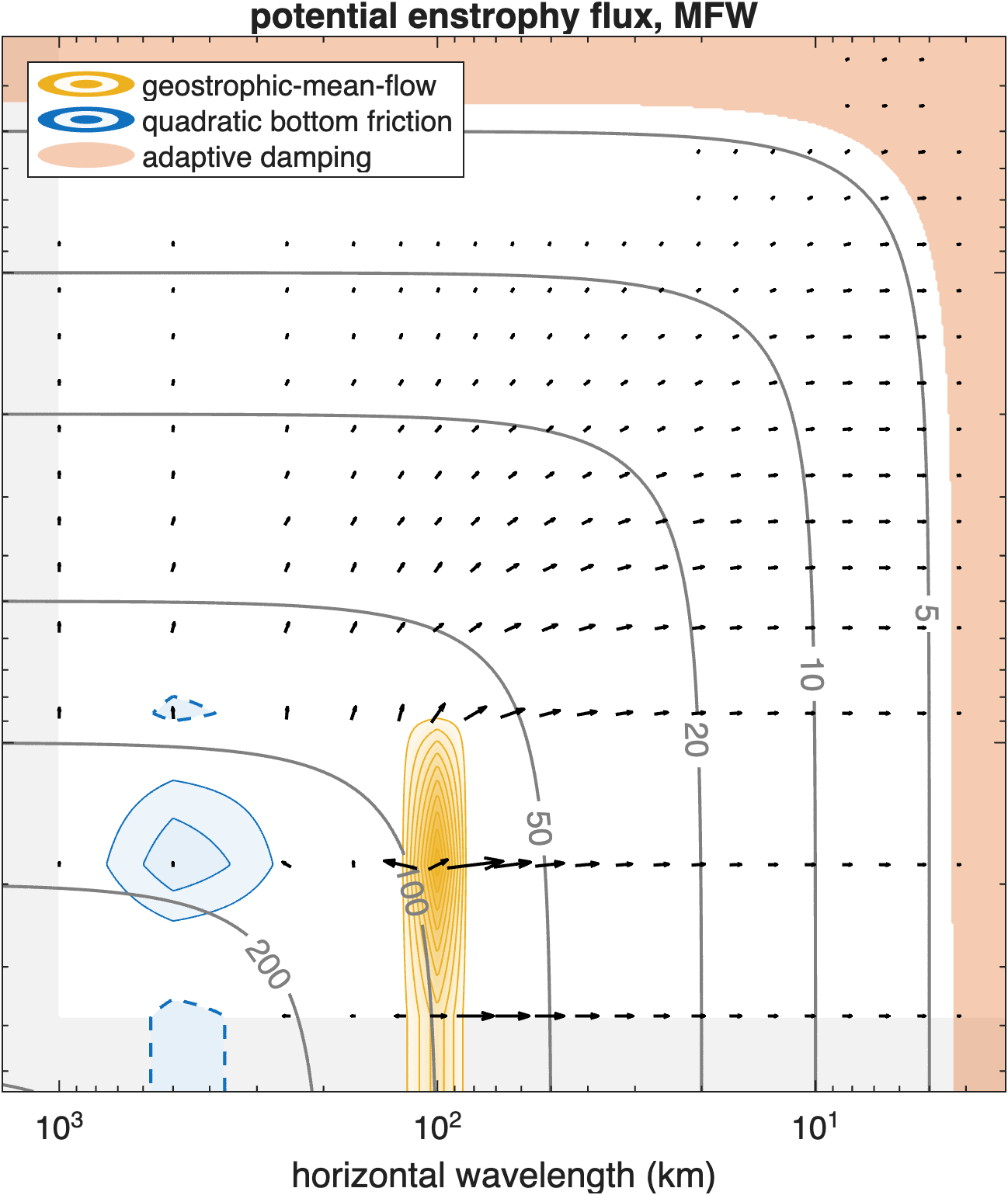}
\caption{Available potential enstrophy sources (solid contours), sinks (dashed contours), and advective flux (arrows) in the MF (left) and MFW (right) simulations. }
\label{fig:enstrophy_flux_2D_flow}
\end{figure}

The available potential enstrophy spectral fluxes are shown in Figure~\ref{fig:enstrophy_flux_2D_flow}. The figure shows that the available potential enstrophy flux for two simulations are almost identical, both following the standard quasigeostrophic potential enstrophy picture where potential enstrophy cascades from mean flow forcing to smaller scales. The most notable difference between the two simulations is the weaker flux in MFW. Note also that bottom friction has a net positive potential enstrophy flux, consistent with classic picture that topography and friction generate potential enstrophy.

\section{Linearization}
\label{sec:linearization}
%
%%%%%%%%%%%%%%%%%%%%%%%%%

To complete the sources, sinks and reservoirs diagram in Figure~\ref{fig:sources_sinks} requires that we decompose the fluid into wave and geostrophic parts, and derive energy and potential enstrophy conservation statements that include those reservoirs. To proceed, we linearize the equations of motions and determine the associated conservation statements in section~\ref{subsec:eqns-of-motion}. A key feature of our approach is that the quadratically conserved quantities for the linear dynamical equations follow directly from their nonlinear counterparts---the approximations made here are central to understanding the limitations of the method. Appendix~\ref{sec:orthogonal-solutions} formally derives the projection operators that take the fluid state and project onto the wave and vortex solutions, the result of which are summarized in section~\ref{subsec:wave-vortex-decomposition}. Section~\ref{sec:energy-fluxes} applies the projection operators to the nonlinear equations of motion in order to derive the energy fluxes, and thereby complete the analysis used in Figure~\ref{fig:sources_sinks}.

%%%%%%%%%%%%%%%%%%%%%%%%
\subsection{Perturbation equations and their conservation laws}
\label{subsec:eqns-of-motion}
%%%%%%%%%%%%%%%%%%%%%%%%

Perturbation equations are constructed by expressing the equations of motion \eqref{eqn:boussinesq} relative to the no-motion solution. It is helpful to express the buoyancy as $b= -g \rho_\textrm{e}/\rho_0=-N^2 \etae $, with total density and pressure given by
\begin{subequations}
\label{rho-p-etae}
\begin{align}
\rho_\textrm{tot}(\mathbf{x},t) =& \rho_\textrm{nm}(z) + \frac{\rho_0}{g} N^2(z) \etae(\mathbf{x},t) \\
p_\textrm{tot}(\mathbf{x},t) =& p_\textrm{nm}(z) + p_\textrm{e}(\mathbf{x},t).
\end{align}
\end{subequations}
Using \eqref{rho-p-etae} and expressing the nonlinear equations of motion in terms of state-vector $\psi$ where
\begin{equation}
\label{eqn:uvw-state-vector-simple}
\psi(\vect{x},t)= 
\begin{bmatrix}
    u(\vect{x},t)\\ v(\vect{x},t) \\ w(\vect{x},t)\\ \etae(\vect{x},t) \\ p_\textrm{e}(\vect{x},t)
\end{bmatrix},
\end{equation}
we arrive at a particularly convenient form of the nonlinear equations,
\begin{equation}
\label{eqn:nonlinear-eom-vect}
    \left( \timeop  + \op{\linop} \right) \psi +\op{\nonlinop} \left[ \psi \right]  =  \vect{\mathcal{S}},
\end{equation}
where we have defined
\begin{equation}
\label{eqn:eom-operators}
\op{\linop} =
    \begin{bmatrix}
    0 & - f & 0 & 0 & \frac{1}{\rho_0} \partial_x \\
    f & 0 & 0 & 0 & \frac{1}{\rho_0} \partial_y \\
    0 & 0 & 0 & N^2 & \frac{1}{\rho_0} \partial_z \\
    0 & 0 & -1 & 0 & 0\\
    \partial_x & \partial_y & \partial_z & 0 & 0
    \end{bmatrix},
\,
    \op{\nonlinop}   \left[ \psi \right] =
    \begin{bmatrix}
        \vect{u} \cdot \nabla u\\
        \vect{u} \cdot \nabla v \\
        \vect{u} \cdot \nabla w \\
        \vect{u} \cdot \nabla \etae + w \etae \partial_z \ln N^2 \\
        0 
    \end{bmatrix} \textrm{ and }  \vect{\mathcal{S}} =
    \begin{bmatrix}
    \mathcal{S}_u \\
    \mathcal{S}_v \\
    \mathcal{S}_w \\
    \mathcal{S}_{\etae} \\
    0
    \end{bmatrix}
\end{equation}
and rescaled the buoyancy forcing so that $\mathcal{S}_{\etae} = - \mathcal{S}_\rho/\partial_z \rho_\textrm{nm}$
The operator $\timeop$ is simply the time derivative along the diagonal, except in the entry for pressure: $\timeop \equiv  \partial_t \circ \left( \delta_{ij} - \delta_{i5}\delta_{j5} \right)  $ where $\delta_{ij}$ is the Kronecker delta. In this notation, the unforced linear equations of motion are simply $ \left( \timeop  + \op{\linop} \right) \psi=0$ and do not contain any quadratic combinations of the variables.

% \begin{subequations}
% \label{boussinesq-tilde}
% \begin{align}
% \label{x-momentum-eta}
%  \frac{du}{dt}  - f v  =& - \frac{1}{\rho_0} \partial_x p_\textrm{e } + \mathcal{S}_u\\ \label{y-momentum-eta}
%  \frac{dv}{dt}  + f u  =& -\frac{1}{\rho_0} \partial_y p_\textrm{e} + \mathcal{S}_v \\ \label{z-momentum-eta}
% \frac{dw}{dt}   =& - \frac{1}{\rho_0}  \partial_z p_\textrm{e} - N^2 \etae + \mathcal{S}_w\\ \label{thermodynamic-eta}
% \frac{d}{dt} N^2 \etae - N^2 w=& \frac{g}{\rho_0}\mathcal{S}_\rho \\ \label{continuity-eta}
% \partial_x u + \partial_y v + \partial_z w =&  0.
% \end{align}
% \end{subequations}

% Linearizing \eqref{boussinesq-tilde} requires discarding any quadratic combinations of $(u,v,w,\etae,p_\textrm{e})$
% \begin{subequations}
% \begin{align}
% \label{x-momentum-eta-lin}
% \partial_t u - f v =& - \frac{1}{\rho_0} \partial_x p_\textrm{e}\\ \label{y-momentum-eta-lin}
% \partial_t v + f u =& - \frac{1}{\rho_0}  \partial_y p_\textrm{e}  \\ \label{z-momentum-eta-lin}
% \partial_t w  =& - \frac{1}{\rho_0}  \partial_z p_\textrm{e}
% - N^2
% \etae \\
% \label{thermodynamic-eta-lin}
% \partial_t \etae  =& w  \\ \label{continuity-eta-lin}
% \partial_x u + \partial_y v + \partial_z w =& 0,
% \end{align}
% \label{eqn:boussinesq-eta-lin}
% \end{subequations}
% where \emph{all} variables are now lower order approximations of the nonlinear versions.

The linearization of APE follows exactly the approach laid out in \citet{holliday1981-jfm}. Starting from the definition of APE in \eqref{eqn:KE-APE-definition}, one may use integration by parts to rewrite APE as
\begin{equation}
\textrm{APE}
%=& - \int_0^\etat g \xi \partial \tilde{\rho}_\textrm{nm}(z- \xi) d \xi\\
= - \frac{g}{2} \etat^2 \partial \tilde{\rho}_\textrm{nm}(z- \etat) + \frac{g}{2}\int_0^\etat \xi^2 \partial^{(2)} \tilde{\rho}_\textrm{nm}(z- \xi) d \xi.
\end{equation}
Next, by using the relations $\tilde{\rho}_\textrm{nm}(z-\eta) = \tilde{\rho}_\textrm{nm}(z) + \rho_\textrm{e}$, $\rho_\textrm{e} = \rho_0 g^{-1} N^2 \eta_\textrm{e}$ together with the expansion
\begin{equation}
\label{eta-expansion-explicit}
    \etat = \etae + \frac{1}{2} \partial_z \log \left( N^2(z) \right) \etae^2 + O\left( \etae^3 \right),
\end{equation}
one arrives at the linearized version of APE 
\begin{equation}
    \textrm{APE}_\textrm{lin} \equiv \frac{1}{2} N^2 \etae^2,
\end{equation}
where terms of order $O\left(\etae^3\right)$ are neglected. From this we define the volume-integrated energy $\mathcal{E}$ as 
\begin{equation}
    \label{eqn:energy-volume-integral}
    \mathcal{E} = \frac{1}{2 L_x L_y}  \int (u^2 +v^2 + w^2 + N^2 \etae^2) \, dV,
\end{equation}
which is the quadratic counterpart to the exact volume integrated energy defined in \eqref{eqn:KE-APE-definition} and \eqref{eqn:volume-integrated-exact-energy}.

Linearizing APV in \eqref{eqn:APV-definition} is achieved by discarding quadratic combinations of $(u,v,w,\eta)$ and expanding $\etat$ in terms of $\etae=-\rho_\textrm{e} \left( \partial_z \tilde{\rho}_\textrm{nm}\right)^{-1}$ using \eqref{eta-expansion-explicit}. This consistency results in
\begin{equation}
\label{eqn:qgpv-eta}
\textrm{QGPV} \equiv \partial_x v - \partial_y u - f \partial_z \etae,
\end{equation}
as shown in \citet{early2022-arxiv}. From this we define QGPV enstrophy $\mathcal{Z}$ as 
\begin{equation}
    \label{eqn:enstrophy-volume-integral}
    \mathcal{Z} = \frac{1}{2 L_x L_y}  \int \left( \partial_x v - \partial_y u - f \partial_z \etae \right)^2\; dV.
\end{equation}
which is the quadratic counterpart to the volume integral of exactly conserved available potential enstrophy defined in \eqref{eqn:exact-available-potential-enstrophy-def}. Importantly, volume-integrated conservation of mass and potential density are satisfied exactly (matching their nonlinear counterparts), a requirement for physically realizable states, which must not create mass or internal energy.

\begin{table*}
    \centering
    \begin{tabular}{l|l|l} %\rowcolor{LightGray}
    & differential & volume integrated \\ \hline
    mass & $\nabla \cdot \mathbf{u} = 0$  & $\int \nabla \cdot \mathbf{u} \, dV = 0$ \\ 
    potential density & $\frac{d}{dt} \etae = w$  & $\int N^2 \etae \,dV = 0$ \\
    potential vorticity & $\frac{d}{dt} \textrm{QGPV} = 0$ & $\int \textrm{QGPV} \,dV = 0$ \\
    energy & $\frac{\partial}{\partial t} \left( \textrm{KE} + \textrm{APE}_\textrm{lin} \right)=  - \mathbf{u} \cdot \nabla p_\textrm{e}$ & $\frac{\partial}{\partial t} \mathcal{E} = 0$ \\
    potential enstrophy & $\frac{\partial}{\partial t} \frac{1}{2} \textrm{QGPV}^2 = 0$ & $\frac{\partial}{\partial t} \mathcal{Z} = 0$
    \end{tabular}
    \caption{The quadratic conservation laws satisfied by the individual eigen-solutions of $ \left( \timeop  + \op{\linop} \right) \psi=0$, shown in Table~\ref{tab:solutions}. Note that the partial and total derivatives are intentional---all equations (except mass conservation) are quadratic.}
    \label{tab:conservation-laws}
\end{table*}

When the forcing is included, the quadratic conservation laws are
\begin{equation}
     \frac{\partial}{\partial t} \left( \textrm{KE} + \textrm{APE}_\textrm{lin} \right)=  - \mathbf{u} \cdot \nabla p_\textrm{e} + \vect{u} \cdot \mathcal{S}_\mathbf{u} + N^2 \etae \mathcal{S}_{\etae}
\end{equation}
and
\begin{equation}
    \frac{d}{dt} \textrm{QGPV} = \left( \partial_x \mathcal{S}_v - \partial_y \mathcal{S}_u - f \partial_z \mathcal{S}_{\etae} \right).
\end{equation}
The corresponding volume integrated fluxes are,
\begin{equation}
\label{eqn:volume-integrated-energy-flux-quadratic}
     \frac{\partial}{\partial t} \mathcal{E} = \frac{1}{2 L_x L_y}  \int  \left( \vect{u} \cdot \mathcal{S}_\mathbf{u} + N^2 \etae \mathcal{S}_{\etae} \right) \; dV.
\end{equation}
and
\begin{equation}
     \frac{\partial}{\partial t} \mathcal{Z} = \frac{1}{2 L_x L_y}  \int \textrm{QGPV} \left( \partial_x \mathcal{S}_v - \partial_y \mathcal{S}_u - f \partial_z \mathcal{S}_{\etae} \right) \; dV.
\end{equation}
These quadratic fluxes are approximations to \eqref{eqn:volume-integrated-exact-energy} and \eqref{volume-integrated-exact-enstrophy} respectively, and are used to compute fluxes to the individual reservoirs in Figure~\ref{fig:sources_sinks}.

%%%%%%%%%%%%%%%%%%%%%%%%
\subsection{The wave-vortex decomposition}
\label{subsec:wave-vortex-decomposition}
%%%%%%%%%%%%%%%%%%%%%%%%

The primary tool that enables the analysis are the complete set of energetically orthogonal solutions to linearized equations of motion $\left( \timeop  + \op{\linop} \right) \psi=0$ with their associated projection operators. Mathematical details are relegated to appendix~\ref{sec:orthogonal-solutions}, so here we summarize the essential features.

\begin{figure}
    \centering
    \includegraphics[width=1.0\textwidth]{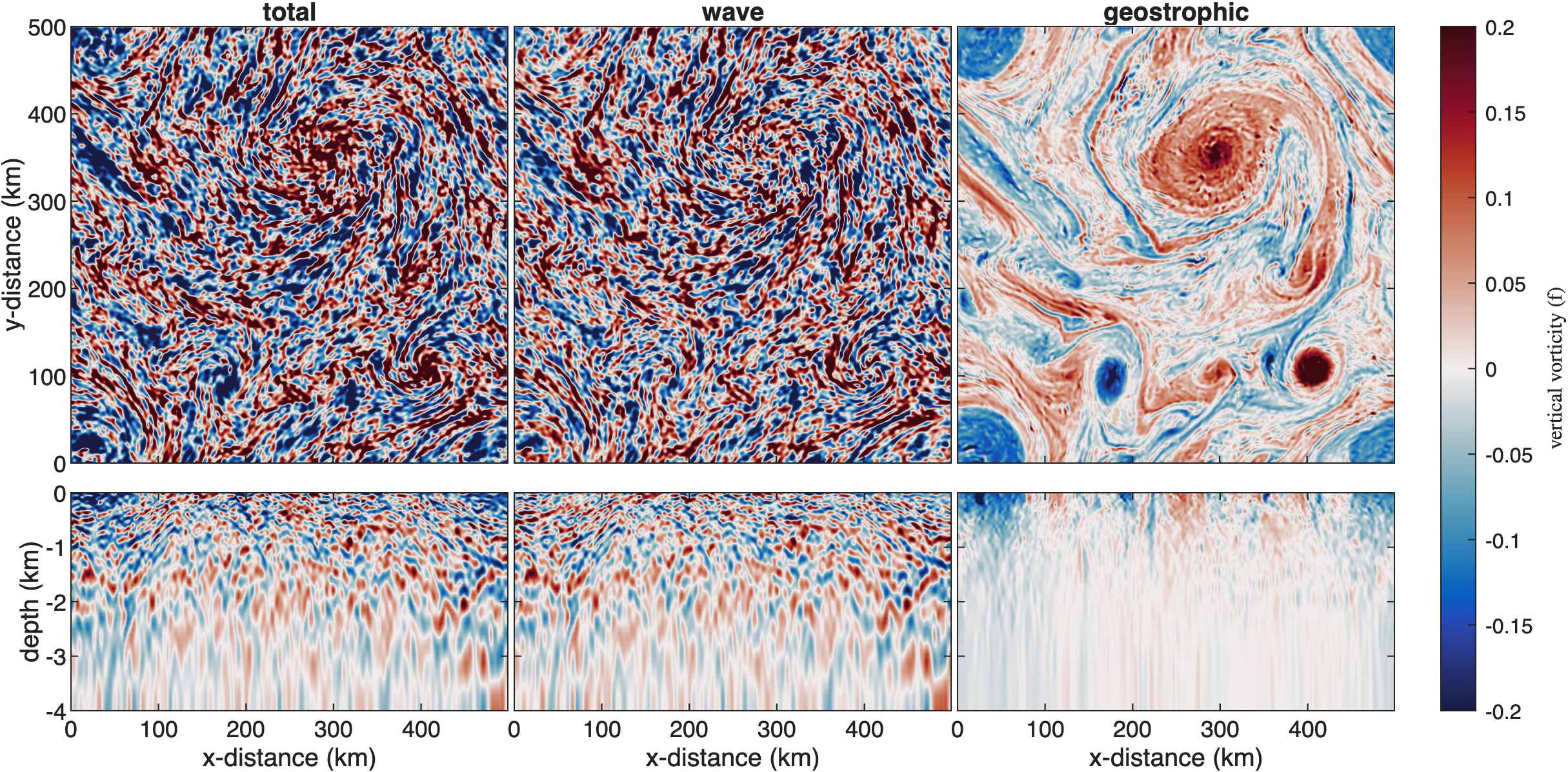}
    \caption{Decomposition of the MFW simulation vertical vorticity $\zeta$ (left) into wave (center) and geostrophic (right) components.  The top row shows surface vorticity and the bottom row shows a vertical cross-section at $y=0$.}
    \label{fig:fluid_decomposition}
\end{figure}

For our numerical simulations with a bounded domain and finite resolution there is a countable set of eigenmode solutions to $ \left( \timeop  + \op{\linop} \right) \psi = 0$ which are indexed by horizontal wavenumbers $k, \ell$ and vertical mode $j$ that correspond to geostrophic, internal gravity wave, inertial oscillation, and mean density anomaly solutions. This solution set is \emph{complete}, which means that we can represent any ocean state as a sum of the individual modes
\begin{equation}
    \psi = \sum_{k \ell j} A_0^{k\ell j}(t) \Psi_0^{k\ell j} + A_+^{k\ell j}(t) \Psi_+^{k\ell j} + A_-^{k\ell j}(t) \Psi_-^{k\ell j}.
\end{equation}
The $\Psi_0$ modes are the geostrophic and mean-density anomaly (mda) solutions with a non-zero signature of QGPV, the $\Psi_\pm$ modes are the internal gravity wave and inertial oscillations solutions with no QGPV, and $A_0^{k\ell j}$(t), $A_\pm^{k\ell j}(t)$ are the coefficients. While any complete basis can be used to represent the ocean state, two key features make this base useful:
\begin{enumerate}
    \item each solution satisfies all boundary conditions and the conservations laws in Table~\ref{tab:conservation-laws} and is thus a physically realizeable state of the fluid and,
    \item the solutions are all energetically orthogonal.
\end{enumerate}
Energy orthogonality means that the volume integral of the total quadratic energy is the sum of the squares of the energy of the individual solutions,
\begin{equation}
    \frac{1}{2 L_x L_y}  \int (u^2 +v^2 + w^2 + N^2 \etae^2) \, dV = \sum_{k \ell j} \epsilon_g^{k\ell j} \left| A_0^{k\ell j} \right|^2 +  \epsilon_w^{k\ell j} \left| A_+^{k\ell j} \right|^2 +  \epsilon_w^{k\ell j} \left| A_-^{k\ell j} \right|^2
\end{equation}
where $\epsilon_g$ and $\epsilon_w$ are the total energy of each mode noted in the second column of Table~\ref{tab:solution-projection} and combined such that,
\begin{equation}
\label{eqn:energy-scaling-factor}
    \epsilon^{k\ell j}_\textrm{g} \equiv
    \begin{cases}
        \frac{1}{2}   \left(\kappa^2 + L^{-2} \right)^{-1} h_g^j & k > 0, j \geq 0 \\
        \frac{g}{2} & k = 0, j \geq 1
    \end{cases}, \quad
    \epsilon^{k\ell j}_\textrm{w} \equiv \begin{cases}
        h_\kappa^j & k > 0, j \geq 1 \\
        h_\textrm{io}^j & k = 0, j \geq 0
    \end{cases}.
\end{equation}
It is worth emphasizing that any complete basis could be used to partition the fluid flow (a linear partitioning). However, the fact that it is energetically orthogonal (a quadratic partitioning) is what gives this decomposition meaning.

Energy orthogonality provides the prescription for projecting the fluid-state $\psi$ onto the orthogonal modes $\Psi_0$, $\Psi_\pm$ and recovering the coefficients $A_0$, $A_\pm$. The projection operators are summarized in Table~\ref{tab:solution-projection} and require $u$, $v$, $\eta_\textrm{e}$---both pressure and vertical velocity are determined diagnostically for this system. Other combinations of the state variables are not sufficient for a complete description of the fluid.

Figure~\ref{fig:fluid_decomposition} shows the result of applying the decomposition to the MFW simulation on the last day of the simulation. The geostrophic portion of the field resembles the MF simulation in Figure~\ref{fig:two_sim_comparison}, although with additional smaller scale features. With the decomposed wave and geostrophic fields, the energies of each reservoir in Figure~\ref{fig:sources_sinks} can now be computed using the coefficients from the energy column in Table~\ref{tab:solution-projection}. Recall that the energy computed in the decomposed fields is the quadratic approximation \eqref{eqn:energy-volume-integral} to the exact total energy defined in \eqref{eqn:KE-APE-definition} and \eqref{eqn:volume-integrated-exact-energy}. That the sum of the energies in the wave and geostrophic reservoirs in Figure~\ref{fig:sources_sinks} add to the total exact energy indicate that the approximation is good.

% \begin{figure}
%     \centering
%     \includegraphics[width=1.0\linewidth]{figures/energy_spectrum_simple.png}
%     \caption{Depth integrated quadratic energy spectrum as a function of radial wavelength (left) and vertical mode radius of deformation (right).}
%     \label{fig:energy_spectra}
% \end{figure}

\begin{figure}[t]
    \includegraphics[width=1.0\textwidth]{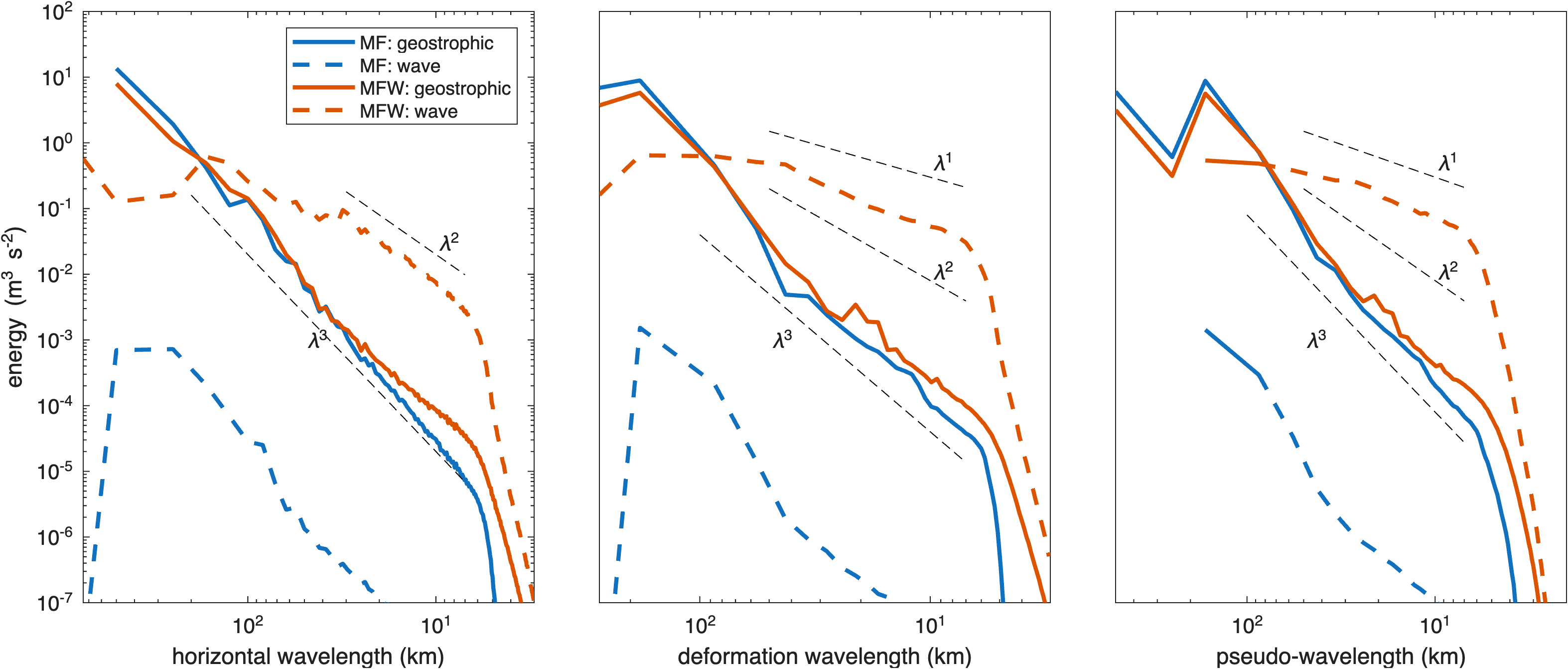}
    \caption{Decomposed depth-integrated quadratic energy spectrum on the last day of the simulation as a function of horizontal wavelength (left) and deformation wavelength (middle) and pseudo-wavelength (right).}
    \label{fig:energy_spectra}
\end{figure}

The depth-integrated energy spectra for the wave and geostrophic portions of the flow for the simulations are shown in Figure~\ref{fig:energy_spectra}. The geostrophic energy spectra are consistent at scales larger than 15~km, below which the MFW simulation shows increased energy, similar to Figure~7 in \citet{hernandez2021-jpo} and Figure~2 in \citet{thomas2021-jfm}. We will also note a discrepancy around this scale between APV and QGPV in section~\ref{sec:discussion} below, and attribute it to mis-projection (error in the decomposition) of the vortex signal.

\begin{figure}
    \centering
    \includegraphics[width=1.0\linewidth]{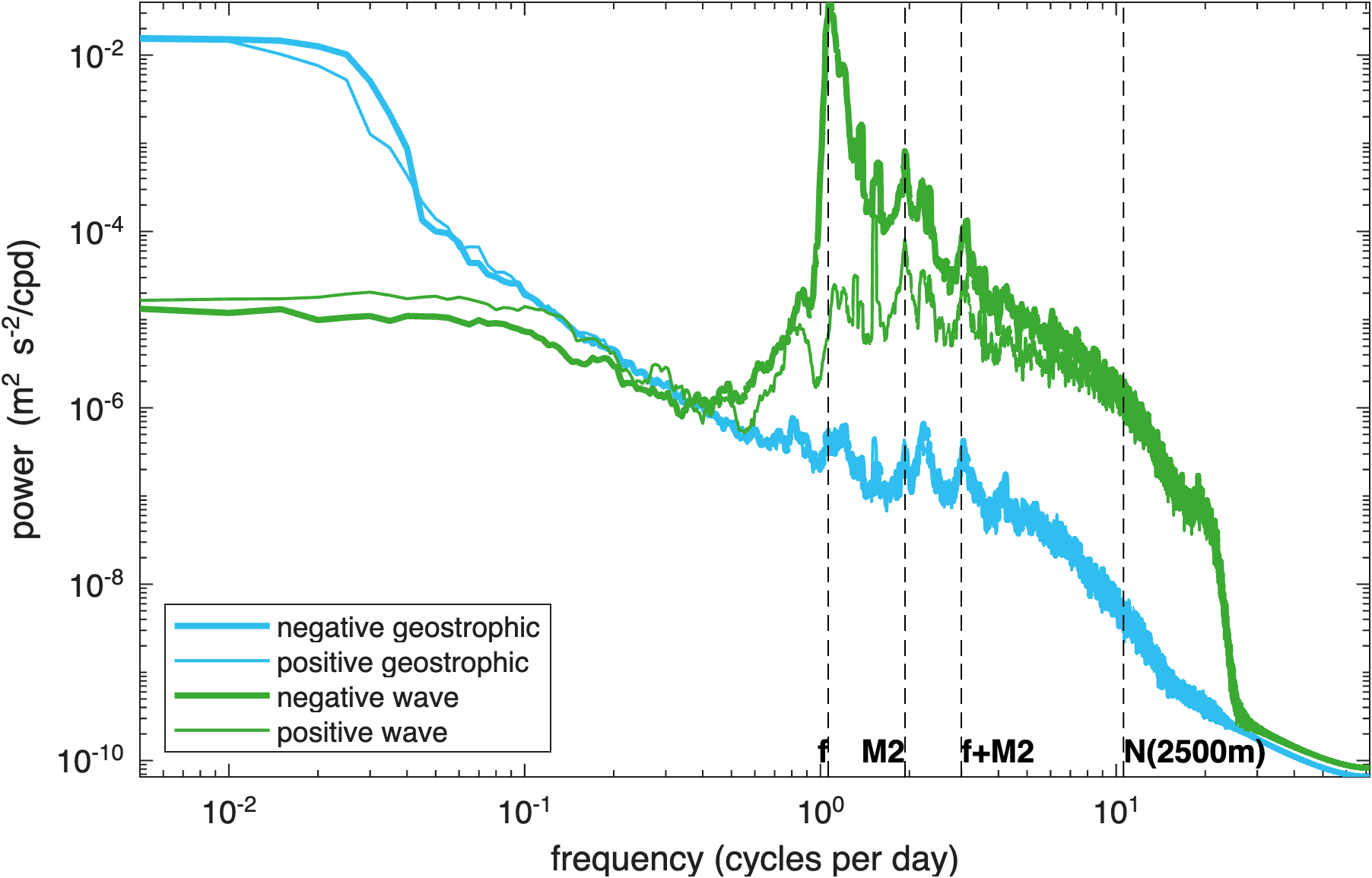}
    \caption{Rotary spectra of the decomposed velocity fields in the MFW simulation, sampled at mooring at 2500m deep.}
    \label{fig:moorings_decomposed}
\end{figure}

The wave-vortex decomposition does not use any temporal information, and thus the rotary spectrum constructed from the synthetic moorings in Figure~\ref{fig:mooring} can be decomposed to assess the fidelity of the decomposition. Figure~\ref{fig:moorings_decomposed} shows the rotary spectrum of the decomposed fields. The wave component shows the characteristic asymmetry between the positive and negative sides of the spectrum at frequencies above $f$. The geostrophic component also behaves as expected, and dominants at lower frequencies. The small peaks above $f$ show no asymmetry in the rotary spectrum, and are thus likely an imprint of the advection of the geostrophic flow by the waves.

We now proceed to compute energy fluxes by projecting the nonlinear equations of motion onto wave-vortex space.

%%%%%%%%%%%%%%%%%%%%%%%%%%%%%%%%%%%%%%%%%%
%
\section{Energy fluxes and the wave-vortex decomposition}
\label{sec:energy-fluxes}
%
%%%%%%%%%%%%%%%%%%%%%%%%%%%%%%%%%%%%%%%%%%

The key result from section~\ref{sec:linearization} is the projection operators which take the fluid state $(u,v,\eta)$ and return the fields separated into their wave and geostrophic parts. However, to compute the energy fluxes between the wave-vortex modes, we must express the nonlinear equations of motion in terms of the wave-vortex basis using the projection operators $\mathcal{P}^0_\textrm{g}$, $\mathcal{P}_\textrm{w}^+$, and $\mathcal{P}_\textrm{w}^-$ as defined in appendix~\ref{sec:generalized-projection-operators}. In terms of the state-vector $\psi$, the equations-of-motion were expressed as
\begin{equation}
\tag{\ref{eqn:nonlinear-eom-vect}}
    \left( \timeop  + \op{\linop} \right) \psi +\op{\nonlinop} \left[ \psi \right]  =  \vect{\mathcal{S}}.
\end{equation}
Starting by projecting the forcing, we define the forcing terms in wave-vortex space as,
\begin{align*}
        F_0^{k\ell j}  \equiv& \mathcal{P}^0_\textrm{g}\left[ \mathcal{S}\right] \\
        F_+^{k\ell j}  \equiv& \mathcal{P}^+_\textrm{w}\left[ \mathcal{S}\right] \\
        F_-^{k\ell j}  \equiv& \mathcal{P}^-_\textrm{w}\left[ \mathcal{S}\right]
\end{align*}
and then use a shorthand to express the nonlinear advection terms,
\begin{align*}
    \left[ \vect{u} \nabla \vect{u} \right]_\textrm{g}^{k\ell j} \equiv& \mathcal{P}^0_\textrm{g} \left[ \vect{u} \cdot \nabla u, \vect{u} \cdot \nabla v, \vect{u} \cdot \nabla w, \vect{u} \cdot \nabla \etae + w \etae \partial_z \ln N^2\right] \\
    \left[ \vect{u} \nabla \vect{u} \right]_+^{k\ell j} \equiv& \mathcal{P}^+_\textrm{w} \left[ \vect{u} \cdot \nabla u, \vect{u} \cdot \nabla v, \vect{u} \cdot \nabla w, \vect{u} \cdot \nabla \etae + w \etae \partial_z \ln N^2\right] \\
    \left[ \vect{u} \nabla \vect{u} \right]_-^{k\ell j} \equiv& \mathcal{P}^-_\textrm{w} \left[ \vect{u} \cdot \nabla u, \vect{u} \cdot \nabla v, \vect{u} \cdot \nabla w, \vect{u} \cdot \nabla \etae + w \etae \partial_z \ln N^2\right]
\end{align*}
resulting in the nonlinear equations of motion,
\begin{subequations}
\label{eqn:nonlinear-wave-vortex-eqn}
    \begin{align}
        \partial_t A_0^{k\ell j} =&\left[ \vect{u} \nabla \vect{u} \right]_\textrm{g}^{k\ell j} + F_0^{k\ell j},  \\ \label{eqn:nonlinear-wave-vortex-eqn-wave}
        \partial_t A_+^{k\ell j} =& \left[ \vect{u} \nabla \vect{u} \right]_+^{k\ell j} + F_+^{k\ell j}, \\
        \partial_t A_-^{k\ell j} =& \left[ \vect{u} \nabla \vect{u} \right]_-^{k\ell j} + F_-^{k\ell j},
    \end{align}
\end{subequations}
expressed in wave-vortex space. Equation \eqref{eqn:nonlinear-wave-vortex-eqn} is identical to the original equations of motion \eqref{eqn:boussinesq} expressed in the physical domain, but now has an interpretation as coupled forced QGPV and wave equations. This is presented in `pseudospectral' form, where the nonlinear terms are expressed in terms of $(u,v,\etae,w)$ and are then transformed using the projection operators from Table~\ref{tab:solution-projection}. Of course, the components $(u,v,\etae,w)$ in \eqref{eqn:nonlinear-wave-vortex-eqn} could be further expressed in terms of their constituent parts, e.g., $u=u_\textrm{g} + u_\textrm{w}$ in order to create the classic triads as we will do below in section~\ref{subsec:triads}.

%%%%%%%%%%%%%%%%%%%%%%%%%%%%%%%%%
\subsection{Energy fluxes}
\label{sec:energy fluxes}
%%%%%%%%%%%%%%%%%%%%%%%%%%%%%%%%%

With the nonlinear equations of motion expressed in wave-vortex space \eqref{eqn:nonlinear-wave-vortex-eqn}, it is now possible to compute the energy flux into the wave and geostrophic reservoirs, completing Figure~\ref{fig:sources_sinks}. Constructing the energy equation requires multiplying \eqref{eqn:nonlinear-wave-vortex-eqn} by the conjugate of the coefficients, and then scaling by the energy factors defined in \eqref{eqn:energy-scaling-factor} so that
\begin{subequations}
\label{eqn:energy-flux-wave-vortex}
\begin{align}
\epsilon^{k\ell j}_\textrm{g}  \partial_t \left| A_0^{k\ell j} \right|^2 =& \Re \left[ \epsilon^{k\ell j}_\textrm{g} \bar{A}_0^{k\ell j} \left[ \vect{u} \nabla \vect{u} \right]^{k\ell j}_\textrm{g}  \right] +  \Re \left[ \epsilon^{k\ell j}_\textrm{g} \bar{A}_0^{k\ell j} F_0^{k\ell j}  \right] \\
    \epsilon^{k\ell j}_\textrm{w} \partial_t \left| A_\pm^{k\ell j} \right|^2 =& \Re \left[ \epsilon^{k\ell j}_\textrm{w} \bar{A}_\pm^{k\ell j} \left[ \vect{u} \nabla \vect{u} \right]^{k\ell j}_\textrm{w}  \right] +  \Re \left[ \epsilon^{k\ell j}_\textrm{w} \bar{A}_\pm^{k\ell j} F_\pm^{k\ell j}  \right]
\end{align}
\end{subequations}
where we have combined the two sets of wave coefficients into one ($\pm$ and w). These expressions can be compared to equation~2.48 in \citet{frisch1995-book} written for the Navier-Stokes equations and, because both expressions use an energetically orthogonal basis, the same analysis techniques will apply.

There are two quasi-conservation laws related to \eqref{eqn:energy-flux-wave-vortex} that are helpful for interpreting the results. First, the nonlinear advection terms are inertial---they have no net energy flux and thus must sum to zero for exact total energy \eqref{eqn:volume-integrated-exact-energy}, and approximately so for quadratic total energy,
\begin{equation}
\label{eqn:inertial-energy-flux-conservation}
    \sum_{k \ell j}  \Re \left[ \epsilon^{k\ell j}_\textrm{g} \bar{A}_0^{k\ell j} \left[ \vect{u} \nabla \vect{u} \right]^{k\ell j}_\textrm{g}  \right] + \Re \left[ \epsilon^{k\ell j}_\textrm{w} \bar{A}_\pm^{k\ell j} \left[ \vect{u} \nabla \vect{u} \right]^{k\ell j}_\textrm{w}  \right] \approx 0.
\end{equation}
As a result, this implies that that total change in quadratic total energy should be entirely from the forcing,
\begin{equation}
\label{eqn:inertial-forcing-flux-conservation}
    \sum_{k \ell j} \epsilon^{k\ell j}_\textrm{g}  \partial_t | A_0^{k\ell j}|^2 + \epsilon^{k\ell j}_\textrm{w} \partial_t | A_\pm^{k\ell j}|^2 \approx \Re \left[ \epsilon^{k\ell j}_\textrm{g} \bar{A}_0^{k\ell j} F_0^{k\ell j}  \right] +  \Re \left[ \epsilon^{k\ell j}_\textrm{w} \bar{A}_\pm^{k\ell j} F_\pm^{k\ell j}  \right].
\end{equation}

The arrows from the sources and sinks to the reservoirs in Figure~\ref{fig:sources_sinks} are computed from the total of each forcing flux term in \eqref{eqn:energy-flux-wave-vortex}. Energy transfers \emph{between} the wave and geostrophic reservoirs can only come from the inertial terms in \eqref{eqn:inertial-energy-flux-conservation}. Thus, the final energy flux arrow in Figure~\ref{fig:sources_sinks} is computed from either one of the two terms in \eqref{eqn:inertial-energy-flux-conservation}.

%%%%%%%%%%%%%%%%%%%%%%%%
\subsection{Triads}
\label{subsec:triads}
%%%%%%%%%%%%%%%%%%%%%%%%

The inertial fluxes can be decomposed into their constituent parts  $\vect{u}=\vect{u}_\textrm{g} + \vect{u}_\textrm{w}$ such that,
\begin{subequations}
    \begin{align}
       \left[ \vect{u} \nabla \vect{u} \right]_\textrm{g} =& \underbrace{\left[ \vect{u}_\textrm{g} \nabla \vect{u}_\textrm{g} \right]_\textrm{g}}_\textrm{ggg} + \underbrace{\left[ \vect{u}_\textrm{w} \nabla \vect{u}_\textrm{g} \right]_\textrm{g} + \left[ \vect{u}_\textrm{g} \nabla \vect{u}_\textrm{w} \right]_\textrm{g}}_\textrm{ggw} + \underbrace{\left[ \vect{u}_\textrm{w} \nabla \vect{u}_\textrm{w} \right]_\textrm{g}}_\textrm{wwg} \\
        \left[ \vect{u} \nabla \vect{u} \right]_\textrm{w} =& \underbrace{\left[ \vect{u}_\textrm{w} \nabla \vect{u}_\textrm{w} \right]_\textrm{w}}_\textrm{www} + \underbrace{\left[ \vect{u}_\textrm{w} \nabla \vect{u}_\textrm{g} \right]_\textrm{w} + \left[ \vect{u}_\textrm{g} \nabla \vect{u}_\textrm{w} \right]_\textrm{w}}_\textrm{wwg} + \underbrace{\left[ \vect{u}_\textrm{g} \nabla \vect{u}_\textrm{g} \right]_\textrm{w}}_\textrm{ggw}
    \end{align}
\end{subequations}
where we have grouped terms according to which triad it belongs. 
As noted above the total quadratic energy flux must approximately vanish (equation~\ref{eqn:inertial-energy-flux-conservation}), but additionally, the closed triads must also vanish. All quadratic interactions shift energy between exactly three modes and the sum all of interactions involving modes of a specfic triad must vanish, e.g., the wwg triad conserves
\begin{equation}
\label{eqn:wwg-energy-flux-conservation}
    \sum_{k \ell j}  \Re \left[ \epsilon^{k\ell j}_\textrm{g} \bar{A}_0^{k\ell j} \left[ \vect{u}_\textrm{w} \nabla \vect{u}_\textrm{w} \right]^{k\ell j}_\textrm{g}  \right] + \Re \left[ \epsilon^{k\ell j}_\textrm{w} \bar{A}_\pm^{k\ell j} \left[ \vect{u}_\textrm{w} \nabla \vect{u}_\textrm{g} + \vect{u}_\textrm{g} \nabla \vect{u}_\textrm{w} \right]^{k\ell j}_\textrm{w}  \right] \approx 0
\end{equation}
and similarly for ggg, ggw, and www as described in \citet{kraichnan1973-jfm} for triads in the Navier-Stokes equations. 

\begin{table}[]
    \centering
    \begin{tabular}{c|c|c|c}
         & geostrophic cascade & wave cascade & transfer \\ \hline
        ggg & \cmark & \xmark & \xmark \\
        ggw & \cmark & \xmark & \cmark \\
        wwg & \xmark & \cmark & \cmark \\
        www & \xmark & \cmark & \xmark \\ \hline
    \end{tabular}
    \caption{The four energetically closed triads. A cascade requires at least two components of the triads be of the same type, while a transfer between wave and geostrophic reservoirs is possible only when the triad has both wave and geostrophic components.}
    \label{tab:triads}
\end{table}

Each triad type may be involved in energy cascades and/or energy transfers, as summarized in Table~\ref{tab:triads}. Triads involving two or more of the same flow component can support a cascade within that reservoir. For example, the wwg triad can shift energy between wave modes, and thus cascade energy within the wave reservoir (we use the term cascade loosely here, as the transfers may still be non-local). Additionally, the wwg triad can transfer energy between wave and geostrophic reservoirs, but cannot cascade energy between geostrophic modes.
The wwg and ggw triads can transfer energy between the flow components, while www and ggg can only cause energy cascade within their respective flow components.
Below we will show that the wwg triad appears to be a pathway from geostrophic energy to wave energy, but ggw triad does not have any significant transfer for our simulations.

% In contrast, any gain or loss of geostrophic energy is determined from the triad component $\left[ \vect{u}_\textrm{w} \nabla \vect{u}_\textrm{w} \right]_\textrm{g}$ which isolates the transfer from waves into the geostrophic reservoir from wwg triad. Similarly, the triad component $\left[ \vect{u}_\textrm{g} \nabla \vect{u}_\textrm{g} \right]_\textrm{w}$ isolates the transfer from geostrophic modes into the wave reservoir from the ggw triad. These are the only two mechanisms for transferring energy between wave and geostrophic modes. Below we will show that the wwg triad appears to be a pathway from geostrophic energy to wave energy, but ggw triad does not have any significant transfer for our simulations.

The term $\left[ \vect{u}_\textrm{w} \nabla \vect{u}_\textrm{w} \right]_\textrm{g}$ (with implicit indices $k\ell j$ of the geostrophic modes) tells us which geostrophic modes gain energy from the interaction of two wave modes. But how do we determine which wave modes lost energy to the geostrophic modes? Here we use an idea from \citet{frisch1995-book} and systematically `mask' the wave modes to include modes below certain values. Specifically, let $\vect{u}_\textrm{w}^{<k \ell j}$ denote all wave modes with mode numbers less than $k\ell j$. Thus, the quantity
\begin{equation}
    \Pi_\textrm{w}^{<}(k,\ell,j) \equiv \sum_{k^\prime \ell^\prime j^\prime}\Re \left[ \epsilon^{k^\prime \ell^\prime j^\prime}_\textrm{g} \bar{A}_0^{k^\prime \ell^\prime j^\prime} \left[ \vect{u}_\textrm{w}^{<k \ell j} \nabla \vect{u}_\textrm{w}^{<k \ell j} \right]_\textrm{g}^{k^\prime \ell^\prime j^\prime} \right]
\end{equation}
is the total amount of energy fluxed into the geostrophic reservoir from two waves with scales less than $k \ell j$. This is the most computationally expensive operation of this manuscript, as it requires considering all combinations of $k \ell j$ (in practice we restrict to logarithmically-spaced $k_h$ and $j$ suitable for analysis and visualization). The source wave modes for $\left[ \vect{u}_\textrm{w} \nabla \vect{u}_\textrm{w} \right]_\textrm{g}$ is the derivative of $\Pi_\textrm{w}^{<}(k,\ell,j)$ which we call the `mirror flux',
\begin{equation}
    \mathcal{M}\left[ \left[ \vect{u}_\textrm{w} \nabla \vect{u}_\textrm{w} \right]_\textrm{g} \right] \equiv \partial_k \partial_\ell \partial_j \Pi_\textrm{w}^{<}(k,\ell, j)/\left( \epsilon^{k\ell j}_\textrm{w} \bar{A}_\pm^{k\ell j} \right)
\end{equation}
where $\partial_k \Pi(k,\ell,j) \equiv \Pi(k,\ell,j)-\Pi(k-1,\ell,j)$ is the first order difference and $ \Pi_\textrm{w}^{<}(k,\ell, j)=0$ for any $k,\ell,j<0$. Just as $\left[ \vect{u}_\textrm{w} \nabla \vect{u}_\textrm{w} \right]_\textrm{g}$ has implicit indices $k\ell j$ of the geostrophic modes, $\mathcal{M}\left[ \left[ \vect{u}_\textrm{w} \nabla \vect{u}_\textrm{w} \right]_\textrm{g} \right]$ has implicit indices $k\ell j$ of the wave modes---with the property that both the flux term and its mirror sum to the same total energy. The mirror operation $\mathcal{M}$ can also be applied to the $\left[ \vect{u}_\textrm{g} \nabla \vect{u}_\textrm{g} \right]_\textrm{w}$ triad component to produce a flux from the geostrophic modes. This construction allows us to rewrite the triads in their most useful form, which separates the effects of cascades within the fields from transfers between the two fields:
\begin{tabular}{rll}
        geostrophic ggg-cascade: & $F_\textrm{ggg}^{k \ell j} \equiv \left[ \vect{u}_\textrm{g} \nabla \vect{u}_\textrm{g} \right]_\textrm{g}$ \\
        geostrophic ggw-cascade: & $F_\textrm{ggw}^{k \ell j} \equiv \left[ \vect{u}_\textrm{w} \nabla \vect{u}_\textrm{g} + \vect{u}_\textrm{g} \nabla \vect{u}_\textrm{w} \right]_\textrm{g} + \mathcal{M} \left[ \left[ \vect{u}_\textrm{g} \nabla \vect{u}_\textrm{g} \right]_\textrm{w} \right]$ \\
    geostrophic transfer: & $F_\textrm{g-tx}^{k \ell j} \equiv \left[ \vect{u}_\textrm{w} \nabla \vect{u}_\textrm{w} \right]_\textrm{g} -  \mathcal{M} \left[ \left[ \vect{u}_\textrm{g} \nabla \vect{u}_\textrm{g} \right]_\textrm{w} \right]$  \\
    wave www-cascade: & $F_\textrm{www}^{k \ell j} \equiv \left[ \vect{u}_\textrm{w} \nabla \vect{u}_\textrm{w} \right]_\textrm{w}$ \\
    wave wwg-cascade: & $F_\textrm{wwg}^{k \ell j} \equiv  \left[ \vect{u}_\textrm{w} \nabla \vect{u}_\textrm{g}+ \vect{u}_\textrm{g} \nabla \vect{u}_\textrm{w} \right]_\textrm{w} + \mathcal{M} \left[ \left[ \vect{u}_\textrm{w} \nabla \vect{u}_\textrm{w} \right]_\textrm{g} \right]$ \\
    wave transfer: & $F_\textrm{w-tx}^{k \ell j} \equiv -  \mathcal{M} \left[ \left[ \vect{u}_\textrm{w} \nabla \vect{u}_\textrm{w} \right]_\textrm{g} \right] + \left[ \vect{u}_\textrm{g} \nabla \vect{u}_\textrm{g} \right]_\textrm{w} $. 
\end{tabular}

The four cascade terms inherit the triad flux conservation laws from \eqref{eqn:wwg-energy-flux-conservation} that
\begin{equation}
    \sum \epsilon^{k\ell j}_\textrm{g} \bar{A}_0^{k\ell j} F_\textrm{ggg}^{k \ell j}\approx 0,\quad \sum \epsilon^{k\ell j}_\textrm{g} \bar{A}_0^{k\ell j} F_\textrm{ggw}^{k \ell j}\approx 0,\quad \sum \epsilon^{k\ell j}_\textrm{w} \bar{A}_\pm^{k\ell j} F_\textrm{www}^{k \ell j}\approx 0 \quad \textrm{and} \quad \sum \epsilon^{k\ell j}_\textrm{w} \bar{A}_\pm^{k\ell j} F_\textrm{wwg}^{k \ell j}\approx 0
\end{equation}
and, of course, the transfer terms are opposite and equal,
\begin{equation}
\label{eqn:wave-geo-transfer-conservation}
    \sum \epsilon^{k\ell j}_\textrm{g} \bar{A}_0^{k\ell j} F_\textrm{g-tx}^{k \ell j}= - \sum \epsilon^{k\ell j}_\textrm{w} \bar{A}_\pm^{k\ell j} F_\textrm{w-tx}^{k \ell j} 
\end{equation}
thereby conserving energy.

%%%%%%%%%%%%%%%%%%%%%%%%%%%%%%%%%%%%%%%%%%
%
\section{Results}
\label{sec:results}
%
%%%%%%%%%%%%%%%%%%%%%%%%%%%%%%%%%%%%%%%%%%

The flux computations in section~\ref{sec:energy fluxes} are carried out in three-dimensions $(k,\ell,j)$. Here we present the results in three forms: 1) time-averaged one-dimensional fluxes, computed along pseudo-wavelength, kinetic energy fraction, and intrinsic frequency axes, 2) two-dimensional time-averaged fluxes as a function of horizontal wavelength and deformation wavelength, and 3) spatio-temporal averages summarized by the sources, sinks and reservoirs box plots in Figure~\ref{fig:sources_sinks}. 

\begin{figure}[t]
\includegraphics[width=\textwidth]{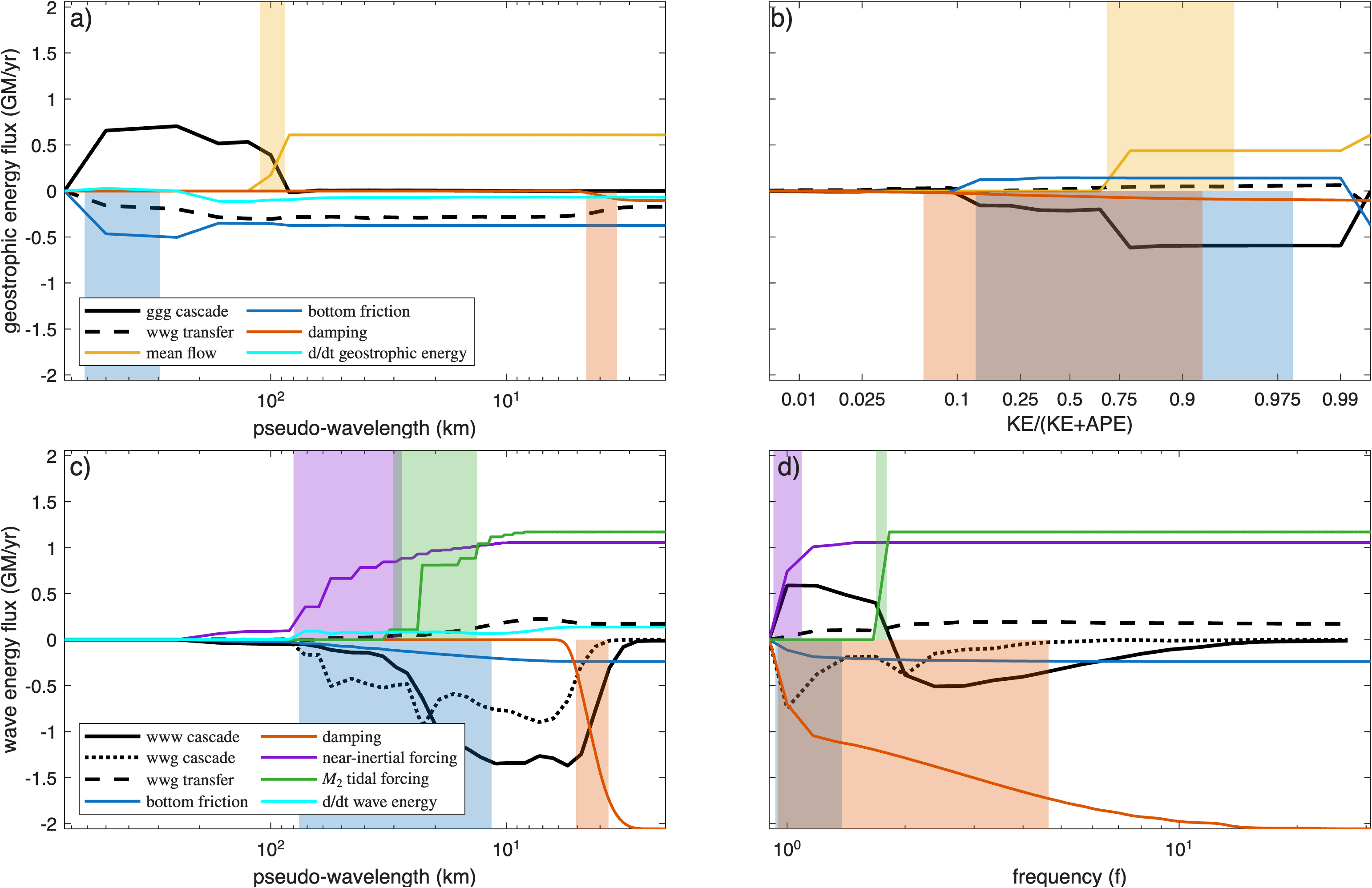}
\caption{Cumulative depth-integrated energy flux into the geostrophic (a, b) and wave (c, d) reservoirs. Fluxes are shown as a function of pseudo-wavelength (a, c), kinetic energy fraction (b), and intrinsic wave frequency (d). Colored lines show mean flow forcing (yellow), near-inertial forcing (purple), tidal forcing (green), bottom friction (blue), and small scale dissipation (red) with dominant forcing and damping scales shaded. In (b, d), friction and dissipation scales overlap. Black lines show nonlinear triad fluxes. Positive (negative) slopes indicate energy is being added (removed). For the cascade terms, positive (negative) values indicate inverse (forward) energy cascade. The dominant triad fluxes are $\left[ \vect{u}_\textrm{g} \nabla \vect{u}_\textrm{g} \right]_\textrm{g}$ and $\left[ \vect{u}_\textrm{w} \nabla \vect{u}_\textrm{w} \right]_\textrm{w}$ (solid); other fluxes redistribute energy within reservoirs (wwg cascade, dotted) or exchange energy between reservoirs (wwg transfer, dashed). The rate of change of the spectrum is also shown (cyan).}
\label{fig:energy_flux_1d}
\end{figure}

%%%%%%%%%%%%%%%%%%%%%%%%%%%%%%%%%%%%%%%%%%
\subsection{One-dimensional fluxes}
\label{sec:1d-fluxes}
%%%%%%%%%%%%%%%%%%%%%%%%%%%%%%%%%%%%%%%%%%

With no- or constant-stratification the linear, energetically orthogonal solutions are comprised of sines and cosines with clearly defined vertical scales, and it is easy to define a total wavelength. However, in variable stratification, the energetically orthogonal solutions have a vertical structure with variable length scales, as described by the vertical modes in Table~\ref{tab:vertical-mode-projection} of appendix~\ref{sec:orthogonal-solutions}, and a total wavelength is harder to define. For example, an internal wave packet initially localized to the surface, will change its apparent length scale as it propagates downward entirely under linear dynamics. Thus, we can no longer rely on changes in length scale to indicate nonlinearity. One approach is to stretch the vertical modes on a WKB scaled vertical coordinate \cite{charney1971-jas}, but this method only works for hydrostatic modes, and fails for the non-hydrostatic wave modes used here.

For variable stratification, we define a length scale combining vertical mode and horizontal scales. Just as the horizontal wavelength is $2 \pi/\kappa$, the \emph{deformation wavelength} is $\lambda_d = 2\pi \sqrt{g h} / f$ for both the geostrophic modes (using $h_g^j$) and the wave modes (using $h_\kappa^j$). Note that the name `radius of deformation' is reserved for the geostrophic eigenvalues $\sqrt{g h_g} / f$ with units of length per radian. With these definitions we thus define the pseudo-wavelength as $\lambda_p \equiv 2 \pi/k_p$ where $k_p^2 \equiv k^2 + \ell^2 + f^2/(gh)$.

To visualize and interpret the fluxes (both forced and inertial) we use pseudo-wavelength to collapse the three-dimensional values $(k,\ell,j)$ onto the one-dimensional value,
\begin{equation}
\label{eqn:cumsum-flux-kp}
    \Pi(k_p) \equiv \sum_{k,\ell,j=0}^{k_p} \epsilon^{k \ell j} A^{k \ell j} F^{k \ell j}
\end{equation}
where $F^{k \ell j}$ represents any of the inertial and forcing fluxes. The interpretation of this quantity is such that positive (negative) slopes indicate energy is being deposited (removed) at those wavelengths. For inertial fluxes, whose sums over all $k$ are approximately zero, $\Pi(k_p) > 0$ ($\Pi(k_p)<0$) indicates a cascade to larger (smaller) pseudo-wavelengths, under the assumption that the triad fluxes are local. Note this is opposite to the usual definition where positive slopes indicate energy is being removed, but is necessary to match the sign of the forcing fluxes. This visualization has the important caveat that for the inertial fluxes transfers may not be local, and could jump across wavelength, making the local cascade interpretation incorrect.

The geostrophic fluxes can also be visualized in terms of kinetic energy fraction, the ratio of kinetic energy to total energy, and similarly the wave fluxes can be visualized in terms of intrinsic wave frequency
\begin{equation}
\label{eqn:cumsum-flux-omega}
    \Pi(\omega) \equiv \sum_{k,\ell,j=0}^{\omega^{k\ell j}<\omega} \epsilon_w^{k \ell j} A_\pm^{k \ell j} F^{k \ell j}
\end{equation}
with the same interpretation as described for pseudo-wavelength. These axes provide somewhat orthogonal information, as can be seen in the two-dimensional fluxes below. However, note that the intrinsic wave frequency---the wave frequency predicted from linear theory---is not a measured frequency.

The geostrophic energy fluxes are shown in terms of both pseudo-wavelength (Figure~\ref{fig:energy_flux_1d}a) and kinetic energy fraction (Figure~\ref{fig:energy_flux_1d}b), from which a fairly simple story emerges. Using approximate numbers,
\begin{enumerate}
    \item geostrophic energy is input by the mean-flow forcing at ($\lambda_p \approx 100\,\mathrm{km}$ at 0.61\,GM/yr) and by bottom friction ($\lambda_p \approx 200\,\mathrm{km}$ at 0.13\,GM/yr),
    \item geostrophic energy is removed by direct transfer to waves from the wwg triad ($\lambda_p > 150\,\mathrm{km}$ at 0.30\,GM/yr) and bottom friction ($\lambda_p \approx 250\,\mathrm{km}$ at 0.50\,GM/yr),
    \item geostrophic energy cascades to longer length scales (an inverse cascade) and higher kinetic energy fraction with ggg, while the ggw triad plays no appreciable role. 
\end{enumerate}
In these schematic numbers, this sums to a net loss of 0.07\,GM/yr, a typical amount of temporal variation for the geostrophic energy. 

The wave energy fluxes are shown in terms of both pseudo-wavelength (Figure~\ref{fig:energy_flux_1d}c) and intrinsic frequency (Figure~\ref{fig:energy_flux_1d}d). Here we report in either frequency $\omega$ or pseudo-wavelength, depending on which is more narrow-band. In summary,
\begin{enumerate}
    \item wave energy is input by near-inertial forcing ($\omega \approx f$ at 1.07\,GM/yr), $M_2$ tidal forcing ($\omega \approx 2f$ at 1.17\,GM/yr), and by direct transfer from geostrophic energy from the wwg triad ($\lambda_p < 30\,\mathrm{km}$ at 0.30\,GM/yr),
    \item wave energy is removed by small scale damping ($\lambda_p < 10\,\mathrm{km}$ at 2.15\,GM/yr) and bottom friction ($\omega \approx f$ at 0.25\,GM/yr),
    \item wave energy cascades to shorter length scales (a forward cascade) with both www and wwg, and also shows forward cascade toward smaller wave periods, except from the www triad which fluxes wave energy toward longer periods when $\omega < 2 f$.
\end{enumerate}
The wave energy flux of the www triad as a function of frequency in Figure~\ref{fig:energy_flux_1d}d should be compared to Figure~5c in \citet{wu2023-jfm} which shows the same features in an isolated wave model.

The inertial ranges of geostrophic and wave energy flux are not very broad. The inverse geostrophic cascade, at length scales above 100 km, has both bottom friction and a transfer to waves preventing it from being purely inertial, and is thus also difficult to discern in the geostrophic energy spectrum in Figure~\ref{fig:energy_spectra}c. The inertial range for wave energy is somewhat less contaminated, with relatively weak forcing from bottom friction and geostrophic energy transfer at scales below 30\,km. The result is a nearly constant slope in that range in Figure~\ref{fig:energy_spectra}. The most prominent inertial range is the forward potential enstrophy cascade, discussed in section~\ref{sec:potential-enstrophy-fluxes} below.

%%%%%%%%%%%%%%%%%%%%%%%%
\subsection{Two-dimensional fluxes}
\label{subsec:2d-fluxes}
%%%%%%%%%%%%%%%%%%%%%%%%

To visualize how the energy fluxes from its sources to its sinks, we can use any of the conserved inertial fluxes to construct a vector field with divergence matching the inertial fluxes. Thus, if $F^\textrm{cas}_{jk}$ is a cascade term, then we can find a vector field $(r,s)$ such that,
\begin{equation}
    \frac{\partial r_{jk}}{\partial k} + \frac{\partial s_{jk}}{\partial j} = - F^\textrm{cas}_{jk}
\end{equation}
by defining $\phi_{jk}$ so that $ r_{jk} = \frac{\partial \phi_{jk}}{\partial k}$, $ s_{jk} = \frac{\partial \phi_{jk}}{\partial j}$. The flux must vanish at boundaries, and thus if we express $\phi$ in terms of a cosine basis,
\begin{equation}
    \phi(k,j) = \sum_{m=0}^{N_k} \sum_{n=0}^{N_j} \phi_{mn} \cos(\pi m k) \cos(\pi n j)
\end{equation}
we automatically satisfy the correct boundary conditions. The two-dimensional vector field that results exactly matches the value in the more traditional flux visualization \eqref{eqn:cumsum-flux-kp} when collapsed along one dimension and also suffers from the same limitation that it assumes all transfers are local.

\begin{figure}
\centering
\includegraphics[height=7cm]{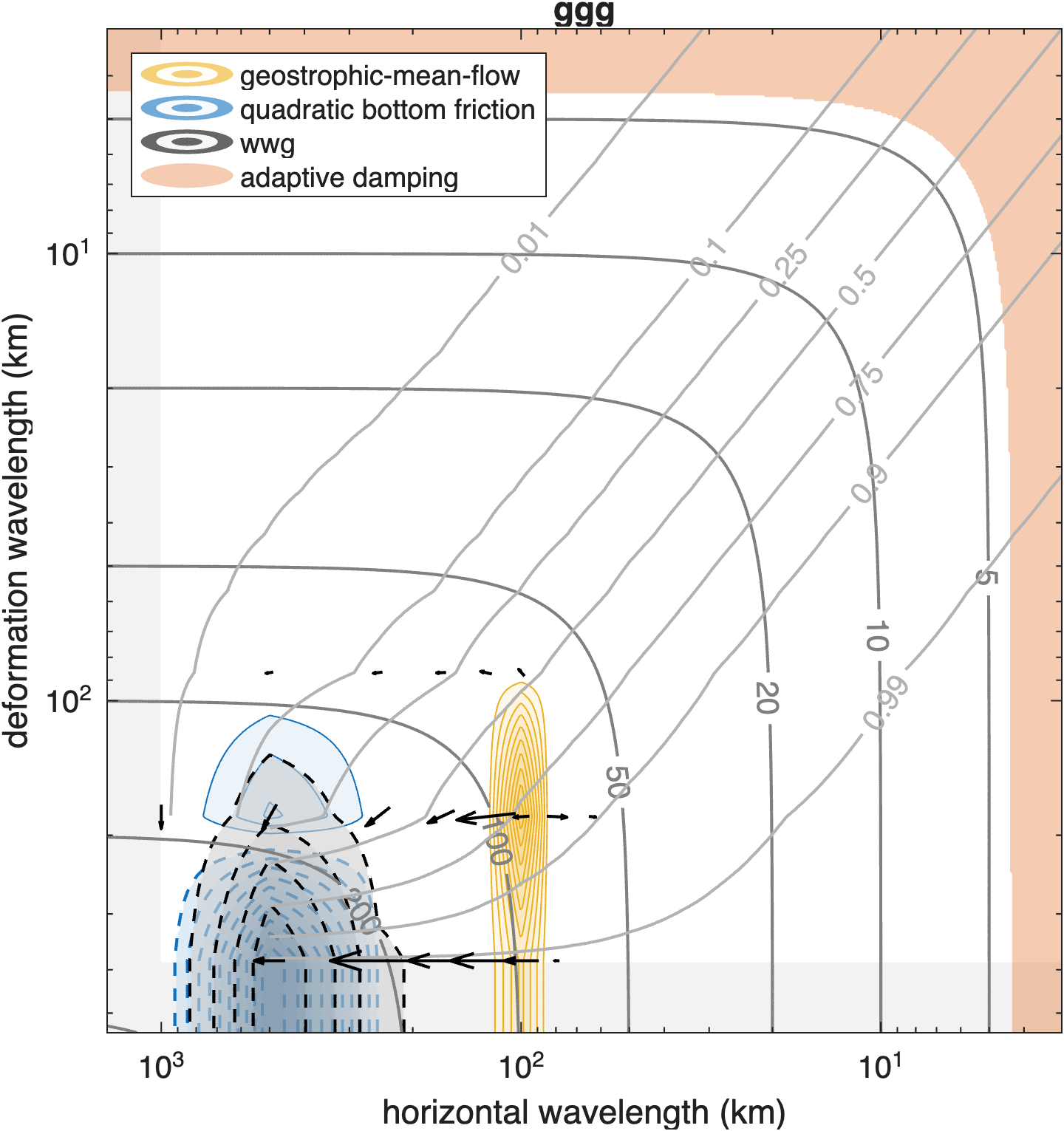}
\caption{ Energy flux of the ggg triad (arrows) and forcing (colored contours) in the geostrophic reservoir. Dark gray contours show pseudo-wavelength while light gray contours show kinetic energy fraction, $\mathrm{KE}/(\mathrm{KE+APE})$. The damping region is shaded light red.}
\label{fig:energy-flux-geostrophic-2d}
\end{figure}

Figure~\ref{fig:energy-flux-geostrophic-2d} clarifies the story that was deduced from the 1-dimensional fluxes in Figure~\ref{fig:energy_flux_1d}.  The bottom friction forcing (blue) is active at long horizontal length scales, removing barotropic energy, and adding a small amount of baroclinic energy (the process is nonlinear, and can thus redistribute energy within the modes). The light dashed gray contours indicate that the wwg triad is removing energy at these same scales that the bottom friction is acting.  The energy flux from the ggg triad moves barotropic energy directly to longer length scales, and also moves barolinic energy toward the longest, barotropic mode. Notably, this ggg triad flux in the geostrophic domain of the MFW simulation looks almost identical to the exact energy flux of the MF experiment in Figure~\ref{fig:energy_flux_2D_flow}. One feature of visualizing the fluxes in two-dimensions is that fluxes are spread over a large number of modes at shorter pseudo-wavelength scales. As a result, the damping term at each individual scale is relatively weak, and falls below the threshold for visualization. We have thus highlighted the damping region in light red.

\begin{figure}[H]
\includegraphics[height=7cm]{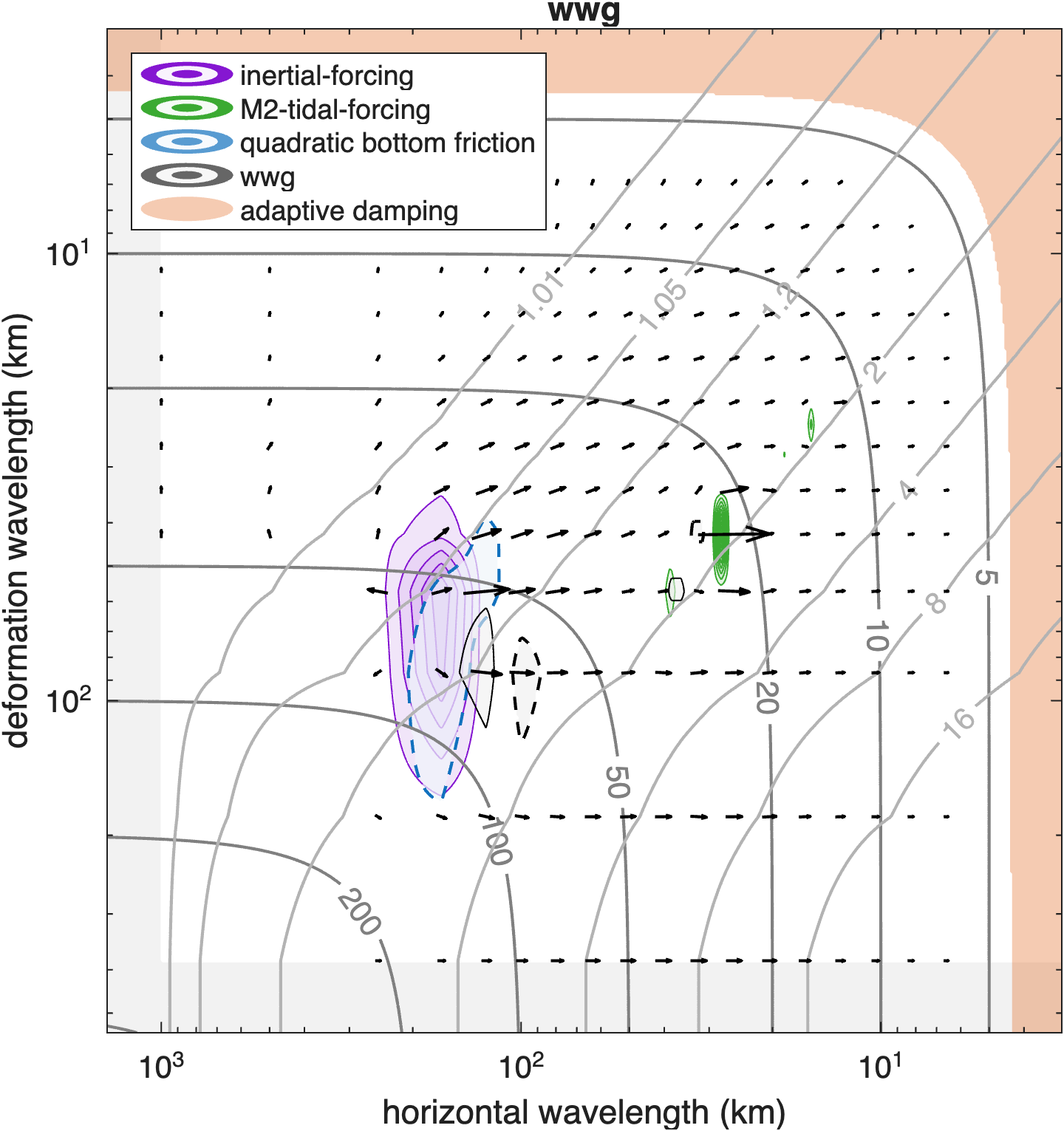}
\hspace{0.5cm}
\includegraphics[height=7cm]{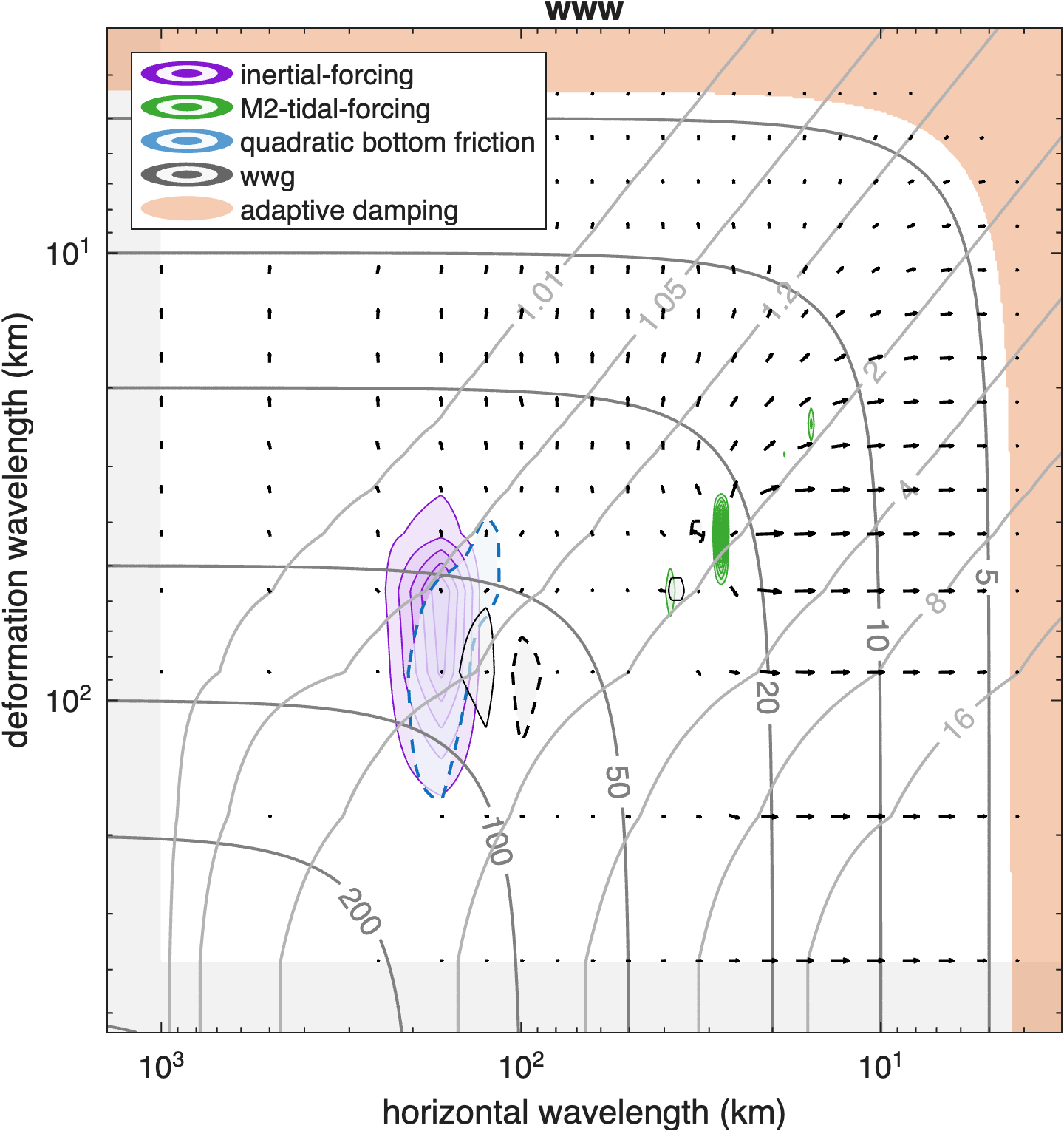}
\caption{ Energy flux of the wwg triad (arrows, left panel) and the www triad (arrows, right panel) and forcing (colored contours) in the wave reservoir. Dark gray contours show pseudo-wavelength (km) and light gray contours show intrinsic frequency (in units of $f$). The damping region is shaded light red (pink). }
\label{fig:energy-flux-wave-2d}
\end{figure}

The wave energy fluxes for the wwg and www triads are shown in Figure~\ref{fig:energy-flux-wave-2d} which reveals several notable features. First, the wwg triad appears to strongly flux energy out of the forcing modes---its largest fluxes are right at and between the inertial (purple) and $M_2$ (green) forcing modes. In contrast, the www triad is strongest in the region $\omega > 2 f$ (above $M_2$), and appears to `pick up' the energy fluxed out of $M_2$ by the wwg triad and flux it towards smaller horizontal wavelength. Second, the www flux appears to move energy almost perfectly orthogonal to the pseudo-wavelength contours. This has the effect of moving energy to lower frequencies for modes below about $2 f$, but moving energy to high frequency otherwise, explaining the inverse frequency cascaded seen in Figure~\ref{fig:energy_flux_1d}. In contrast, the wwg triad appears to mostly flux energy along the lines of constant frequency in the region $\omega < 2 f$

The sum of the ggg, wwg and www triads in Figure~\ref{fig:energy-flux-geostrophic-2d} and \ref{fig:energy-flux-wave-2d} should be comparable to the exact energy flux in Figure~\ref{fig:energy_flux_2D_flow} from the same simulation. The only significant difference comes from the assumption of locality in the flux vector flow. For the quadratic fluxes the energy transfer via the wwg triad was found to be very non-local, transferring large scale geostrophic energy to relatively small scale wave energy.

%%%%%%%%%%%%%%%%%%%%%%%%
\subsection{Separating the closure scales}
\label{subsec:closure}
%%%%%%%%%%%%%%%%%%%%%%%%

The inertial fluxes in Figure~\ref{fig:energy_flux_1d} show a very dramatic change in behavior at the length scales where the damping is strongest. In particular, the wwg transfer term for the geostrophic energy shows no significant transfer at scales below 100\,km, until the damping scales below 6\,km, and shows increasing total transfer to wave energy until the same scales. This suggests that, rather than partition the fluid into two reservoirs, wave and geostrophic, we should consider additional reservoirs that separate the physics of the closure scales from the inertial scales. Formally then, we decompose the fluid into,
\begin{equation}
    u = u_g + u_w + u_{\tilde{g}} + u_{\tilde{w}}
\end{equation}
where $u_g$ and $u_w$ are the the geostrophic and wave part of the fluid at scales above the damping scale of 6\,km, and $u_{\tilde{g}}$, $u_{\tilde{w}}$ are the geostrophic and wave part of the flow at scales below the damping scale. Separating the flow into three or more energy reservoirs introduces yet another tedious calculation which we relegate to appendix~\ref{appendix:transfers-from-triad-fluxes}.

The complete the sources, sinks, reservoirs diagram in Figure~\ref{fig:sources_sinks} clarifies the energy pathways of the these simulations. The most notable result is that there is no geostrophic forward cascade directly to small scales. Instead, geostrophic energy transfers to wave energy, where it is then fluxed to damped geostrophic and wave modes.

\begin{figure}[t]
\centering % [height=6.0cm]
\includegraphics[width=1.0\textwidth]{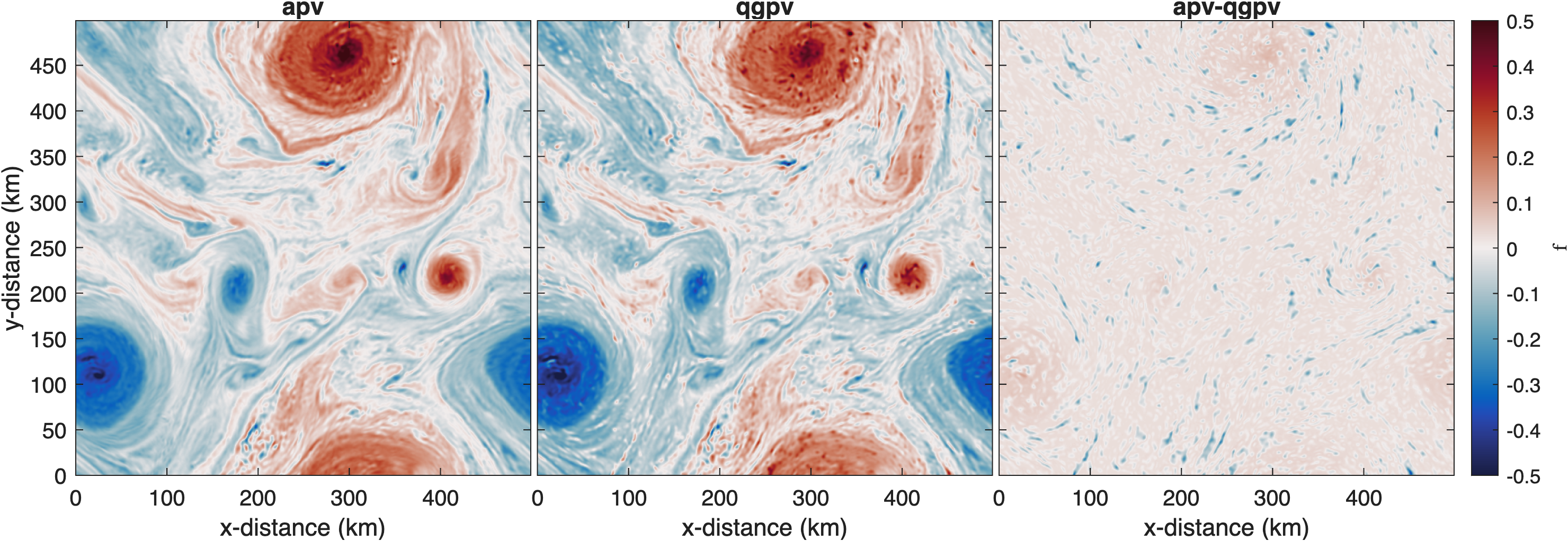}
\caption{Quasigeostrophic potential vorticity (left) and available potential vorticity (right) at the surface from the last day of the MFW simulation.}
\label{fig:qgpv-apv-comparison}
\end{figure}

%%%%%%%%%%%%%%%%%%%%%%%%%%%%%%%%%
%
\section{Discussion}
\label{sec:discussion}
%
%%%%%%%%%%%%%%%%%%%%%%%%%%%%%%%%%

As discussed in the introduction, estimating energy fluxes within and between wave and geostrophic flows requires two conditions: 1) a method for separating waves and vortices and 2) energy orthogonality between the two flow components. Here we examine our proposed approach in the context of these two conditions.

%%%%%%%%%%%%%%%%%%%%%%%%%%%%%%%%%%%%%%%%%%
\subsection{Waves and APV}
\label{sec:waves-and-apv}
%%%%%%%%%%%%%%%%%%%%%%%%%%%%%%%%%%%%%%%%%%

An interpretation of the wave-vortex decomposition as fundamentally about `PV-inversion' emerges from the form of the projection operator for internal-gravity waves (fourth row in Table~\ref{tab:solution-projection}). To separate the internal gravity waves from the rest of the flow, one must \emph{first} compute the potential vorticity, deduce the associated geostrophic streamfunction (a PV-inversion \cite{hoskins1985-qjrms}), and only then can the internal gravity wave structures be deduced. In the case of constant stratification or hydrostatic flows, the vertical mode projection operators all commute, and this fact is quickly obscured in the derivation. At least conceptually then, it is helpful to think about the wave-vortex decomposition as originating from a definition of potential vorticity.

Our theoretical framework is that we define a wave as flow that has no Eulerian signature of available potential vorticity. The strength of this approach is that it does not rely on assumptions about the frequency of oscillation of flow features, but can be diagnosed at any instant in time. The decomposition, however, does not use APV \eqref{eqn:APV-definition}, but the lower order approximation of QGPV \eqref{eqn:qgpv-eta} to separate wave and vortical flow. Figure~\ref{fig:qgpv-apv-comparison} shows that the QGPV is fairly good approximation to APV, although some of the small-scale features are artifacts of the approximation. To quantify fidelity of QGPV as a diagnostic for APV we compare the spectrum of APV and QGPV with respect to the orthogonal QGPV modes as defined in appendix~\ref{appendix:modal-spectra}.

\begin{figure}[t]
\centering
\includegraphics[width=1.0\textwidth]{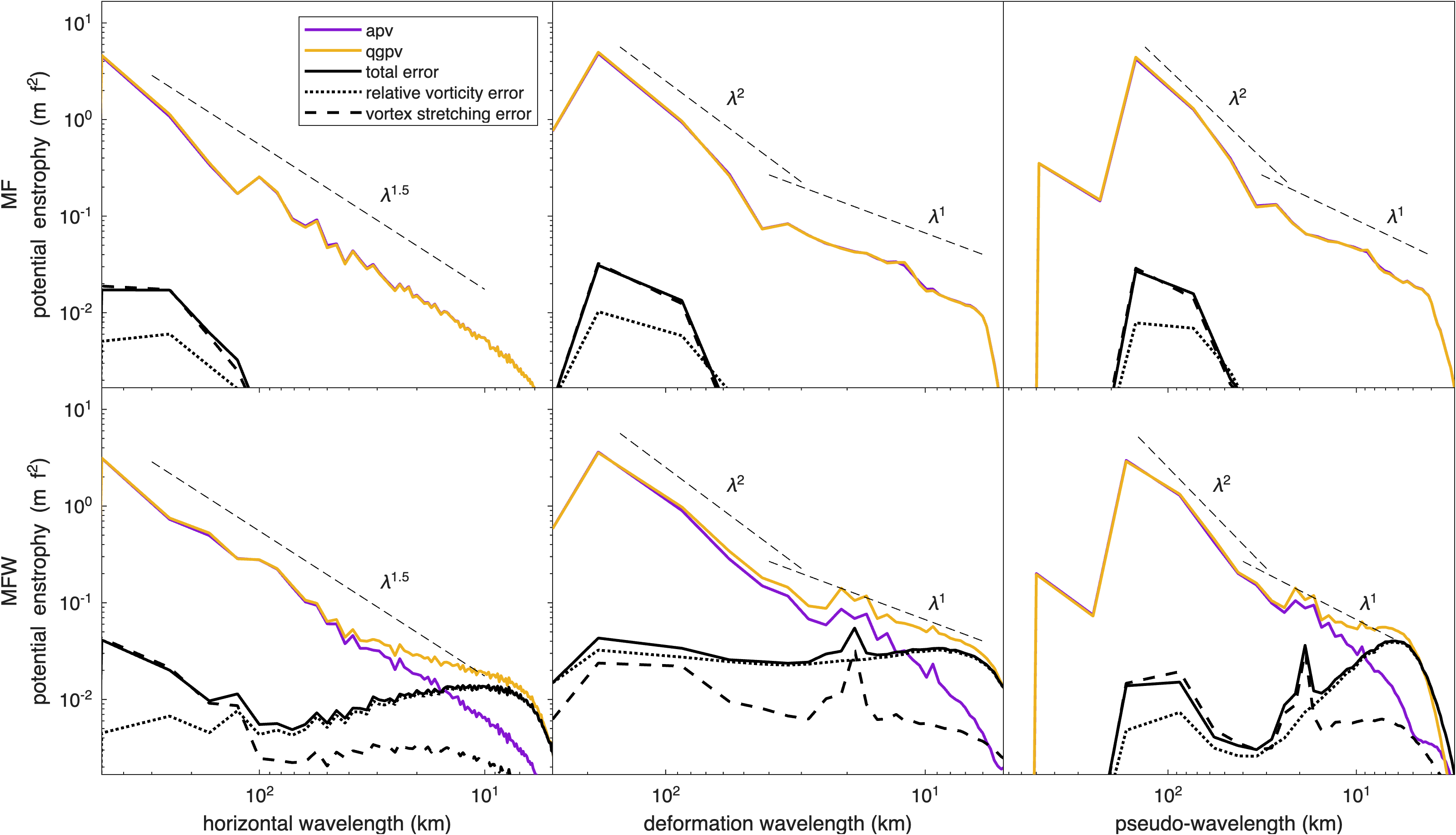}
\caption{Potential enstrophy spectra for the MF (top) and the MFW (bottom) simulations as a function of horizontal wavelength (left), deformation wavelength (middle), and pseudo-wavelength (right). The APV and QGPV spectrum are plotted in blue and red, respectively. The total error (solid black) is divided into the contribution from relative vorticity (dashed black) and vortex stretching (solid black).}
\label{fig:enstrophy_spectra}
\end{figure}

Figure~\ref{fig:enstrophy_spectra} shows the spectra of APV and QGPV in the two simulations. The error spectrum is defined as the spectrum of the difference of these two scalar quantity. From Figure~\ref{fig:enstrophy_spectra} it is clear that QGPV works nearly flawlessly as proxy for APV in the MF simulation largely devoid of waves. However, in the MFW simulation with a significant wave field, QGPV has an $O(1)$ error at wavelengths below about 15\,km. Recall that this is the same scale which the slope of the geostrophic energy spectrum changed in Figure~\ref{fig:energy_spectra}. Along the vertical mode axis the error reaches $O(1)$ by the 15th mode, with a deformation wavelength of 12\,km. Importantly: although the error in potential enstrophy is large at scales below 15 km, the total energy this represents is well below the $0.01$ GM threshold reported in Figure~\ref{fig:sources_sinks}, as seen in Figure~\ref{fig:energy_spectra}.

There are two sources of error in the approximation of APV, equation~\ref{eqn:APV-definition}, to QGPV: dropping the contribution of relative vorticity to vortex stretching, $\left(\nabla \times \vect{u} \right) \cdot \nabla \eta$, and the low order approximation to planetary vortex stretching, $f \partial_z \eta \approx f \partial_z \etae$. The relative vorticity error scales with Rossby number, and becomes a significant error at the submeoscales. The planetary vortex stretching error is a generalization of the height nonlinearity terms identified by \citet{anderson1979-dsr} and \citet{charney1981-book}, applicable to mesoscale eddies. This nonlinearity was explored in \citet{early2024-arxiv} and shown to be very significant within a mesoscale eddy. 

A higher order correction to QGPV inversion can be found using optimal balance, an approach that starts from the linear eigenmodes used here, and uses an iterative procedure to improve decomposition of the potential vorticity part of the flow  \cite{masur2020-gafd,chouksey2023-jfm}. Application of a single backward-forward iteration of optimal balance over one inertial period reduces the APV error of the balanced field by an order of magnitude compared to the linear geostrophic modes, but appears to come at the cost of energy orthogonality.

%%%%%%%%%%%%%%%%%%%%%%%%%%%%%%%%%%%%%%%%%%
\subsection{Quadratic energy}
\label{sec:discussion-quadratic-energy}
%%%%%%%%%%%%%%%%%%%%%%%%%%%%%%%%%%%%%%%%%%

The second major condition for measuring energy and energy fluxes is orthogonality between the energy content of the wave and vortex fields. The method described here has formal depth-integrated orthogonality with respect to quadratic energy, and not the fully conserved non-linear energy, equation~\eqref{total-energy-hm}. However, for the particular problems considered here, we find no significant differences between these two quantities. Based on the linearization of APE, we would only expect significant discrepancies when the vertical displacements are large. 

%%%%%%%%%%%%%%%%%%%%%%%%%%%%%%%%%%%%%%%%%%
\subsection{Potential enstrophy fluxes}
\label{sec:potential-enstrophy-fluxes}
%%%%%%%%%%%%%%%%%%%%%%%%%%%%%%%%%%%%%%%%%%

\begin{figure}[t]
\centering
\includegraphics[width=1.0\textwidth]{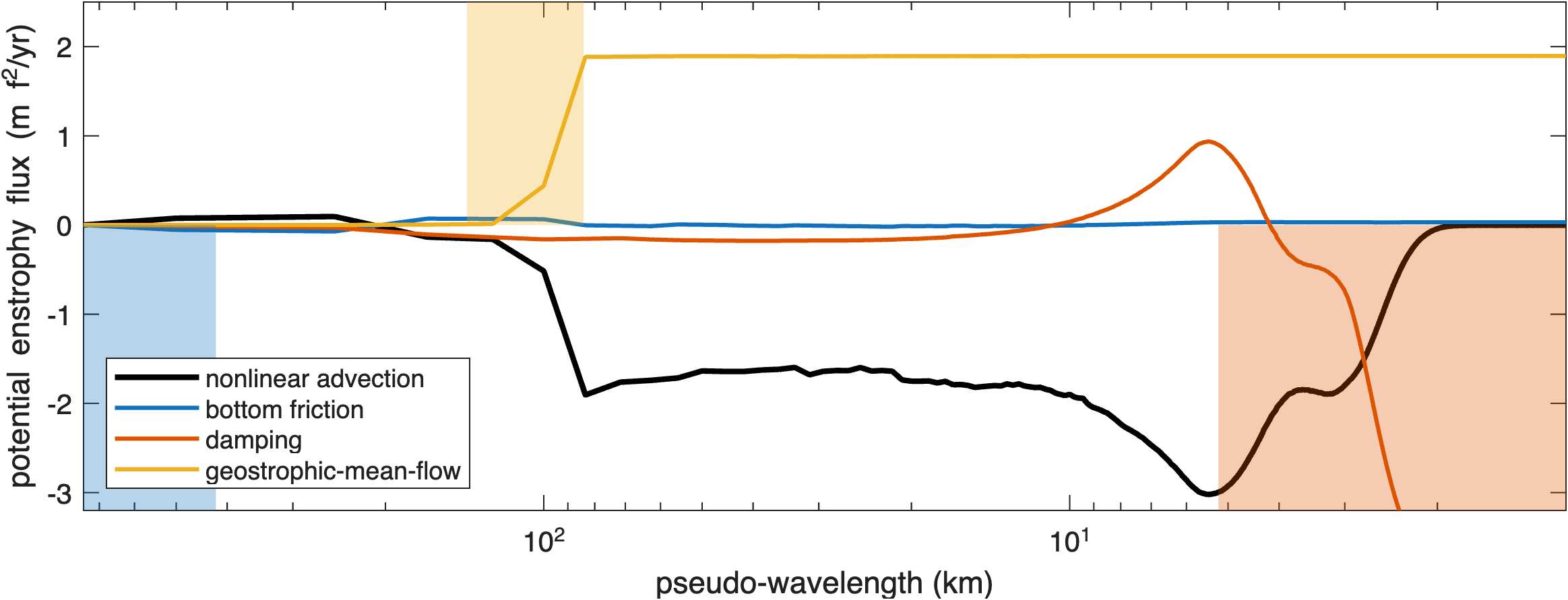}
\caption{Available potential enstrophy flux for the MFW simulation as a function of pseudo-wavelength. }
\label{fig:enstrophy_flux1D}
\end{figure}

The potential enstrophy fluxes are mostly a simple story, as seen in the similarity between the simulations in Figure~\ref{fig:enstrophy_flux_2D_flow}. However, if we produce the one-dimensional potential enstrophy fluxes from equation \eqref{volume-integrated-exact-enstrophy} using the modal cross-spectrum \eqref{eqn:modal-cross-spectrum-f} found in Appendix~\ref{appendix:modal-spectra}, then a slightly more interesting picture emerges in Figure~\ref{fig:enstrophy_flux1D} (which is an integration along lines of constant pseudo-wavelength in Figure~\ref{fig:enstrophy_flux_2D_flow}). The primary story is that mean flow forcing produces potential enstrophy which is fluxed to small scales. Additionally the small-scale damping produces potential enstrophy just outside the damping scales, the rate of which increases dramatically in the higher resolution simulation. Recall that the damping mechanism here uses spectral vanishing viscosity, and thus there is identically zero damping at modes outside these scales. However, because the available potential enstrophy computed here is a nonlinear quantity, its effect is non-local. The non-locality and sensitivity of potential enstrophy production to resolution was shown in a DNS simulation by \citet{waite2013-atm} and, although this simulation is not a DNS, this is the same effect.

%%%%%%%%%%%%%%%%%%%%%%%%%%%%%%%%%%%%%%%%%%
\subsection{Helmholtz decomposition}
\label{sec:discussion-helmholtz-decomposition}
%%%%%%%%%%%%%%%%%%%%%%%%%%%%%%%%%%%%%%%%%%

When presenting this work we are often asked how the wave-vortex decomposition relates to the Helmholtz decomposition of the horizontal velocities. This wave-vortex decomposition reduces to a Helmholtz decomposition if you neglect the buoyancy anomaly, as can be seen by zeroing out $\etae$ in the projection operators shown in Table~\ref{tab:solution-projection}. The consequence of this approximation is that one now requires additional assumptions to make a decomposition because waves have both rotational and divergent components and, simply by degrees-of-freedom counting, a straight decomposition is not possible without more information. In both \citet{barkan2024-jpo} and \citet{shaham2025-james} a temporal filter is applied which relies on a temporal scale separation between the waves and geostrophic flow. This provides an orthogonal decomposition of horizontal kinetic energy in the time domain and, having neglected buoyancy, total energy is no longer decomposed, and the full equations of motion cannot be projected. Notably the same temporal filter could be applied to the wave-vortex decomposition here, as no temporal information is used in this decomposition.

%%%%%%%%%%%%%%%%%%%%%%%%%%%%%%%%%%%%%%%%%%
\subsection{Hydrostatics}
\label{sec:hydrostatics}
%%%%%%%%%%%%%%%%%%%%%%%%%%%%%%%%%%%%%%%%%%

In the course of this work, the MFW simulation was also run under hydrostatic dynamics (not shown). The results were quite similar, with only a few key differences. First, the hydrostatic simulation produced fluxes from the NIO, $M_2$, and damping that are 22\%, 25\%, and 17\% greater than the non-hydrostatic fluxes, respectively. The differences are statistically significant given our precision. Second, and somewhat unexpectedly, the hydrostatic and non-hydrostatic simulations showed nearly perfectly overlapping wave spectra of depth-integrated total energy. The key difference between the hydrostatic and non-hydrostatic simulations appears in the mooring spectra, where the distribution of wave energy above the local buoyancy frequency is significantly higher under hydrostatic dynamics, an effect obscured by the depth-integrated spectrum. We conjecture that the inertial energy cascade preserves total energy fluxed in both the hydrostatic and non-hydrostatic simulations, hence the spectra match, even though the vertical spatial distribution does not.

\section{Conclusions}
\label{sec:conclusions}
%
%%%%%%%%%%%%%%%%%%%%%%%%%%%%%%%%%

This work establishes a framework for computing total energy and potential enstrophy fluxes between orthogonal reservoirs of a fluid in variable stratification. 
A key aspect is that the wave–vortex decomposition is derived from the fully conserved available potential vorticity, enabling its fidelity to be tested.
Secondly, because the decomposition is complete, the full nonlinear equations of motion are expressed in this basis without approximation. 
Within this framework there are two wave cascade mechanisms, two geostrophic cascade mechanisms, and two mechanisms for transferring between waves and geostrophic motions.

Application of the decomposition to mid-ocean simulations with mean-flow, tidal, and near-inertial forcing shows a robust geostrophic inverse cascade with significant transfer of large-scale geostrophic energy to small-scale wave energy, along with forward cascade of wave energy. Further decomposing the fluid into damped (small-scale) and un-damped wave and geostrophic reservoirs, we find no evidence for a forward geostrophic cascade in this forcing regime. In the broadest sense this agrees with other work where large-scale low-frequency motions flux energy to small-scale motions \cite{taylor2016-jpo,taylor2020-jpo,thomas2021-jfm,shaham2025-james}. The work here points to transfers from the wwg triad as the primary energetic pathway from large-scale geostrophic energy to small-scale damped motions as seen in Figure~\ref{fig:sources_sinks}. 

% Errors in the decomopsition stem from two sources, the height nonlinearity in vertical vortex stretching at large scales, and high Rossby number effects at horizontal scales below 15 km.  The amount of energy associated with the potential vorticity field at those scales is insignificant as seen in figure~\ref{fig:energy_spectra}, and thus the conclusions here are on firm footing.

% \begin{enumerate}
%     \item We are in a different regime than \citet{thomas2021-jfm}---our simulations have four times as much geostrophic energy as wave energy, whereas the closest that \citet{thomas2021-jfm} gets to is comparable levels of energy. In that case they some direct transfer of energy, and no-forward cascade. So that is similar.
%     \item Comparing with \citet{shaham2025-james} is difficult as they consider only horizontal kinetic energy which is not a conserved quantity and hence does not lend itself to the same interpretations.
% \end{enumerate}

%%%%%%%%%%%%%%%%%%%%%%%%%%%%%%%%%%%%%%%%%%%%%%%%%%%%%%%%%%%%%%%%%%%%%
% ACKNOWLEDGMENTS
%%%%%%%%%%%%%%%%%%%%%%%%%%%%%%%%%%%%%%%%%%%%%%%%%%%%%%%%%%%%%%%%%%%%%
% \section{Acknowledgements}
\section*{Acknowledgments}
We thank Jonathan Lilly and Roger Samelson for their careful reading and comments on the draft manuscript. Leslie Smith gratefully acknowledges the support by the Deutsche Forschungsgemeinschaft (DFG) through the Research Unit FOR5528. J.~Early, C.~Wortham and M.P.~Lelong were supported by NSF grant OCE-2123740. J.~Early was additionally support by NSF grant OCE-2048552 and NASA award 80NSSC21K1823. G. H-D was supported, in part, by grants UNAM-DGAPA-PAPIIT IN115925 and Conahcyt A1-S-17634. G. H-D would like to thank the hospitality of NorthWest Research Associates and the support of UNAM-PASPA-DGAPA during his sabbatical visit. The authors report no conflicts of interest.

%%%%%%%%%%%%%%%%%%%%%%%%%%%%%%%%%%%%%%%%%%%%%%%%%%%%%%%%%%%%%%%%%%%%%
% Data Availability Statement
%%%%%%%%%%%%%%%%%%%%%%%%%%%%%%%%%%%%%%%%%%%%%%%%%%%%%%%%%%%%%%%%%%%%%

\section*{Data Availability Statement}

Scripts to rerun the model simulations, create the diagnostics, and recreate the figures are found here \cite{james2026-repo}. The scripts depend on two Matlab software packages. The numerical simulations were run using the WaveVortexModel, a fully spectral non-hydrostatic model \cite{gloceankit2025-repo}. The model output was then diagnosed using the wave-vortex-diagnostic tools \cite{wvdiagnostics2025-repo}, which produce energy decompositions and compute the triad fluxes.

%%%%%%%%%%%%%%%%%%%%%%%%%%%%%%%%%%%%%%%%%%%%%%%%%%%%%%%%%%%%%%%%%%%%%
% Bibliography
%%%%%%%%%%%%%%%%%%%%%%%%%%%%%%%%%%%%%%%%%%%%%%%%%%%%%%%%%%%%%%%%%%%%%

\bibliographystyle{agufull08}
\bibliography{references}

@software{wvdiagnostics2025-repo,
  author       = {Jeffrey J. Early and
                  Cimarron Wortham},
  title        = {Energy-Pathways-Group/wave-vortex-model-
                   diagnostics: Initial release
                  },
  month        = nov,
  year         = 2025,
  publisher    = {Zenodo},
  version      = {v1.0.0},
  doi          = {10.5281/zenodo.17671732},
}

@software{james2026-repo,
  author       = {Cimarron Wortham and
                  Jeffrey J. Early},
  title        = {Energy-Pathways-Group/JAMES2026\_EnergyFlux
                  },
  month        = nov,
  year         = 2025,
  publisher    = {Zenodo},
  version      = {v1.0.0},
  doi          = {10.5281/zenodo.17671636},
}

@software{gloceankit2025-repo,
  author       = {Jeffrey J. Early and
                  Leticia Fabre-Lima and
                  Bailey J. Remy and
                  Cimarron Wortham and
                  Miles A. Sundermeyer},
  title        = {Energy-Pathways-Group/GLOceanKit: Non-hydrostatic  wave-vortex model},
  month        = aug,
  year         = 2025,
  publisher    = {Zenodo},
  version      = {v4.0},
  doi          = {10.5281/zenodo.16749006},
}

@article{masur2020-gafd, 
year = {2020}, 
title = {{Optimal balance for rotating shallow water in primitive variables}}, 
author = {Masur, G. T. and Oliver, M.}, 
journal = {Geophysical \& Astrophysical Fluid Dynamics}, 
issn = {0309-1929}, 
doi = {10.1080/03091929.2020.1745789}, 
pages = {429--452}, 
number = {4-5}, 
volume = {114}
}

@article{leboyer2021-jpo, 
year = {2021}, 
title = {{Variability and sources of the internal wave continuum examined from global moored velocity records}}, 
author = {{Le Boyer}, Arnaud and Alford, Matthew H}, 
journal = {Journal of Physical Oceanography}, 
issn = {0022-3670}, 
doi = {10.1175/jpo-d-20-0155.1}
}

@article{wunsch2024-po, 
year = {2024}, 
title = {{A time-average ocean: Thermal wind and flow spirals}}, 
author = {Wunsch, Carl}, 
journal = {Progress in Oceanography}, 
issn = {0079-6611}, 
doi = {10.1016/j.pocean.2024.103206}, 
pages = {103206}, 
volume = {221}
}

@article{dar2001-physd, 
year = {2001}, 
title = {{Energy transfer in two-dimensional magnetohydrodynamic turbulence: formalism and numerical results}}, 
author = {Dar, Gaurav and Verma, Mahendra K. and Eswaran, V.}, 
journal = {Physica D: Nonlinear Phenomena}, 
issn = {0167-2789}, 
doi = {10.1016/s0167-2789(01)00307-4}, 
eprint = {nlin/0109004}, 
pages = {207--225}, 
number = {3}, 
volume = {157}
}

@article{yassin2021-jmp, 
year = {2021}, 
title = {{Normal modes with boundary dynamics in geophysical fluids}}, 
author = {Yassin, Houssam}, 
journal = {Journal of Mathematical Physics}, 
issn = {0022-2488}, 
doi = {10.1063/5.0048273}, 
pages = {093102}, 
number = {9}, 
volume = {62}
}

@article{tailleux2018-jfm, 
year = {2018}, 
title = {{Local available energetics of multicomponent compressible stratified fluids}}, 
author = {Tailleux, Rémi}, 
journal = {Journal of Fluid Mechanics}, 
issn = {0022-1120}, 
doi = {10.1017/jfm.2018.196}, 
eprint = {1712.01051}, 
pages = {R1}, 
volume = {842}
}

@article{lorenz1955-tellus, 
year = {1955}, 
title = {{Available Potential Energy and the Maintenance of the General Circulation}}, 
author = {Lorenz, Edward N.}, 
journal = {Tellus}, 
issn = {0040-2826}, 
doi = {10.1111/j.2153-3490.1955.tb01148.x}, 
pages = {157--167}, 
number = {2}, 
volume = {7}
}

@article{kraichnan1967-pof, 
year = {1967}, 
title = {{Inertial Ranges in Two-Dimensional Turbulence}}, 
author = {Kraichnan, Robert H}, 
journal = {The Physics of Fluids}, 
issn = {0031-9171}, 
doi = {10.1063/1.1762301}, 
pages = {1417--1423}, 
number = {7}, 
volume = {10}
}

@article{charney1971-jas, 
year = {1971}, 
title = {{Geostrophic Turbulence}}, 
author = {Charney, Jule G}, 
journal = {Journal of the Atmospheric Sciences}, 
issn = {0022-4928}, 
doi = {10.1175/1520-0469(1971)028<1087:gt>2.0.co;2}, 
pages = {1087--1095}, 
number = {6}, 
volume = {28}
}

@article{garrett1972-gfd, 
year = {1972}, 
title = {{Space-Time scales of internal waves}}, 
author = {Garrett, Christopher and Munk, Walter}, 
journal = {Geophysical Fluid Dynamics}, 
doi = {10.1080/03091927208236082}, 
pages = {225 -- 264}, 
number = {1}, 
volume = {3}, 
language = {English}
}

@article{gonella1972-dsr, 
year = {1972}, 
title = {{A rotary-component method for analysing meteorological and oceanographic vector time series}}, 
author = {Gonella, J.}, 
pages = {833 -- 846}, 
number = {12}, 
volume = {19}
}

@article{kraichnan1973-jfm, 
year = {1973}, 
title = {{Helical turbulence and absolute equilibrium}}, 
author = {Kraichnan, Robert H.}, 
journal = {Journal of Fluid Mechanics}, 
issn = {1469-7645}, 
doi = {10.1017/s0022112073001837}, 
pages = {745--752}, 
number = {4}, 
volume = {59}
}

@article{mccomas1977-jgr, 
year = {1977}, 
title = {{Resonant interaction of oceanic internal waves}}, 
author = {McComas, C. Henry and Bretherton, Francis P.}, 
journal = {Journal of Geophysical Research}, 
issn = {0148-0227}, 
doi = {10.1029/jc082i009p01397}, 
pages = {1397--1412}, 
number = {9}, 
volume = {82}
}

@article{anderson1979-dsr, 
year = {1979}, 
title = {{Nonlinear propagation of long Rossby waves}}, 
author = {Anderson, David L. T. and Killworth, Peter D.}, 
journal = {Deep Sea Research Part I: Oceanographic Research Papers}, 
doi = {10.1016/0198-0149(79)90046-3}, 
pages = {1033 -- 1049}, 
number = {9}, 
volume = {26}, 
language = {English}
}

@article{salmon1980-gafd, 
year = {1980}, 
title = {{Baroclinic instability and geostrophic turbulence}}, 
author = {Salmon, R}, 
journal = {Geophysical \& Astrophysical Fluid Dynamics}, 
doi = {10.1080/03091928008241178}, 
pages = {167 -- 211}, 
number = {1}, 
volume = {15}, 
language = {English}
}

@article{holliday1981-jfm, 
year = {1981}, 
title = {{On potential energy density in an incompressible, stratified fluid}}, 
author = {Holliday, Dennis and Mc{I}ntyre, Michael E.}, 
journal = {Journal of Fluid Mechanics}, 
issn = {1469-7645}, 
doi = {10.1017/s0022112081001742}, 
pages = {221--225}, 
number = {-1}, 
volume = {107}
}

@incollection{charney1981-book, 
year = {1981}, 
title = {{Oceanic analogues of large-scale atmospheric motions}}, 
author = {Charney, J.G. and Flierl, G.R.}, 
urldate = {0}, 
pages = {504 -- 548}, 
series = {Evolution of Physical Oceanography}
}

@article{hoskins1985-qjrms, 
year = {1985}, 
title = {{On the use and significance of isentropic potential vorticity maps}}, 
author = {Hoskins, B. J. and McIntyre, M. E. and Robertson, A. W.}, 
journal = {Quarterly Journal of the Royal Meteorological Society}, 
issn = {0035-9009}, 
doi = {10.1002/qj.49711147002}, 
pages = {877--946}, 
number = {470}, 
volume = {111}
}

@incollection{olbers1986-igw, 
year = {1986}, 
title = {{Internal gravity waves}}, 
author = {Olbers, Dirk}, 
booktitle = {Landolt-Börnstein - Numerical data and functional relationships in science and technology}, 
pages = {37--82}, 
volume = {3a}, 
publisher = {Springer-Verlag}, 
address = {Berlin}
}

@article{lelong1991-jfm, 
year = {1991}, 
title = {{Internal wave—vortical mode interactions in strongly stratified flows}}, 
author = {Lelong, M -Pascale and Riley, James J}, 
journal = {Journal of Fluid Mechanics}, 
issn = {0022-1120}, 
doi = {10.1017/s0022112091003609}, 
pages = {1}, 
number = {-1}, 
volume = {232}
}

@article{lien1992-jpo, 
year = {1992}, 
title = {{Normal-Mode Decomposition of Small-Scale Oceanic Motions}}, 
author = {Lien, Ren-Chieh and Müller, Peter}, 
journal = {Journal of Physical Oceanography}, 
issn = {0022-3670}, 
doi = {10.1175/1520-0485(1992)022<1583:nmdoss>2.0.co;2}, 
pages = {1583--1595}, 
number = {12}, 
volume = {22}
}

@article{zakharov1992-book, 
year = {1992}, 
title = {{Kolmogorov Spectra of Turbulence I, Wave Turbulence}}, 
author = {Zakharov, Vladimir E. and L’vov, Victor S. and Falkovich, Gregory}, 
journal = {Springer Series in Nonlinear Dynamics}, 
issn = {0940-2535}, 
doi = {10.1007/978-3-642-50052-7}
}

@article{shepherd1993-ao, 
year = {1993}, 
title = {{A unified theory of available potential energy}}, 
author = {Shepherd, Theodore G.}, 
journal = {Atmosphere-Ocean}, 
issn = {0705-5900}, 
doi = {10.1080/07055900.1993.9649460}, 
pages = {1--26}, 
number = {1}, 
volume = {31}
}

@book{frisch1995-book, 
year = {1995}, 
title = {{Turbulence: The Legacy of A. N. Kolmogorov}}, 
author = {Frisch, Uriel}, 
isbn = {9780521457132}, 
publisher = {Cambridge University Press}, 
address = {Cambridge}, 
doi = {10.1017/cbo9781139170666}
}

@article{warn1995-qjrms, 
year = {1995}, 
title = {{Rossby number expansions, slaving principles, and balance dynamics}}, 
author = {Warn, T and Bokhove, O and Shepherd, TG}, 
journal = {Quarterly Journal of the Royal Meteorological Society}, 
doi = {10.1002/qj.49712152313}, 
pages = {723 -- 739}, 
number = {523}, 
volume = {121}, 
language = {English}
}

@article{bartello1995-jas, 
year = {1995}, 
title = {{Geostrophic adjustment and inverse cascades in rotating stratified turbulence}}, 
author = {Bartello, P}, 
journal = {Journal of the Atmospheric Sciences}, 
doi = {10.1175/1520-0469(1995)052<4410:gaaici>2.0.co;2}, 
pages = {4410 -- 4428}, 
number = {24}, 
volume = {52}, 
language = {English}
}

@article{young1997-jmr, 
year = {1997}, 
title = {{Propagation of near-inertial oscillations through a geostrophic flow}}, 
author = {Young, W. R. and Ben Jelloul, Mahdi}, 
journal = {Journal of Marine Research}, 
issn = {0022-2402}, 
doi = {10.1357/0022240973224283}, 
pages = {735--766}, 
number = {4}, 
volume = {55}
}

@book{salmon1998-book, 
year = {1998}, 
title = {{Lectures on Geophysical Fluid Dynamics}}, 
author = {Salmon, Rick}, 
publisher = {Oxford University Press}, 
address = {New York}, 
doi = {10.1093/oso/9780195108088.001.0001}
}

@article{smith1999-pof, 
year = {1999}, 
title = {{Transfer of energy to two-dimensional large scales in forced, rotating three-dimensional turbulence}}, 
author = {Smith, Leslie M. and Waleffe, Fabian}, 
journal = {Physics of Fluids}
}

@article{riley2000-arfm, 
year = {2000}, 
title = {{Fluid motions in the presence of strong stable stratification}}, 
author = {Riley, J J and Lelong, M Pascale}, 
journal = {Annual Reviews in Fluid Mechanics}, 
doi = {10.1146/annurev.fluid.32.1.613}, 
pages = {613 -- 657}, 
number = {1}, 
volume = {32}, 
language = {English}
}

@article{smith2002-jfm, 
year = {2002}, 
title = {{Generation of slow large scales in forced rotating stratified turbulence}}, 
author = {Smith, Leslie M. and Waleffe, Fabian}, 
journal = {Journal of Fluid Mechanics}, 
doi = {10.1017/s0022112001006309}, 
pages = {145--168}
}

@article{wunsch2004-arfm, 
year = {2004}, 
title = {{Vertical Mixing, Energy, and the General Circulation of the Oceans}}, 
author = {Wunsch, Carl and Ferrari, Raffaele}, 
issn = {0066-4189}, 
doi = {10.1146/annurev.fluid.36.050802.122121}, 
pages = {281--314}, 
number = {1}, 
volume = {36}
}

@article{waite2006b-jfm, 
year = {2006}, 
title = {{The transition from geostrophic to stratified turbulence}}, 
author = {Waite, Michael L. and Bartello, Peter}, 
journal = {Journal of Fluid Mechanics}, 
issn = {1469-7645}, 
doi = {10.1017/s0022112006002060}, 
pages = {89--108}, 
volume = {568}
}

@article{waite2006-jfm, 
year = {2006}, 
title = {{Stratified turbulence generated by internal gravity waves}}, 
author = {Waite, Michael L. and Bartello, Peter}, 
journal = {Journal of Fluid Mechanics}, 
issn = {1469-7645}, 
doi = {10.1017/s0022112005007111}, 
pages = {313--339}, 
volume = {546}, 
month = {1}
}

@article{roullet2008-jfm, 
year = {2008}, 
title = {{Available potential energy diagnosis in a direct numerical simulation of rotating stratified turbulence}}, 
author = {Roullet, Guillaume and Klein, Patrice}, 
journal = {Journal of Fluid Mechanics}, 
issn = {0022-1120}, 
doi = {10.1017/s0022112008004473}, 
pages = {45--55}, 
volume = {624}
}

@article{remmel2009-jfm, 
year = {2009}, 
title = {{New intermediate models for rotating shallow water and an investigation of the preference for anticyclones}}, 
author = {Remmel, Mark and Smith, Leslie}, 
journal = {Journal of Fluid Mechanics}, 
issn = {0022-1120}, 
doi = {10.1017/s0022112009007897}, 
pages = {321--359}, 
volume = {635}
}

@article{gertz2009-jpo, 
year = {2009}, 
title = {{Near-Inertial Oscillations and the Damping of Midlatitude Gyres: A Modeling Study}}, 
author = {Gertz, Aaron and Straub, David N.}, 
journal = {Journal of Physical Oceanography}, 
issn = {0022-3670}, 
doi = {10.1175/2009jpo4058.1}, 
pages = {2338--2350}, 
number = {9}, 
volume = {39}
}

@article{polzin2011-rg, 
year = {2011}, 
title = {{Toward Regional Characterizations of the Oceanic Internal Wavefield}}, 
author = {Polzin, K L and Lvov, Y V}, 
journal = {Reviews of Geophysics}, 
doi = {10.1029/2010rg000329}, 
pages = {894 -- 57}, 
number = {4}, 
volume = {49}, 
language = {English}, 
month = {11}
}

@article{chelton2011-pio, 
year = {2011}, 
title = {{Global observations of nonlinear mesoscale eddies}}, 
author = {Chelton, Dudley B. and Schlax, Michael G. and Samelson, Roger M.}, 
journal = {Progress In Oceanography}, 
doi = {10.1016/j.pocean.2011.01.002}, 
pages = {167 -- 216}, 
number = {2}, 
volume = {91}, 
language = {English}
}

@article{vanneste2013-arfm, 
year = {2013}, 
title = {{Balance and Spontaneous Wave Generation in Geophysical Flows}}, 
author = {Vanneste, J.}, 
journal = {Fluid Mechanics}, 
issn = {0066-4189}, 
doi = {10.1146/annurev-fluid-011212-140730}, 
pages = {147--172}, 
number = {1}, 
volume = {45}
}

@article{winter2013-jfm, 
year = {2013}, 
title = {{Available potential energy density for Boussinesq fluid flow}}, 
author = {Winters, Kraig B. and Barkan, Roy}, 
journal = {Journal of Fluid Mechanics}, 
issn = {0022-1120}, 
doi = {10.1017/jfm.2012.493}, 
pages = {476--488}, 
volume = {714}
}

@article{smith2013-jpo, 
year = {2013}, 
title = {{A Surface-Aware Projection Basis for Quasigeostrophic Flow}}, 
author = {Smith, K. Shafer and Vanneste, Jacques}, 
journal = {Journal of Physical Oceanography}, 
doi = {10.1175/jpo-d-12-0107.1}, 
pages = {548 -- 562}, 
number = {3}, 
volume = {43}, 
language = {English}
}

@article{hernandez2014-jfm, 
year = {2014}, 
title = {{Investigation of Boussinesq dynamics using intermediate models based on wave–vortical interactions}}, 
author = {Hernandez-Duenas, Gerardo and Smith, Leslie M and Stechmann, Samuel N}, 
journal = {Journal of Fluid Mechanics}, 
issn = {0022-1120}, 
doi = {10.1017/jfm.2014.138}, 
pages = {247--287}, 
volume = {747}
}

@article{wagner2015-jfm, 
year = {2015}, 
title = {{Available potential vorticity and wave-averaged quasi-geostrophic flow}}, 
author = {Wagner, G. L. and Young, W. R.}, 
journal = {Journal of Fluid Mechanics}, 
issn = {0022-1120}, 
doi = {10.1017/jfm.2015.626}, 
pages = {401--424}, 
volume = {785}
}

@article{xie2015-jfm, 
year = {2015}, 
title = {{A generalised-Lagrangian-mean model of the interactions between near-inertial waves and mean flow}}, 
author = {Xie, J.-H. and Vanneste, J.}, 
journal = {Journal of Fluid Mechanics}, 
issn = {0022-1120}, 
doi = {10.1017/jfm.2015.251}, 
eprint = {1411.3748}, 
pages = {143--169}, 
volume = {774}
}

@article{kelly2016-jpo, 
year = {2016}, 
title = {{The Vertical Mode Decomposition of Surface and Internal Tides in the Presence of a Free Surface and Arbitrary Topography}}, 
author = {Kelly, Samuel M.}, 
journal = {Journal of Physical Oceanography}, 
doi = {10.1175/jpo-d-16-0131.1}, 
pages = {3777 -- 3788}, 
number = {12}, 
volume = {46}, 
language = {English}
}

@article{rocha2016-jpo, 
year = {2016}, 
title = {{On Galerkin approximations of the surface active quasigeostrophic equations}}, 
author = {Rocha, C B and Young, W R and Grooms, Ian}, 
journal = {Journal of Physical Oceanography}, 
doi = {10.1175/jpo-d-15-0073.1}, 
pages = {125--139}, 
volume = {46}
}

@article{taylor2016-jpo, 
year = {2016}, 
title = {{Forced Near-Inertial Motion and Dissipation of Low-Frequency Kinetic Energy in a Wind-Driven Channel Flow}}, 
author = {Taylor, Stephanne and Straub, David}, 
journal = {Journal of Physical Oceanography}, 
issn = {0022-3670}, 
doi = {10.1175/jpo-d-15-0060.1}, 
pages = {79--93}, 
number = {1}, 
volume = {46}
}

@article{wagner2016-jfm, 
year = {2016}, 
title = {{A three-component model for the coupled evolution of near-inertial waves, quasi-geostrophic flow and the near-inertial second harmonic}}, 
author = {Wagner, G. L. and Young, W. R.}, 
journal = {Journal of Fluid Mechanics}, 
issn = {0022-1120}, 
doi = {10.1017/jfm.2016.487}, 
pages = {806--837}, 
volume = {802}
}

@article{barkan2017-jpo, 
year = {2017}, 
title = {{Stimulated Imbalance and the Enhancement of Eddy Kinetic Energy Dissipation by Internal Waves}}, 
author = {Barkan, Roy and Winters, Kraig B and McWilliams, James C}, 
journal = {Journal of Physical Oceanography}, 
issn = {0022-3670}, 
doi = {10.1175/jpo-d-16-0117.1}, 
pages = {181--198}, 
number = {1}, 
volume = {47}
}

@article{eden2019-jpo, 
year = {2019}, 
title = {{Numerical evaluation of energy transfers in internal gravity wave spectra of the ocean}}, 
author = {Eden, Carsten and Pollmann, Friederike and Olbers, Dirk}, 
journal = {Journal of Physical Oceanography}, 
issn = {0022-3670}, 
doi = {10.1175/jpo-d-18-0075.1}, 
pages = {737--749}, 
number = {3}, 
volume = {49}
}

@article{eden2020-jpo, 
year = {2020}, 
title = {{Towards a global spectral energy budget for internal gravity waves in the ocean}}, 
author = {Eden, Carsten and Pollmann, Friederike and Olbers, Dirk}, 
journal = {Journal of Physical Oceanography}, 
issn = {0022-3670}, 
doi = {10.1175/jpo-d-19-0022.1}, 
pages = {935--944}, 
number = {4}, 
volume = {50}
}

@article{zagar2020-book, 
year = {2020}, 
title = {{Modal View of Atmospheric Variability, Applications of Normal-Mode Function Decomposition in Weather and Climate Research}}, 
author = {Zagar, Nedjeljka}, 
journal = {Mathematics of Planet Earth}, 
issn = {2524-4264}, 
doi = {10.1007/978-3-030-60963-4}
}

@article{taylor2020-jpo, 
year = {2020}, 
title = {{Effects of Adding Forced Near-Inertial Motion to a Wind-Driven Channel Flow}}, 
author = {Taylor, Stephanne and Straub, David}, 
journal = {Journal of Physical Oceanography}, 
issn = {0022-3670}, 
doi = {10.1175/jpo-d-19-0299.1}, 
pages = {2983--2996}, 
number = {10}, 
volume = {50}
}

@article{xie2020-jfm, 
year = {2020}, 
title = {{Downscale transfer of quasigeostrophic energy catalyzed by near-inertial waves}}, 
author = {Xie, Jin-Han}, 
journal = {Journal of Fluid Mechanics}, 
issn = {0022-1120}, 
doi = {10.1017/jfm.2020.709}, 
pages = {A40}, 
volume = {904}
}

@article{hernandez2021-jpo, 
year = {2021}, 
title = {{Impact of Wave-Vortical Interactions on Oceanic Submesoscale Lateral Dispersion}}, 
author = {Hernández-Dueñas, Gerardo and Lelong, M -Pascale and Smith, Leslie M}, 
journal = {Journal of Physical Oceanography}, 
issn = {0022-3670}, 
doi = {10.1175/jpo-d-20-0299.1}
}

@article{thomas2021-jfm, 
year = {2021}, 
title = {{Forward flux and enhanced dissipation of geostrophic balanced energy}}, 
author = {Thomas, Jim and Daniel, Don}, 
journal = {Journal of Fluid Mechanics}, 
issn = {0022-1120}, 
doi = {10.1017/jfm.2020.1026}, 
pages = {A60}, 
volume = {911}
}

@article{vasy2021-qjrms, 
year = {2021}, 
title = {{A high‐accuracy global prognostic model for the simulation of Rossby and gravity wave dynamics}}, 
author = {Vasylkevych, Sergiy and Žagar, Nedjeljka}, 
journal = {Quarterly Journal of the Royal Meteorological Society}, 
issn = {0035-9009}, 
doi = {10.1002/qj.4006}, 
pages = {1989--2007}, 
number = {736}, 
volume = {147}
}

@article{early2021-jfm, 
year = {2021}, 
title = {{A generalized wave-vortex decomposition for rotating Boussinesq flows with arbitrary stratification}}, 
author = {Early, Jeffrey J. and Lelong, M.P. and Sundermeyer, M.A.}, 
journal = {Journal of Fluid Mechanics}, 
issn = {0022-1120}, 
doi = {10.1017/jfm.2020.995}, 
pages = {A32}, 
volume = {912}
}

@article{early2022-arxiv, 
year = {2022}, 
title = {{Exact expressions for available potential energy and available potential vorticity}}, 
author = {Early, Jeffrey J and Hernández-Dueñas, Gerardo and Smith, Leslie M and Lelong, M -Pascale}, 
journal = {arXiv}, 
doi = {10.48550/arxiv.2212.07405}, 
eprint = {2212.07405}
}

@article{waite2013-atm, 
year = {2023}, 
title = {{Potential Vorticity Generation in Breaking Gravity Waves}}, 
author = {Waite, Michael L. and Richardson, Nicholas}, 
journal = {Atmosphere}, 
doi = {10.3390/atmos14050881}, 
pages = {881}, 
number = {5}, 
volume = {14}
}

@article{wu2023-jfm, 
year = {2023}, 
title = {{Energy cascade in the Garrett–Munk spectrum of internal gravity waves}}, 
author = {Wu, Yue and Pan, Yulin}, 
journal = {Journal of Fluid Mechanics}, 
issn = {0022-1120}, 
doi = {10.1017/jfm.2023.862}, 
eprint = {2305.13110}, 
pages = {A11}, 
volume = {975}
}

@article{chouksey2023-jfm, 
year = {2023}, 
title = {{A comparison of methods to balance geophysical flows}}, 
author = {Chouksey, Manita and Eden, Carsten and Masur, Gökce Tuba and Oliver, Marcel}, 
journal = {Journal of Fluid Mechanics}, 
issn = {0022-1120}, 
doi = {10.1017/jfm.2023.602}, 
pages = {A2}, 
volume = {971}
}

@article{jlab2024, 
year = {2024}, 
title = {{jLab: A data analysis package for Matlab, v.1.7.3}}, 
author = {Lilly, Jonathan M.}, 
journal = {Zenodo}, 
doi = {10.5281/zenodo.4547006}
}

@article{barkan2024-jpo, 
year = {2024}, 
title = {{Eddy–Internal Wave Interactions: Stimulated Cascades in Cross-Scale Kinetic Energy and Enstrophy Fluxes}}, 
author = {Barkan, Roy and Srinivasan, Kaushik and McWilliams, James C}, 
journal = {Journal of Physical Oceanography}, 
issn = {0022-3670}, 
doi = {10.1175/jpo-d-23-0191.1}, 
pages = {1309--1326}, 
number = {6}, 
volume = {54}
}

@article{early2024-arxiv, 
year = {2024}, 
title = {{Available potential vorticity and the wave-vortex decomposition for arbitrary stratification}}, 
author = {Early, Jeffrey J and Hernández-Dueñas, Gerardo and Smith, Leslie M and Lelong, M -Pascale}, 
journal = {arXiv}, 
doi = {10.48550/arxiv.2403.20269}, 
eprint = {2403.20269}
}

@article{shaham2025-james, 
year = {2025}, 
title = {{Spectral Flux Decomposition in a Wind‐Driven Channel Flow With Near‐Inertial Waves}}, 
author = {Shaham, Michal and Barkan, Roy}, 
journal = {Journal of Advances in Modeling Earth Systems}, 
issn = {1942-2466}, 
doi = {10.1029/2023ms004036}, 
number = {1}, 
volume = {17}
}

%%%%%%%%%%%%%%%%%%%%%%%%%%%%%%%%%%%%%%%%%%%%%%%%%%%%%%%%%%%%%%%%%%%%%
% Appendixes
%%%%%%%%%%%%%%%%%%%%%%%%%%%%%%%%%%%%%%%%%%%%%%%%%%%%%%%%%%%%%%%%%%%%%

\appendix

%%%%%%%%%%%%%%%%%%%%%%%%%%%%%%%%%%%%%%%%%%
%
\section{Orthogonal solutions}
\label{sec:orthogonal-solutions}
%
%%%%%%%%%%%%%%%%%%%%%%%%%%%%%%%%%%%%%%%%%%
%%%%%%%%%%%%%%%%%%%%%%%%%%%%%%%%%%%%

Here we express the energetically and enstrophically orthogonal solutions and derive the projection operators that follow from orthogonality which are required to express the equations of motion with respect to the wave-vortex basis. Appendix~\ref{sec:proof-of-orthogonality} proves orthogonality for the modes.

%%%%%%%%%%%%%%%%%%%%%%%%
%
\subsection{Inner product space and bra-ket notation}
\label{sec:inner-product}
%
%%%%%%%%%%%%%%%%%%%%%%%%%

The energy in equation \eqref{eqn:energy-volume-integral} takes the form of an inner-product on the sum of the squares of the 4 variables $(u,v,w,\etae)$. The enstrophy in equation \eqref{eqn:enstrophy-volume-integral} is an inner-product on the square of one variable,  the vertical component of vorticity $\partial_x v - \partial_y u -f \partial_z \etae$. Conceptually then, these two inner-products have the same general form if we imagine a two step process: first we produce an $m$-dimensional vector from the dynamical variables (e.g. velocity or vorticity), and second we dot the vector with some inner-product operator. A good notion for this idea is the bra-ket notation.

% These are useful notes from John Wheeler:
%https://www.physics.usu.edu/Wheeler/QuantumMechanics/QM11ChangeOfBasis.pdf

The linear equations of motion and conservation laws expressed in section~\ref{subsec:eqns-of-motion} can be written in bra-ket notation. Solutions of the system are represented as kets, e.g. $\ket{\psi}$ replaces the vector $\psi(\vect{x},t)$ in equation~\eqref{eqn:uvw-state-vector-simple}, while matrix and differential operators are still expressed as, e.g., $\op{L}$. We start by defining time-dependent state-vector $\ket{\psi(t)}$, such that
\begin{equation}
\label{eqn:uvw-state-vector}
\ket{\psi(t)}_{\mathcal{U}} = 
\begin{bmatrix}
    u(\vect{x},t)\\ v(\vect{x},t) \\ w(\vect{x},t)\\ \etae(\vect{x},t) \\ p_\textrm{e}(\vect{x},t)
\end{bmatrix}
\end{equation}
where $\mathcal{U} = \left\{ \ket{u(\vect{x})}, \ket{v(\vect{x})}, \ket{w(\vect{x})}, \ket{\etae(\vect{x})}, \ket{p_\textrm{e}(\vect{x})} \right\}$ is an ordered basis for these observable flow features, such that, e.g., $u(\vect{x},t) \equiv \qprod{u(\vect{x})}{\psi(t)}$. In general we will drop the explicit dependence on $\vect{x}$ and $t$, as well as the subscript $\mathcal{U}$ until we consider a basis transformation. With this notation, the linear system of equations can be written as,
\begin{equation}
\label{eqn:linear-momentum-braket}
    \timeop \ket{\psi} + \op{\linop} \ket{\psi} = 0
\end{equation}
using the same operators defined in section~\ref{subsec:eqns-of-motion}. 
Energy operator $\op{H}$ and enstrophy operator $\op{Z}$ are expressed as
\begin{equation}
\op{H} \equiv = \frac{1}{2}
    \begin{bmatrix}
        1 & 0 & 0 & 0   & 0\\
        0 & 1 & 0 & 0   & 0\\
        0 & 0 & 1 & 0   & 0\\
        0 & 0 & 0 & N^2 & 0 \\
        0 & 0 & 0 & 0   & 0
    \end{bmatrix}, \;
\op{Z} \equiv \frac{1}{2}, \;
    \textrm{and} \;
 \op{Q} \equiv 
    \begin{bmatrix}
        -\partial_y & \partial_x & 0 & - f \partial_z & 0
    \end{bmatrix}
\end{equation}
where $\op{Q}$ produces the vertical component of potential vorticity. With these definitions the volume 
integrated energy of \eqref{eqn:energy-volume-integral} is now
\begin{equation}
\label{eqn:energy-braket}
   \mathcal{E} = \braopket{\psi}{\op{H}}{\psi}
\end{equation}
and volume integrated enstrophy from \eqref{eqn:enstrophy-volume-integral}
\begin{equation}
\label{eqn:enstrophy-braket}
\mathcal{Z} = \braopket{\op{Q} \psi}{\op{Z}}{\op{Q}\psi}.
\end{equation}
It is the inner-products of energy \eqref{eqn:energy-braket} and enstrophy \eqref{eqn:enstrophy-braket} that motivate the use of bra-ket notation for this problem. When considering purely geostrophic flows, e.g. \cite{smith2013-jpo}, both inner-products can be expressed in terms of a single scalar field (a streamfunction). In the full Boussinesq flow considered here, the energy inner-product involves a 4-component vector and the enstrophy inner-product involves the scalar potential vorticity.

Linear momentum evolution in \eqref{eqn:linear-momentum-braket} can be equivalently expressed as,
\begin{subequations}
\label{eqn:braket-linear-eom}
    \begin{align}
        \op{H} \timeop \ket{\psi} + \op{H} \op{\linop} \ket{\psi} =& 0 \\
        \partial_t \op{H} \ket{\psi} + \op{H} \op{\linop} \ket{\psi} =& 0
    \end{align}
\end{subequations}
while the linear quasigeostrophic potential vorticity evolution equation reduces to,
\begin{equation}
    \partial_t \ket{\op{Q} \psi} = 0
\end{equation}
because $\op{Q} \op{\linop} \ket{\psi} = 0$.

%%%%%%%%%%%%%%%%%%%%%%%%%%%%%%%%%%%%%%%%%%
%
\subsection{The solutions}
%
%%%%%%%%%%%%%%%%%%%%%%%%%%%%%%%%%%%%%%%%%%

The central idea is that any observable state-vector, $\ket{\psi}$, is composed energetically and enstrophically orthogonal observable solutions. The four major solution types, summarized in Table~\ref{tab:solutions}, are internal gravity waves $\ket{\wmode}$, inertial oscillations $\ket{\iomode}$, geostrophic motions $\ket{\gmode}$, and the mean-density anomaly $\ket{\mdamode}$. Thus, any observable state-vector can be expressed as,
\begin{equation}
\label{eqn:psi-expanded}
    \ket{\psi} = \sum_{k\ell j} \left( A_0^{k\ell j} \ket{\gmode} + A_\pm^{k\ell j} \ket{\wmode}  + \textrm{c.c.} \right) + \sum_{j} \left( A_-^{00j} \ket{\iomode} + \textrm{c.c.} \right) + \sum_{j} A_0^{00j} \ket{\mdamode}
\end{equation}
where $A_0^{k\ell j}$, $A_+^{k\ell j}$, and $A_-^{k\ell j}$ are time-dependent coefficient matrices. The notation here is such that the $A_0^{k\ell j}$ coefficients multiply modes with a potential vorticity signature, while the $A_\pm^{k\ell j}$ coefficients multiply wave modes with frequencies $\pm \omega_\kappa^j$. To recover a coefficient, e.g., $A_0^{k\ell j}$, we dot the state-vector $\ket{\psi}$ against an orthogonal solution and normalize, e.g.,
\begin{equation}
\label{eqn:wmode-coefficient}
    A_\pm^{k\ell j}(t) = \frac{\braopket{\wmode}{\op{H}}{\psi}}{\braopket{\wmode}{\op{H}}{\wmode}}
\end{equation}
recovers the coefficient from \eqref{eqn:psi-expanded} for the wave solution $\ket{\wmode}$ by exploiting energy orthogonality. The projection operators, energy and enstrophy are summarized in Table~\ref{tab:solution-projection}, and detailed in the sections that follow.

That this basis is complete means that we can express the fluid state in terms of the wave-vortex eigenmodes, i.e.,
\begin{equation}
\label{eqn:wave-vortex-vector}
 \ket{ \psi(t) }_{\mathcal{A}} \equiv
    \begin{bmatrix}
        A_0^{00j}(t) \\
        A_0^{k\ell j}(t) \\
        A_-^{00j}(t) \\
        A_\pm^{k\ell j} (t)
    \end{bmatrix}
\end{equation}
where $\mathcal{A} = \left\{  \ket{\mdamode}, \ket{\gmode}, \ket{\iomode}, \ket{\wmode} \right\}$ is an ordered basis, exactly as was done in  $\eqref{eqn:uvw-state-vector}$. An additional consequence is that the combination of the forward and inverse projection form the identity matrix, analogous to writing $P P^{-1}$ for some matrix operator. In the notation here, this means that the wave-vortex projection operator $\mathcal{P}$ can be defined as
\begin{equation}
\label{eqn:wave-vortex-projection}
    \mathcal{P} \equiv \sum_{k\ell j}  \frac{ \ket{\gmode} \bra{\gmode}\op{H}}{\braopket{\gmode}{\op{H}}{\gmode}}  + \frac{ \ket{\wmode} \bra{\wmode}\op{H}}{\braopket{\wmode}{\op{H}}{\wmode}} + \sum_{j} \frac{ \ket{\iomode} \bra{\iomode}\op{H}}{\braopket{\iomode}{\op{H}}{\iomode}} +\frac{ \ket{\mdamode} \bra{\mdamode}\op{H}}{\braopket{\mdamode}{\op{H}}{\mdamode}}
\end{equation}
such that $\mathcal{P} \ket{\psi} = \ket{\psi}_{\mathcal{A}}$. This will be written out explicitly in section~\ref{sec:generalized-projection-operators} and is used in section~\ref{sec:energy-fluxes} to project the equations of motion into wave-vortex space.

Once the fluid state $\ket{\psi}$ is projected onto the energetically and enstrophically orthogonal solutions, the total energy and enstrophy are computed with
\begin{align} \nonumber
      \braopket{\psi}{\op{H}}{\psi} =& \sum_{k\ell j} 2 \left|A_0^{k\ell j} \right|^2 \braopket{\gmode}{\op{H}}{\gmode} + 2\left|A_\pm^{k\ell j} \right|^2 \braopket{\wmode}{\op{H}}{\wmode} + \sum_{j} 2 \left|A_-^{00j}\right|^2 \braopket{\iomode}{\op{H}}{\iomode} + \left|A_0^{00j}\right|^2 \braopket{\mdamode}{\op{H}}{\mdamode}    \\
    \braopket{\op{Q} \psi}{\op{Z}}{\op{Q}\psi} =& \sum_{k\ell j} 2 \left|A_0^{k\ell j} \right|^2 \braopket{\op{Q}\gmode}{\op{Z}}{\op{Q}\gmode}  + \sum_{j}\left|A_0^{00j}\right|^2 \braopket{\op{Q}\mdamode}{\op{Z}}{\op{Q}\mdamode}
\end{align}
where the factor of $2$ accounts for the conjugate (the exact summation limits are detailed later in equation~\ref{eqn:parsevals-theorem}) . The solutions are eigenvectors of the operators $\timeop$ and equivalently $-\op{\linop}$ such that,
\begin{equation}
        \timeop \ket{\gmode} = 0 \ket{\gmode}, \, 
        \timeop \ket{\wmode} = \pm \omega \ket{\wmode}, \, 
        \timeop \ket{\iomode} = - f \ket{\iomode}, \, \textrm{and }
        \timeop \ket{\mdamode} = 0 \ket{\mdamode}
\end{equation}
where the vectors and operators are expressed relative to the basis $\left\{ \ket{u(\vect{x})}, \ket{v(\vect{x})}, \ket{w(\vect{x})}, \ket{\etae(\vect{x})} \right\}$, specifically excluding pressure from the basis $\mathcal{U}$.

\begin{table*}[t]
\centering
\[
\ket{\wmode}
= \frac{1}{\omega_\kappa^j \kappa}
\begin{bmatrix}
(k\omega_\kappa^j \mp f i \ell) F^j_\kappa(z) \\
(\ell \omega_\kappa^j \pm f i k)F^j_\kappa(z) \\
- i \kappa^2 \omega_\kappa^j h_\kappa^j G^j_\kappa(z) \\
\mp \kappa^2 h_\kappa^j G^j_\kappa(z)\\
\mp \rho_0 g \kappa^2 h_\kappa^j F^j_\kappa(z)
\end{bmatrix} e^{i k x + i \ell y\pm i\omega_\kappa^j t}, \, \ket{\iomode}
= 
\begin{bmatrix}
F^j_\textrm{io}(z) \\
i F^j_\textrm{io}(z) \\
0 \\
0\\
0
\end{bmatrix} e^{i f t}, \,
\ket{\gmode} =  \frac{-1}{\kappa^2 + \lambda_j^{-2}}
\begin{bmatrix}
- i \ell F^j_\textrm{g}(z) \\
i k F^j_\textrm{g}(z) \\
0\\
\frac{f}{g} G^j_\textrm{g}(z) 
\\
\rho_0 f F^j_\textrm{g}(z)
\end{bmatrix} e^{i k x + i \ell y}, \,
\ket{\mdamode} = 
\begin{bmatrix}
0 \\
0 \\
0\\
G^j_{\textrm{g}}(z) \\
\rho_0 g F^j_\textrm{g}(z)
\end{bmatrix}
\]
\caption{Summary of energetically and enstrophically orthogonal solutions. These are the `half-complex' solutions, and must be added to their complex-conjugate for the full, physically-realizeable solution.}
\label{tab:solutions}
\end{table*}

% \begin{table}
%     \centering
%     \begin{tabular}{l|l|l} %\rowcolor{LightGray}
%          Energy $\braopket{\psi}{\op{H}}{\psi}$& Enstrophy $\qprod{\op{Q}\psi}{\op{Q}\psi}$ & Range \\ \hline
%          $\frac{g}{2}   \left(\kappa^2   + \linop_j^{-2} \right)\linop_j^2$ & $\frac{g}{2} \left( \kappa^2  + \linop_j^{-2} \right)^2 \linop_j^2$ & $k > 0, j \geq 1$ \\
%          $\frac{g}{2} \kappa^2 \linop_0^2$ & $\frac{g}{2}  \kappa^4 \linop_0^2 $ & $k > 0, j = 0$ \\
%          $\frac{g}{2}$ & $\frac{g}{2} \linop_j^{-2}$ & $k = 0, j \geq 1$ \\
%          $h_w^j$ & $0$ & $k > 0, j \geq 1$ \\
%          $h_\textrm{io}^j$ & $0$ & $k = 0, j \geq 0$
%     \end{tabular}
%     \caption{Energy, enstrophy, and valid range for the five primary solution types.}
%     \label{tab:energy-enstrophy-ranges}
% \end{table}

%%%%%%%%%%%%%%%%%%%%%%%%%%%%%%%%%%%%%%%%%%
\subsection{Projection and notation}
\label{subsection-notation}
%%%%%%%%%%%%%%%%%%%%%%%%%%%%%%%%%%%%%%%%%%

The ultimate outcome of this work are a set of orthogonal solutions and projection operators, which take the real-value observable dynamical fields $(u,v,\etae)$ and project them onto real-valued observable orthogonal solutions. The other two dynamical variables, $w$ and $p$, are diagnostically determined from the continuity equation and the vertical momentum equation. In fact, it is worth noting that the full state cannot be recovered from other combinations of the variables, only $(u,v,\etae)$ contain the complete information of the state. Thus, before even solving the equations of motion, we describe the projection operators.

The projection operators onto the eigenmode solutions are also composed of two parts: the horizontal projections, which take the form of Fourier transforms, and the vertical projections, which result from various Sturm-Liouville problems.

The solutions and projection operators require specification of $\tilde{\rho}_\textrm{nm}(z)$ and there is a strict physical requirement that $\etae$ must adhere to the condition that
\begin{equation}
     \tilde{\rho}_\textrm{nm}(z) >     \partial_z \tilde{\rho}_\textrm{nm}(z)  \eta_\textrm{e}(\mathbf{x}) > \tilde{\rho}_\textrm{nm}(z) - \tilde{\rho}_\textrm{nm}\left(-D \right),
\end{equation}
to maintain the adiabatic constraint. Although not a physical requirement, for the work in this manuscript we make the additional simplifying assumption that $\etae(0)=0=\etae(-D)$. In notable contrast to $\etae$, the physical fields $(u,v)$ do not have bounds on their values.

%%%%%%%%%%%%%%%%%%%%%%%%%%%%%%%%%%%%%%%%%%
\subsubsection{Horizontal transforms}
%%%%%%%%%%%%%%%%%%%%%%%%%%%%%%%%%%%%%%%%%%

A simplifying assumption for this problem is that the two horizontal dimensions $(x,y)$ are periodic, which means that we can use a Fourier basis in the horizontal direction for our linear solutions. We now define the operators that move us between the spatial domain $(x,y)$ and wavenumber domain $(k,\ell)$.

The horizontal structure of all real-valued functions in the domain have the form,
% \begin{equation}
%     f(x,y) = \mathcal{D}^{-1}_{xy}\left[ \hat{f}(k_n, \ell_m)\right]  = \hat{f}(0, 0) + \sum_{n=0}^\infty \sum_{\substack{m=-\infty\\m\neq 0}}^\infty \hat{f}(k_n, \ell_m) e^{i k_n x + i \ell_m y} + c.c
% \end{equation}
\begin{equation}
    f(x,y) = \mathcal{D}^{-1}_{xy}\left[ \hat{f}(k_n, \ell_m)\right]  =  \hat{f}(0, 0) + \sum_{n \ge 0, m \in \mathcal{N}(n)} \left[ \hat{f}(k_n, \ell_m) e^{i k_n x + i \ell_m y} + c.c \right]
\end{equation}
where $k_n = 2 \pi n / L_x$ and $\ell_m = 2 \pi m / L_y$ and $\mathcal{D}^{-1}\left[ \cdot \right]$ is the inverse Fourier transform in two-dimensions. We have defined
\begin{equation}
    \mathcal{N}(n) = 
    \begin{cases}
        m=-\infty \ldots \infty & \textrm{for } n \geq 1 \\
        m = 1 .. \infty & \textrm{for } n = 0
    \end{cases}
\end{equation}
so that the summation runs over unique coefficients, exploiting the Hermitian symmetry of real-valued functions that allows us to assume $c.c. = \conj{\hat{f}}(k_n, \ell_m) e^{-i k_n x - i \ell_m y}$. To recover a coefficient we define the forwards Fourier transform projection operator $\mathcal{D}\left[ \cdot \right]$,
\begin{equation}
     \hat{f}(k_n, \ell_m) = \mathcal{D}_{xy}\left[ f(x,y) \right]  = \frac{1}{L_x L_y} \int_A f(x,y) e^{-i k_n x - i \ell_m y} \, dA
\end{equation}
which follows from the orthogonality, i.e., given two solution with wavenumbers $(k_a,\ell_a)$ and $(k_b,\ell_b)$
\begin{equation}
\label{eqn:fourier-orthogonality}
    \frac{1}{L_x L_y} \int_A e^{i k_a x + i \ell_a y} e^{-i k_b x - i \ell_b y} \, dA = \delta_{k_a k_b} \delta_{\ell_a \ell_b}.
\end{equation}
With these definitions, variance is preserved in both domains,
% \begin{equation}
%      \frac{1}{L_x L_y} \int_A f^2(x,y) \, dA = \hat{f}(0, 0) + \sum_{n=0}^\infty \sum_{\substack{m=-\infty\\m\neq 0}}^\infty 2 \left|\hat{f}(k_n,\ell_m) \right|^2 + \sum_{n=-\infty}^\infty .
% \end{equation}
\begin{equation}
\label{eqn:parsevals-theorem}
     \frac{1}{L_x L_y} \int_A f^2(x,y) \, dA = \hat{f}(0, 0) + \sum_{n\ge 0, m \in \mathcal{N}(n)} 2 \left|\hat{f}(k_n,\ell_m) \right|^2 .
\end{equation}
To simplify notation, anytime we write $\hat{f}$, that should be assumed to be shorthand for $\hat{f}(k_n,\ell_m) = \mathcal{D}_{xy}\left[ f(x,y) \right]$ and $\bar{f}$ is shorthand for the $k=\ell=0$ component that is simply a horizontal average.
% $n=1..\infty$, $m=-\infty..\infty$ AND $n=0$,$m=1..\infty$

% $n>0$, all $m$ OR $n=0$, $m>0$

% \begin{equation}
%      \frac{1}{L_x L_y} \int_A f^2(x,y) \, dA = \sum_{n=0}^\infty \sum_{m=-\infty}^\infty c_{mn} \left|\hat{f}(k_n,\ell_m) \right|^2.
% \end{equation}
% where
% \begin{equation}
%     c_{mn} = \begin{cases}
%         2 \\ 
%         1
%     \end{cases}
% \end{equation}

%%%%%%%%%%%%%%%%%%%%%%%%%%%%%%%%%%%%%%%%%%
\subsubsection{Vertical transforms}
%%%%%%%%%%%%%%%%%%%%%%%%%%%%%%%%%%%%%%%%%%

\begin{table*}
    \centering
    \begin{tabular}{l|l|l} %\rowcolor{LightGray}
         EVP & Forward transform & Inverse transform \\ \hline
        $\partial_z \left( \frac{\partial_z F_\textrm{g}}{N^2} \right) = - \frac{1}{g h_\textrm{g}} F_\textrm{g}$ & $\mathcal{F}^j_\textrm{g}[u] \equiv  \frac{1}{h_\textrm{g}^j} \int_{-D}^0 u F^j_\textrm{g} \, dz = u_\textrm{g}^j$ & $\mathcal{F}_\textrm{g}^{-1} [ u_\textrm{g}^j ]  = \sum u_\textrm{g}^j F_\textrm{g}^j= u$ \\
        $\partial_{zz} G_\textrm{g} = - \frac{N^2}{g h_\textrm{g}} G_\textrm{g}$ & $\mathcal{G}^j_\textrm{g}[\eta] \equiv \frac{1}{g} \int N^2 \eta G^j_\textrm{g} \, dz = \eta_g^j$ & $\mathcal{G}_\textrm{g}^{-1} [ \eta_g^j ] = \sum \eta_g^j G_\textrm{g}^j= \eta$ \\
        $\partial_{zz} G_\textrm{mda} = - \frac{N^2}{g h_\textrm{mda}} G_\textrm{mda}$ & $\mathcal{G}^j_\textrm{mda}[\eta] \equiv \frac{1}{g} \int N^2 \eta G^j_\textrm{mda} \, dz = \eta^j$ & $\mathcal{G}_\textrm{mda}^{-1} [ \eta^j ] = \sum \eta^j G_\textrm{mda}^j= \eta$ \\
        $\partial_{zz} G_\kappa -\kappa^2 G_\kappa = - \frac{N^2-f^2}{g h_\kappa} G_\kappa$ & $\mathcal{G}^j_\kappa[\eta] \equiv \frac{1}{g} \int \left( N^2 - f^2 \right) \eta G^j_\kappa \, dz = \eta_\kappa^j$ & $\mathcal{G}_\kappa^{-1} [ \eta_\kappa^j ] = \sum \eta_\kappa^j G_\kappa^j= \eta$ \\
        $\partial_z \left( \frac{\partial_z F_\textrm{io}}{N^2 - f^2} \right) = - \frac{1}{g h_\textrm{io}} F_\textrm{io}$ & $\mathcal{F}^j_\textrm{io} \left[ u \right] \equiv \frac{1}{h_\textrm{io}^j} \int u F^j_\textrm{io} \, dz = u_\textrm{io}^j$ & $\mathcal{F}_\textrm{io}^{-1} [ u_\textrm{io}^j ]  = \sum u_\textrm{io}^j F_\textrm{io}^j= u$ 
    \end{tabular}
    \caption{The second and third column show the vertical mode projection operators and their inverses. In this notation, the function $u$, $\eta$, $F$, $G$ and $N$ are all functions of $z$. All eigenmodes have the relationship $F = h \partial_z G$, but $N^2 G_\textrm{g} = -g \partial_z F_\textrm{g}$ for the geostrophic solution and $(N^2 - \omega_\kappa^2) G_\kappa = -g \partial_z F_\kappa$ for the internal gravity wave solution.}
    \label{tab:vertical-mode-projection}
\end{table*}

To solve the linear system (\ref{eqn:linear-momentum-braket}), it is helpful to notice that the momentum equations imply the same vertical structure for $(u,v,p)$, while the thermodynamic requires the same vertical structure for $(w,\etae)$. In general, the vertical structure will be wavenumber dependent, and we will use $F^{k\ell j}(z)$ for the vertical structure of $\hat{u}^{k\ell}(z),\hat{v}^{k\ell}(z),\hat{p}^{k\ell}(z)$ and $G^{k\ell j}(z)$ for the vertical structure of $\hat{w}^{k\ell}(z),\hat{\etae}^{k\ell}(z)$. For compactness of notation throughout the rest of section~\ref{sec:orthogonal-solutions}, we will use $\partial_z$ to denote ordinary derivatives with respect to $z$.  We will also abbreviate $F^{k\ell j}(z), G^{k\ell j}(z)$ as 
$F^{j}, G^{j}$ except when we need to specify special horizontal wavenumbers.

 All solution types
 will satisfy a Sturm-Liouville ordinary differential equation of the form
\begin{equation}
\partial_z (p(z) \partial_z\phi(z)) + q(z) \phi(z) = - \frac{\sigma(z)}{g h^j} \phi(z)
\label{eqn:sturm-liouville}
\end{equation}
where $ -D < z < 0$ together with separated boundary conditions, where $\phi(z)$ is either $G^j(z)$ or $F^j(z)$, and the Sturm-Liouville eigenvalue is $(gh^j)^{-1}.$ The specific boundary conditions and functions $p(z) > 0$ and $\sigma(z) > 0$ depend on the solution type (see Table~\ref{tab:vertical-mode-projection}).  

There are other noteworthy features of the vertical structure functions $F^j(z)$ and $G^j(z)$.  In particular, all solutions satisfy $F^j(z) = h^j \partial_z G^j(z)$ (although the relation between $G^j(z)$ and $\partial_z F^j(z)$ is different for different solution types, given in 
Table~\ref{tab:vertical-mode-projection}). Furthermore, for all solution types, orthogonality may be expressed in terms of $G^j(z)$, given by 
\begin{equation}
    \int_{-D}^0 \sigma(z) G^i(z) G^j(z) \, dz = g\delta_{ij}
\end{equation}
with either $\sigma(z)=N^2(z)$ or $\sigma(z)=N^2(z) - f^2$, depending on the particular problem.

Just as with the Fourier transform, we will \emph{always} be able to define forward and inverse transforms $\mathcal{G}$ and $\mathcal{G}^{-1}$ for the $G^j(z)$ modes,
\begin{subequations}
    \begin{align}
    \eta(z) =& \mathcal{G}^{-1} [ \eta^j ] =  \sum_{n=0} \eta^j G^j(z) \\
    \eta^j =& \mathcal{G}^j[\eta(z)] = \int_{-D}^0 \sigma(z) \eta(z) G^j(z) \, dz
    \end{align}
\end{subequations}
and \emph{sometimes} be able to define similar projection operators $\mathcal{F}$ and $\mathcal{F}^{-1}$ for the $F^j(z)$ modes. For geostrophic solutions, there is an eigenvalue problem for $F^j(z)$, indicating that we can partition the horizontal kinetic energy independent of the total energy. However, in the case of wave solutions, there is no Sturm-Liouville problem for $F^j(z)$, indicating that the horizontal kinetic energy cannot be separately partitioned.

%%%%%%%%%%%%%%%%%%%%%%%%%%%%%%%%%%%%%%%%%%
\subsection{Geostrophic projection}
%%%%%%%%%%%%%%%%%%%%%%%%%%%%%%%%%%%%%%%%%%

\begin{table*}
    \centering
    \begin{tabular}{l|l|l|l} %\rowcolor{LightGray}
         Projection & Energy $\braopket{{\Psi}}{\op{H}}{{\Psi}}$& Enstrophy $\qprod{\op{Q}{\Psi}}{\op{Q}{\Psi}}$ & Range \\ \hline
         $A_0^{k\ell j} = \mathcal{F}^j_g \left[ \hat{\zeta} \right] - \frac{f}{h_g^j} \mathcal{G}^j_g [ \hat{\eta} ]$ & $\frac{1}{2}   \left(\kappa^2 + \lambda_j^{-2} \right)^{-1} h_g^j$ & $\frac{1}{2} h_g^j$ & $k > 0, j > 0$ \\ 
         $A^{kl0}_0 = \mathcal{F}^j_g \left[ \hat{\zeta} \right]$ & $\frac{1}{2}\frac{D}{\kappa^2} $ & $\frac{1}{2} D $ & $k > 0, j = 0$ \\
         $A_0^{00j} = \mathcal{G}^j_\textrm{mda} [ \bar{\eta}_\textrm{e} ] = \frac{1}{\rho_0 g}\mathcal{F}^j_\textrm{mda} [ \bar{p}_\textrm{e} ]$ & $\frac{g}{2}$ & $\frac{g}{2} \lambda_j^{-2}$ & $k = 0, j \geq 1$ \\
         $A_\pm^{k\ell j} =  \frac{e^{\mp i \omega_\kappa^j t}}{2 \kappa h_\kappa^j}  \left( i \mathcal{G}^j_\kappa \left[\hat{w}(z) \right] \mp \omega_\kappa^j \mathcal{G}^j_\kappa \left[\hat{\eta} - \hat{\eta}_g \right] \right)$ & $h_\kappa^j$ & $0$ & $k > 0, j \geq 1$ \\
         $A_-^{00j} = \frac{e^{-i f t}}{2} \mathcal{F}^j_\textrm{io} \left[ \bar{u} - i  \bar{v}\right]$ & $h_\textrm{io}^j$ & $0$ & $k = 0, j \geq 0$
    \end{tabular}
    \caption{Projection operators, energy, enstrophy, and valid range for the five primary solution types. These projection operators take physical variables $(u,v,\etae)$ and transform them into wave-vortex space.}
    \label{tab:solution-projection}
\end{table*}

The geostrophic solutions have vertical modes $F_\textrm{g}^j(z)$ for the vertical structure of $u$ and $v$, which follows from the eigenvalue problem
\begin{equation}
\label{eqn:f-evp-geostrophic}
    \partial_z\left( \frac{f^2}{N^2} \partial_zF^j_\textrm{g}\right) = - \frac{f^2}{g h^j_\textrm{g}} F^j_\textrm{g}
\end{equation}
with boundary conditions $\partial_z F^j_\textrm{g}(0) = 0 = \partial_z F^j_\textrm{g}(-D)$. As a regular Sturm-Liouville problem the eigenmodes satisfy the orthogonality condition,
\begin{equation}
\label{eqn:f-norm-geostrophic}
    \int_{-D}^0 F_\textrm{g}^i(z) F_\textrm{g}^j(z) \, dz = \gamma^j \delta_{ij} \textrm{where} \; \gamma^j = \begin{cases}
        D & j=0 \\
        h^j_\textrm{g} & \textrm{otherwise}
    \end{cases}
\end{equation}
with chosen normalization $\gamma^j$, and thus can be used to define projection operator $\mathcal{F}_\textrm{g}[u]$ and its inverse $\mathcal{F}^{-1}_\textrm{g}[u_g^j]$ as defined in the first row of Table~\ref{tab:vertical-mode-projection}. The vertical structure of $\etae$ is described by $G^j_\textrm{g} = - \frac{g}{N^2} \partial_z F^j_\textrm{g}$ which also forms an eigenvalue problem
\begin{equation}
\label{eqn:g-evp-geostrophic}
 \partial_{zz}G^j_\textrm{g} = - \frac{N^2}{g h^j_\textrm{g}} G^j_\textrm{g}
\end{equation}
with boundary conditions $G^j_\textrm{g}(0)=0=G^j_\textrm{g}(-D).$ The normalization of the orthogonality condition for the $G_\textrm{g}$ modes
\begin{equation}
\label{eqn:g-norm-geostrophic}
    \int_{-D}^0 N^2(z) G_\textrm{g}^i(z) G_\textrm{g}^j(z) \, dz = g \delta_{ij}
\end{equation}
must be determined by from \eqref{eqn:f-evp-geostrophic} using integration-by-parts. The projection operator $\mathcal{G}_\textrm{g}[\eta]$ and its inverse $\mathcal{G}^{-1}_\textrm{g}[\eta_g^j]$ follow from \eqref{eqn:g-evp-geostrophic} and are defined in the second row of Table~\ref{tab:vertical-mode-projection}. 

It will prove useful to note that the relationships
\begin{subequations}
\label{eqn:gmodes-diff-int}
    \begin{align}
        \mathcal{F}^j_g \left[ \frac{\partial \eta}{\partial z} \right] =&  \frac{1}{h_g^j} \mathcal{G}^j_g \left[ \eta \right] \\
        \mathcal{G}^j_g \left[ \frac{1}{N^2}\frac{\partial u}{\partial z} \right] =& - \frac{1}{g} \mathcal{F}^j_g [ u ]
    \end{align}
\end{subequations}
can be used to integrate and differentiate variables.

To determine the coefficients $A_0^{k\ell j}$ of each geostrophic solution from observed values $(u,v,w,\eta)$, we project onto the geostrophic solution. In practice, this looks like
    \begin{align} \nonumber
       \braopket{\gmode}{\op{H}}{\psi}  =&  \frac{1}{2(\kappa^2 + \lambda_j^{-2})} \int_
        {-D}^0\left( -i \ell F_g^j(z) \hat{u} + i k F_g^j(z) \hat{v}  - \frac{f}{g}N^2 G_g^j(z) \hat{\eta} \right) dz  \\ \label{a0-projection}
        =&   \frac{1}{2(\kappa^2 + \lambda_j^{-2})} \left(  h_g^j \mathcal{F}^j_g \left[ \hat{\zeta} \right] - f \mathcal{G}^j_g [ \hat{\eta} ]  \right)
    \end{align}
where we have defined the vertical component of vorticity with
\begin{equation}
    \hat{\zeta} \equiv i k \hat{v} - i l \hat{u}.
\end{equation}
The total energy of this solution follows,
    \begin{align} \nonumber
       \braopket{\gmode}{\op{H}}{\gmode}  =& \frac{1}{2(\kappa^2 + \lambda_j^{-2})^2} \int_
        {-D}^0\left( \kappa^2 F_g^j(z) F_g^i(z)  + N^2 G_g^j(z) G_g^i(z) \right) dz \\ \label{a0-energy}
        =& \frac{1}{2}  h_g^j \left(\kappa^2   + \lambda_j^{-2} \right)^{-1} \delta^{ij}
    \end{align}
where we have expressed the eigenvalue $h^j_\textrm{g}$ as the squared deformation radius, $\lambda_j^2 \equiv \frac{g h^j_\textrm{g}}{f^2}$. Combining \eqref{a0-projection} with \eqref{a0-energy} means that the coefficient $A_0^{k\ell j}$ is recovered with,
\begin{equation}
A_0^{k\ell j} \ket{\gmode}= \ket{\gmode} \frac{\braopket{\gmode}{\op{H}}{\psi}} {\braopket{\gmode}{\op{H}}{\gmode}}  = \left( \mathcal{F}_g \left[ \hat{\zeta} \right] - \frac{f}{h_g^j} \mathcal{G}_g [ \hat{\eta} ]  \right) \ket{\gmode}.
    \end{equation}
This projection operator, as well as energy and enstrophy for these modes, are shown in the first row of Table~\ref{tab:solution-projection}.

The eigenvalue problem \eqref{eqn:f-evp-geostrophic} admits a $j=0$ mode, $F_\textrm{g}^0(z)=1$, with eigenvalue $\frac{f^2}{g h^0_g} = 0$, known as the barotropic mode. This mode has no density anomaly, $G_\textrm{g}^0(z)=0$, and thus has zero APE and no vortex stretching. As a mode with only vorticity, it is rightfully called a `vortical' mode. Its projection follows exactly the same approach and is the second row of Table~\ref{tab:solution-projection}.

If the fluid state can be described entirely in terms of a geostrophic streamfunction, the projection onto the geostrophic modes simplifies. A geostrophic streamfunction is proportional to the pressure anomaly $\psi_g \equiv \frac{1}{\rho_0 f} p_\textrm{e}$ where then $u=-\partial_y \psi_g$, $v=\partial_x \psi_g$ and $N^2 \etae = - f \partial_z \psi_g$ or, equivalently, $\rho_\textrm{e} = -  \frac{\rho_0 f}{g} \partial_z \psi_g$. This means that the projection operator can now be written as,
\begin{equation}
    A_0^{k\ell j} = - \left( \kappa^2 + \lambda_j^{-2} \right) \mathcal{F}^j_g [ \hat{\psi}_g ].
\end{equation}

%%%%%%%%%%%%%%%%%%%%%%%%%%%%%%%%%%%%%%%%%%
\subsection{Mean density anomaly projection}
\label{sec:mda-solution}
%%%%%%%%%%%%%%%%%%%%%%%%%%%%%%%%%%%%%%%%%%

The mean density anomaly (mda) solution is the horizontally averaged ($\kappa=0$) solution with zero-frequency $\omega=0$.  The mda solution has $(u, v, w) = 0$, and thus the only equation of motion that remains is the horizontally-averaged vertical momentum equation,
\begin{equation}
    N^2 \bar{\eta}_e = - \frac{1}{\rho_0} \partial_z \bar{p}_e,
\label{eqn:mda-constraint}
\end{equation}
where the bar denotes the horizontal average. The mda solutions are exactly the difference between the average density in the fluid, $\bar{\rho}(z)$, and the no-motion density, $\rho_\textrm{nm}(z),$ as noted in section~\ref{subsec:nomotionsoln}. This mode was not included in \cite{early2021-jfm} because there $\bar{\rho}(z)$ was used to define the background state whereas here we use $\rho_\textrm{nm}(z)$ (an anonymous reviewer of \cite{early2021-jfm} first brought this mode to our attention). However, the presence of the mean density anomaly is important for constructing a complete basis when linearizing about the no-motion state. Simply swapping two fluid parcels in the water column results in an mda solution and changes the potential energy and enstrophy of the system.

Depth-integrating \eqref{eqn:mda-constraint},
one finds that
the mda solutions are constrained by global conservation of potential density
\begin{equation}
\label{eqn:mda-density-conservation}
\int_{-D}^{0} N^2 \bar{\eta}_e dz = -\int_{-D}^{0}\frac{1}{\rho_0} \partial_z \bar{p}_e  = \bar{p}_e(0) - \bar{p}_e(-D)=0,
\end{equation}
and they are also constrained by global conservation of QGPV,
\begin{equation}
\label{eqn:mda-qgpv-conservation}
\int_{-D}^{0} \partial_z \bar{\eta}_e dz = \bar{\eta}_e(0) - \bar{\eta}_e(-D) = 0.
\end{equation}
In the rigid-lid problem, \eqref{eqn:mda-density-conservation} tells us that the average pressure anomaly at the two boundaries must be zero, with $\bar{p}_e(-D)=\bar{p}_e(0)=0.$ In addition, \eqref{eqn:mda-qgpv-conservation} says that the average buoyancy anomaly at the boundaries must vanish, with  $\bar{\eta}_e(-D)=\bar{\eta}_e(0)=0$.

Even though the mda solutions have no fluid velocity, their energy and enstrophy are both nonzero, and the $G_\textrm{g}(z)$ eigenmodes found from \eqref{eqn:g-evp-geostrophic} diagonalize both quantities. Thus it is natural to describe the mda solutions using \eqref{eqn:g-evp-geostrophic} together with appropriate boundary conditions
that also satisfy the constraints 
 \eqref{eqn:mda-density-conservation}-\eqref{eqn:mda-qgpv-conservation}. 
 However, if we impose both $p_\textrm{e}(-D)=p_\textrm{e}(0) = 0$ \emph{and} $\eta_\textrm{e}(-D)=\eta_\textrm{e}(0)=0$, then the boundary value problem is overdetermined. 
 Therefore we must ask: what is the correct choice?

Enforcing \eqref{eqn:mda-density-conservation} would disallow diabatic processes, which are included in our formulation \eqref{eqn:boussinesq}. In contrast, if we impose \eqref{eqn:mda-qgpv-conservation}, then no additional restrictions beyond the boundary conditions $\eta_\textrm{e}(-D) = \eta_\textrm{e}(0) = 0$ are imposed. The choice here is clear. Proceeding with the choice $G(-D)=G(0)=0$, projection of $\bar{\eta}_\textrm{e}$ onto the $G_\textrm{g}(z)$ eigenmodes follows immediately. Using integration by parts, it can be shown that
\begin{equation}
    \mathcal{G}_\textrm{g}[\bar{\eta}_\textrm{e}] =  \frac{1}{\rho_0 g}\mathcal{F}_\textrm{g}[\bar{p}_\textrm{e}],
\end{equation}
as noted in the third row of Table~\ref{tab:solution-projection}, where we denote the coefficients of the mda solution $\ket{\mdamode}$ as $A_0^{00j}$.

%%%%%%%%%%%%%%%%%%%%%%%%%%%%%%%%%%%%%%%%%%
\subsection{IGW projection}
%%%%%%%%%%%%%%%%%%%%%%%%%%%%%%%%%%%%%%%%%%

The required relationships between the vertical modes $F_\kappa$ (for $u$, $v$, and $p$) and $G_\kappa$ (for $w$ and $\etae$) are
\begin{equation}
        F_\kappa = h_\kappa \partial_z G_\kappa \textrm{ and } (N^2- \omega_\kappa^2) G_\kappa = -g\partial_z F_\kappa
\end{equation}
which follow from the continuity equation and the vertical momentum equation, respectively. These relationships lead to the eigenvalue problem \begin{equation}
\label{eqn:evp-g-igw}
    \partial_{zz} G_\kappa^j - \kappa^2 G_\kappa^j = - \frac{N^2-f^2}{g h_\kappa^j} G_\kappa^j
\end{equation}
with boundary conditions $G_\kappa(0)=0=G_\kappa(-D)$, where we used that
\begin{equation}
\omega_\kappa^j \equiv \sqrt{ g h^j_\kappa\kappa^2 + f^2}.
\end{equation}
The quantity $\omega_\kappa^j$ should be viewed as a shorthand for the wave frequency that is dependent on the eigenvalue $h_\kappa^j$ and the total wavenumber $\kappa = \sqrt{k^2 + \ell^2}$. 
Unlike constant stratification or hydrostatics, posing the eigenvalue problem in terms of $\omega$ does not lead to a complete basis of orthogonal solutions.

Vertical modes $G_\kappa$ therefore satisfy the orthogonality condition
\begin{equation}
\label{eqn:g-norm-igw}
    \frac{1}{g} \int_{-D}^0 \left( N^2 - f^2\right) G_\kappa^i G_\kappa^j \, dz = \delta_{ij}
\end{equation}
which leads to projection operator $\mathcal{G}^j_\kappa[\eta]$ and its inverse $\mathcal{G}^{-1}_\kappa[\eta_w^j]$ as defined in the fourth row of Table~\ref{tab:vertical-mode-projection}. Unlike the geostrophic modes, there is no eigenvalue problem associated with the $F_\kappa$ modes as discussed in \cite{early2021-jfm}.

To determine the wave coefficients $A_\pm^{k\ell j}$ we again exploit orthogonality and project the observed state $\ket{\psi}$ onto the internal gravity wave mode. This particular calculation is more challenging and we save the details for appendix~\ref{appendix:wave-mode-projection}. The net result is
    \begin{equation}
        A_\pm^{k\ell j} =\frac{\braopket{\wmode}{\op{H}}{\psi}}{\braopket{\wmode}{\op{H}}{\wmode}} \\
        = \frac{e^{\mp i \omega_\kappa^j t}}{2 \kappa h_\kappa^j}  \left( i \mathcal{G}^j_\kappa \left[\hat{w} \right] \mp \omega_\kappa^j \mathcal{G}^j_\kappa \left[\hat{\eta} - \hat{\eta}_g \right] 
        \right)
    \end{equation}
where $\hat{\eta}_g$ is the isopycnal deviation of the geostrophic mode at this same wavenumber. This solution projection operator is shown in the fourth row of Table~\ref{tab:solution-projection}. The vertical velocity can be replaced in favor of the horizontal divergence using that
\begin{equation}
    \hat{w} = -\mathcal{G}_g^{-1} \left[ h_g^j \mathcal{F}^j_g \left[  i k \hat{u} + i l \hat{v} \right] \right],
\end{equation}
% \begin{equation}
%     A_\pm^{k\ell j} =  \frac{e^{\mp i \omega_\kappa^j t}}{2 \kappa h_\kappa^j}  \left( -i \mathcal{G}^j_\kappa \left[\mathcal{G}_g^{-1} \left[ h_g^j \mathcal{F}^j_g \left[  i k \hat{u} + i l \hat{v} \right] \right] \right] \mp \omega_\kappa^j \mathcal{G}^j_\kappa \left[\hat{\eta} - \hat{\eta}_g \right] \right),
% \end{equation}
where we have used the relationships in \eqref{eqn:gmodes-diff-int} and the continuity equation, $-\partial_z \hat{w} = i k \hat{u} + i l \hat{v}$.

%%%%%%%%%%%%%%%%%%%%%%%%%%%%%%%%%%%%%%%%%%
\subsection{IO projection}
%%%%%%%%%%%%%%%%%%%%%%%%%%%%%%%%%%%%%%%%%%

The inertial oscillations exist at the IGW limit where $\kappa=0$ and $\omega=f$, such that $G_\textrm{io}$ satisfies
\begin{equation}
\label{eqn:io-g-evp}
    \partial_{zz} G^j_\textrm{io} = - \frac{N^2-f^2}{g h^j_\textrm{io}} G^j_\textrm{io},
\end{equation}
with $G^j_\textrm{io}(0) = 0 =G^j_\textrm{io}(-D)$. 
The corresponding boundary value problem for $F^j_\textrm{io}$ is given by
\begin{equation}
\label{eqn:io-f-evp}
    \partial_z \left( \frac{\partial_z F^j_\textrm{io}}{N^2 - f^2} \right) = - \frac{1}{g h^j_\textrm{io}} F^j_\textrm{io},
\end{equation}
with $\partial_z F^j_\textrm{io}(0)=0=\partial_z F^j_\textrm{io}(-D).$
Thus for the inertial oscillations with $\kappa=0$, orthogonality of $F^j_\textrm{io}$ may be stated as
\begin{equation}
    \int_{-D}^0 F_\textrm{io}^i F_\textrm{io}^j \, dz = h_\textrm{io}^i \delta^{ij},
\end{equation}
as indicated in the last row of Table~\ref{tab:vertical-mode-projection}. 

Projection onto the inertial oscillation solution follows with,
\begin{equation}
    \braopket{\iomode}{\op{H}}{\psi}  = \frac{e^{-i f t}}{2} \int_{-D}^0 \left( \bar{u} - i  \bar{v} \right) F_\textrm{io}(z) dz = \frac{e^{-i f t}}{2} h_\textrm{io}^j \mathcal{F}^j_\textrm{io} \left[ \bar{u} - i  \bar{v}\right]
\end{equation}
which means that,
\begin{equation}
A_-^{00j} = \frac{\braopket{\iomode}{\op{H}}{\psi}}{\braopket{\iomode}{\op{H}}{\iomode}}= \frac{e^{-i f t}}{2} \mathcal{F}^j_\textrm{io} \left[ \bar{u} - i  \bar{v}\right]
\end{equation}
as shown in the last row of Table~\ref{tab:solution-projection}.

%%%%%%%%%%%%%%%%%%%%%%%%%%%%%%%%%%%%%%%%%%
\subsection{Summary of solutions}
\label{sec:summary-of-solutions}
%%%%%%%%%%%%%%%%%%%%%%%%%%%%%%%%%%%%%%%%%%

These four solutions fall into two obvious categories: 1) wave solutions which include the internal gravity waves $\ket{\wmode}$ and inertial oscillations $\ket{\iomode}$ and 2) potential vorticity (or vortex) solutions which include geostrophic motions $\ket{\gmode}$ and the mean-density anomaly $\ket{\mdamode}$. The key feature separating these solution types being whether or not they have potential enstrophy.

%%%%%%%%%%%%%%%%%%%%%%%%%%%%%%%%%
\subsection{Generalized projection operators }
\label{sec:generalized-projection-operators}
%%%%%%%%%%%%%%%%%%%%%%%%%%%%%%%%%

Inserting the wave-vortex projection operator/identity operator $\mathcal{P}$ from \eqref{eqn:wave-vortex-projection} into the linear momentum equations reduces to the almost trivial statement
\begin{equation}
    \left( \partial_t  + \op{\linop} \right) \mathcal{P} \ket{\psi}_{\mathcal{U}} = \partial_t \ket{\psi}_{\mathcal{A}} = 0,
\end{equation}
which tells us that the coefficients of the linear eigenmodes are constant in time. 
For example, the equation for the linear evolution of the inertia-gravity  waves is given by
\begin{subequations}
    \begin{align}
        \left( \partial_t  + \op{\linop} \right) \frac{ \ket{\wmode} \braopket{\wmode}{\op{H}}{\psi} }{\braopket{\wmode}{\op{H}}{\wmode}} &= 0 \\
        A_\pm^{k\ell j}  \left( \partial_t  + \op{\linop} \right) \ket{\wmode} + \partial_t A_\pm^{k\ell j} \ket{\wmode} &= 0 \\
        \partial_t A_\pm^{k\ell j}  \ket{\wmode} &=0
    \end{align}
\end{subequations}
where we used \eqref{eqn:wmode-coefficient} and $\left( \partial_t  + \op{\linop} \right) \ket{\wmode}=0$ from \eqref{eqn:braket-linear-eom}.

For the purposes of this manuscript we will define two primary projection operators, $\mathcal{P}^0_\textrm{g}$ and $\mathcal{P}_\textrm{w}^\pm$ defined abstractly as,
\begin{subequations}
    \begin{align}
    \mathcal{P}^0_\textrm{g} \equiv \sum_{k\ell j}  \frac{ \ket{\gmode} \bra{\gmode}\op{H}}{\braopket{\gmode}{\op{H}}{\gmode}}  + \sum_{j} \frac{ \ket{\mdamode} \bra{\mdamode}\op{H}}{\braopket{\mdamode}{\op{H}}{\mdamode}} \\
    \mathcal{P}_\textrm{w}^\pm \equiv \sum_{k\ell j}  \frac{ \ket{\wmode} \bra{\wmode}\op{H}}{\braopket{\wmode}{\op{H}}{\wmode}} + \sum_{j} \frac{ \ket{\iomode} \bra{\iomode}\op{H}}{\braopket{\iomode}{\op{H}}{\iomode}}
    \end{align}
\end{subequations}
which isolate the part of the flow with and without qgpv. To be more concrete we unravel the definitions of the inner-product and express the projection operators relative to the wave-vortex basis (indexed by $k \ell j$) acting on a generic vector $\ket{f}_{\mathcal{U}} = \left[f_u(\mathbf{x},t), f_v(\mathbf{x},t), f_w(\mathbf{x},t), f_{\etae}(\mathbf{x},t), f_p(\mathbf{x},t) \right]$. Using the notation from \eqref{eqn:parsevals-theorem} that, e.g., $\hat{f}_u$ is shorthand for $\hat{f}_u(k_n, \ell_m,z,t) = \mathcal{D}_{xy} \left[ f_u(x,y,z,t) \right]$ and $\bar{f}_u(z,t)$ is the horizontal average, then the projection operators are
\begin{equation}
\left[ \mathcal{P}^0_\textrm{g} \left[ f \right] \right]^{k \ell j} =
    \begin{cases}
        \mathcal{G}^j_\textrm{mda} [ \overline{f_{\etae} } ] & \kappa = 0, j \geq 1 \\
        \mathcal{F}^j_g \left[ i k \hat{f}_v - i \ell \hat{f}_u \right] - \frac{f}{h_g^j} \mathcal{G}^j_g [ \hat{f}_{\etae} ] & \kappa > 0, j \geq 0 
    \end{cases}
\end{equation}
and
\begin{equation}
\left[  \mathcal{P}_\textrm{w}^\pm\left[ f \right] \right]^{k \ell j} =
    \begin{cases}
        \frac{e^{-i f t}}{2} \mathcal{F}^j_\textrm{io} \left[ \overline{f_u} - i  \overline{f_v}\right] & \kappa = 0, j \geq 0 \\
        \frac{e^{\mp i \omega_\kappa^j t}}{2 \kappa }  \left(  \mathcal{G}_\kappa \left[ \frac{  N^2 \mathcal{G}^{-1}_g \left[ \mathcal{F}_g [  k \hat{f}_u +  \ell \hat{f}_v ] \right] + i g \kappa^2 \hat{f}_w}{N^2 - f^2} \right] \mp \frac{\omega_\kappa^j}{h_\kappa^j} \mathcal{G}_\kappa \left[\hat{f}_{\etae} - \mathcal{N}_\textrm{g}\left[ f \right] \right] \right) & \kappa > 0, j \geq 1 
    \end{cases}
\end{equation}
where 
\begin{equation}
    \mathcal{N}_\textrm{g}\left[ f \right] = - \frac{1}{\kappa^2 + \lambda_j^{-2}} \frac{f}{g} \mathcal{P}^0_\textrm{g} \left[ f \right] 
\end{equation}
is the fraction of $f_{\etae}$ attributed to the geostrophic field. Note that the inertial oscillation solution at $\kappa=0$ in the definition of $ \mathcal{P}_\textrm{w}^\pm$ is a negative frequency solution and thus only applies to $ \mathcal{P}_\textrm{w}^-$.

%%%%%%%%%%%%%%%%%%%%%%%%%%%%%%%%%%%%%
\section{Modal spectra}
\label{appendix:modal-spectra}

%%%%%%%%%%%%%%%%%%%%%%%%%%%%%%%%%%%%

To compute the spectrum of an arbitrary function $p(x,y,z)$ with boundary conditions $\partial_z p(x,y,0)=0$ and $\partial_z p(x,y,-D)=0$, the transform
\begin{equation}
    \hat{p}_{k\ell j} = \mathcal{F}_g \left[ \mathcal{D}_{xy} \left[ p \right] \right]
\end{equation}
preserves
\begin{equation}
     \frac{1}{L_x L_y} \int_V p^2(x,y,z) \, dV = \sum_{j=0}^\infty \left( h_g^j \left|\hat{p}_{00 j}\right|^2 +\sum_{m,n}^{\mathcal{N}} 2 h_g^j \left|\hat{p}_{k\ell j}\right|^2  \right).
\end{equation}
which defines the `spectrum' as,
\begin{equation}
\operatorname{SF} \left[p, p\right](k_n,\ell_m, j) =
    \begin{cases}
        2 h_g^j \left|\hat{p}_{k\ell j}\right|^2 & n\geq 0, m \in \mathcal{N} \\
        h_g^j \left|\hat{p}_{0 0 j}\right|^2 & \textrm{otherwise}         
    \end{cases}
\end{equation}
where we have used subscripts $k\ell j$ as shorthand for $(k_n,\ell_m,j)$. The cross-spectrum of two real-valued functions $p(x,y,z)$, $q(x,y,z)$ with the same boundary conditions follows with,
\begin{equation}
\label{eqn:modal-cross-spectrum-f}
\operatorname{SF} \left[p, q\right](k_n,\ell_m, j) =
    \begin{cases}
        2 h_g^j \Re\left[ \hat{p}_{k\ell j} \hat{q}^\ast_{k\ell j}\right] & n\geq 0, m  \in \mathcal{N} \\
        h_g^j \Re\left[ \hat{p}_{0 0 j} \hat{q}^\ast_{0 0 j}\right] & \textrm{otherwise.}     
    \end{cases}
\end{equation}

Similarly, if the function $p(x,y,z)$ has boundary conditions $p(x,y,0)=0$ and $p(x,y,-D)=0$, the transform
\begin{equation}
    \hat{p}_{k \ell j} = \mathcal{G}_g \left[ \mathcal{D}_{xy} \left[ p \right] \right]
\end{equation}
preserves
\begin{equation}
     \frac{1}{L_x L_y} \int_V N^2(z) p^2(x,y,z) \, dV  = \sum_{j=0}^\infty \left( g \left|\hat{p}^{0 0 j}\right|^2 + \sum_{m,n}^{\mathcal{N}} 2 g \left|\hat{p}^{k \ell j}\right|^2 \right).
\end{equation}
then the cross-spectrum that results can be defined as,
\begin{equation}
\operatorname{SG} \left[p, q\right](k_n,\ell_m, j) =
    \begin{cases}
        2 g\Re\left[ \hat{p}_{k\ell j} \hat{q}^\ast_{k\ell j}\right] & n\geq 0, m  \in \mathcal{N} \\
        g \Re\left[ \hat{p}_{0 0 j} \hat{q}^\ast_{0 0 j}\right] & \textrm{otherwise.}     
    \end{cases}
\end{equation}

%%%%%%%%%%%%%%%%%%%%%%%%%%%%%%%%%%%%
%
\section{Transfers from triad fluxes}
\label{appendix:transfers-from-triad-fluxes}
%
%%%%%%%%%%%%%%%%%%%%%%%%%%%%%%%%%%%%

Consider a decomposition of the flow into three energetically orthogonal reservoirs of energy, $u = u_a + u_b + u_c$. The energy flux into reservoir $a$ is,
\begin{equation}
    \frac{d}{dt} E_a = \sum_{k\ell j} \Re \left[ \epsilon_a^{k\ell j} A^{k\ell j} [(u_a + u_b + u_c) \nabla (u_a + u_b + u_c) ]_a^{k\ell j} \right]
\end{equation}
where $\epsilon_a$ is the energy multiplier and $A$ is the amplitude for the modes in that reservoir. Using the notation that,
\begin{equation}
    \label{eqn:mixed-triad-component}
    T_{abc}^{(a,b) \mapsto c} \equiv \Re \left[ \epsilon_c^{k\ell j} A^{k\ell j} [u_a \nabla u_b + u_b \nabla u_a]_c^{k\ell j} \right] 
\end{equation}
is the energy transfer from reservoirs a and b to reservoir c, the triad conservation laws can be written as
\begin{subequations}
    \begin{align} \label{eqn:triad-conservation-abc}
       0 =& T_{abc}^{(a,b) \mapsto c} + T_{abc}^{(a,c) \mapsto b} + T_{abc}^{(b,c) \mapsto a}  \\ \label{eqn:triad-conservation-aac}
        0 =& T_{aac}^{a \mapsto c} + T_{aac}^{(a,c) \mapsto a} \\ \label{eqn:triad-conservation-aaa}
        0 =& T_{aaa}^{a \mapsto a}
    \end{align}
\end{subequations}
where $T_{aac}^{a \mapsto c}\equiv T_{aac}^{(a,a) \mapsto c}$ and we have used subscripts label the triad to which that term belongs. Note that in the special case of the $aaa$ triad, equation~\eqref{eqn:triad-conservation-aaa}, the transfer is unambiguously closed---energy stays within the a reservoir. In the special case of the $aac$ triad, equation~\eqref{eqn:triad-conservation-aac}, even though the term $T_{aac}^{(a,c) \mapsto a}$ is ambiguous in its transfer source, the other term $T_{aac}^{a \mapsto c}$ resolves the ambiguity, in contrast to the triad components for three unique reservoirs, equation~\eqref{eqn:triad-conservation-abc}. The total energy fluxes of the three reservoirs can now be written as
\begin{subequations}
\label{eqn:reservoir-conservation}
    \begin{align}
        \frac{d}{dt} E_a =& T_{aaa}^{a \mapsto a} + T_{bba}^{b \mapsto a} - T_{aab}^{a \mapsto b} + T_{cca}^{c \mapsto a}  - T_{aac}^{a \mapsto c} + T_{abc}^{(b,c) \mapsto a}  \\
        \frac{d}{dt} E_b =& T_{aab}^{a \mapsto b} - T_{bba}^{b \mapsto a} + T_{bbb}^{b \mapsto b}  + T_{ccb}^{c \mapsto b}  - T_{bbc}^{b \mapsto c} + T_{abc}^{(a,c) \mapsto b}  \\
        \frac{d}{dt} E_c =& T_{aac}^{a \mapsto c} - T_{cca}^{c \mapsto a} + T_{bbc}^{b \mapsto c} - T_{ccb}^{c \mapsto b}  + T_{ccc}^{c \mapsto c}  + T_{abc}^{(a,b) \mapsto c}  
    \end{align}
\end{subequations}
satisfying the total energy conservation law that
\begin{equation}
    \frac{d}{dt} E_a + \frac{d}{dt} E_b + \frac{d}{dt} E_c= 0.
\end{equation}
With the goal of drawing arrows connecting the reservoirs, the terms in \eqref{eqn:reservoir-conservation} are sorted according to which reservoir they are transferring from. The ambiguity in the $abc$ triad is not resolvable in general \cite{dar2001-physd}, so we make a good practical choice. We would like to express the triad in terms of unambiguous transfers $T^{b \mapsto a}$, $T^{a \mapsto c}$, and $T^{c \mapsto b}$ such that,
\begin{subequations}
    \begin{align}
        T_{abc}^{(b,c) \mapsto a} =& T^{b \mapsto a} - T^{a \mapsto c} \\
        T_{abc}^{(a,c) \mapsto b} =& -T^{b \mapsto a} + T^{c \mapsto b} \\
        T_{abc}^{(a,b) \mapsto c} =& T^{a \mapsto c} - T^{c \mapsto b}
    \end{align}
\end{subequations}
which satisfies the conservation law \eqref{eqn:triad-conservation-abc}. The solution to this equation is
\begin{equation}
        \begin{bmatrix}
        T^{b \mapsto a} \\
        T^{a \mapsto c} \\
        T^{c \mapsto b}
    \end{bmatrix}
    =
    \begin{bmatrix}
        T_{abc}^{(b,c) \mapsto a} - T_{abc}^{(a,c) \mapsto b} \\
        T_{abc}^{(a,b) \mapsto c} + T_{abc}^{(b,c) \mapsto a} \\
        -T_{abc}^{(a,c) \mapsto b} - T_{abc}^{(a,b) \mapsto c}
    \end{bmatrix}
    + \begin{bmatrix}
        T_0 \\
        T_0 \\
        T_0
    \end{bmatrix}
\end{equation}
with the ambiguity up to the arbitrary constant $T_0$.

To choose an appropriate value for $T_0$, it is helpful to consider two limits: the \emph{catalyst limit} and the \emph{equipartition limit}. The catalyst limit occurs when one of the triad terms is zero, e.g. $T_{abc}^{(b,c) \mapsto a} \approx 0$. The usual interpretation here is that the $a$ reservoir is acting as a catalyst, neither gaining or losing energy, but facilitating the transfer between reservoirs $b$ and $c$ and thus $T^{b \mapsto a} = T^{a \mapsto c} = 0$ with only $T^{c \mapsto b} \neq 0$. The equipartition limit occurs when two triad terms are equal, e.g., $T_{abc}^{(b,c) \mapsto a} = T_{abc}^{(a,c) \mapsto b}$, in which case the transfer in or out of reservoirs $a$ and $c$ should be equal, $T^{a \mapsto c} = -T^{c \mapsto b}$. As a consequence, this also requires that $T^{b \mapsto a}=0$, just as in the catalyst limit. We there for make the choice that whenever, e.g., $T_{abc}^{(a,b) \mapsto c}$ has the largest magnitude, one should take $T^{b \mapsto a}=0$ such that
\begin{equation}
        \begin{bmatrix}
        T^{b \mapsto a} \\
        T^{a \mapsto c} \\
        T^{c \mapsto b}
    \end{bmatrix}
    =
    \begin{bmatrix}
        0 \\
        -T_{abc}^{(b,c) \mapsto a} \\
        T_{abc}^{(a,c) \mapsto b}
    \end{bmatrix}
\end{equation}
and similarly when the magnitudes of $T_{abc}^{(a,c) \mapsto b}$ and $T_{abc}^{(b,c) \mapsto a}$ are maximal. Algorithmically, the code to resolve the transfers from the mixed triads is as follows,
\begin{lstlisting}
FUNCTION transfersFromTriads(T_b_c_a, T_a_c_b, T_a_b_c)
  // Inputs: three scalar triad interaction terms
  // Outputs: three scalar transfers T_b_a, T_c_a, T_c_b

  IF |T_a_b_c| >= |T_b_c_a| AND |T_a_b_c| >= |T_a_c_b| THEN
      T_b_a := 0
      T_c_a := T_b_c_a
      T_c_b := T_a_c_b
  ELSE IF |T_a_c_b| >= |T_b_c_a| AND |T_a_c_b| >= |T_a_b_c| THEN
      T_b_a := T_b_c_a
      T_c_a := 0
      T_c_b := -T_a_b_c
  ELSE
      T_b_a := -T_a_c_b
      T_c_a := -T_a_b_c
      T_c_b := 0
  END IF

  RETURN (T_b_a, T_c_a, T_c_b)
END FUNCTION
\end{lstlisting}

The algorithm to compute the unique $n(n-1)/2$ transfers between $n$ reservoirs is listed below, and results in an $n$ by $n$ anti-symmetric matrix, interpreted as a transfer from the \verb|i|-th row to the \verb|j|-the column. The function \verb|triadComponent| computes the triad component from equation~\eqref{eqn:mixed-triad-component}.
\begin{lstlisting}
FUNCTION reservoirTransfers(N)
// Inputs: number of reservoirs, N
// Outputs: N x N anti-symmetric matrix, transfers
  
transfers := N x N matrix filled with 0

FOR k from 1 to N:
  FOR i from 1 to N:
    FOR j from 1 to i:

      IF i == j THEN
        transfers[i,k] := transfers[i,k] + triadComponent(i,i,k)

      ELSE IF i == k THEN
        transfers[i,k] := transfers[i,k] + triadComponent(k,k,j)

      ELSE IF j == k THEN
        transfers[i,k] := transfers[i,k] + triadComponent(k,k,i)

      ELSE
        T_i_j_k = triadComponent(i,j,k)
        T_j_k_i = triadComponent(max(j,k),min(j,k),i)
        T_k_i_j = triadComponent(max(i,k),min(i,k),i)

        (T_j_i,T_k_i,T_k_j) := transfersFromTriads(T_j_k_i,T_k_i_j,T_i_j_k)

        transfers[i,k] := transfers[i,k] - T_k_i
        transfers[j,k] := transfers[j,k] - T_k_j
      END IF

    END FOR
  END FOR
END FOR
END FUNCTION
\end{lstlisting}
The first three cases in the function handle equations \eqref{eqn:triad-conservation-aac} and \eqref{eqn:triad-conservation-aaa}, while the fourth and final case handles \eqref{eqn:triad-conservation-abc}. In practice we do not compute both $T_{abc}^{(a,b) \mapsto c}$ and $T_{bac}^{(b,a) \mapsto c}$ because they are equivalent, so the \verb|min|/\verb|max| in the final case prevents the extra computation.

%%%%%%%%%%%%%%%%%%%%%%%%%%%%%%%%%%%%
%
\section{Wave mode projection}
\label{appendix:wave-mode-projection}
%
%%%%%%%%%%%%%%%%%%%%%%%%%%%%%%%%%%%%

The wave mode projection operator is derived by first noting that because $\braopket{\wmode}{\op{H}}{\gmode}=0$, we can write that $\braopket{\wmode}{\op{H}}{\psi}=\braopket{\wmode}{\op{H}}{\psi- A_0 \gmode}$ where
\begin{equation}
A_0^{k\ell j}  = \frac{\braopket{\gmode}{\op{H}}{\psi}}{\braopket{\gmode}{\op{H}}{\gmode}}= \mathcal{F}_g \left[ \hat{\zeta} \right] - \frac{f}{h_g^j} \mathcal{G}_g [ \hat{\eta} ] .
\end{equation}

In the derivation that follows, we drop all superscripts and subscripts from $\omega_\kappa^j$ and $h_\kappa^j$. The Fourier transformed total field $(\hat{u},\hat{v},\hat{w},\hat{\eta})$, Fourier transformed geostrophic fields $(\hat{u}_g,\hat{v}_g,\hat{\eta}_g)$ as well as $F_\kappa$ and $G_\kappa$ are all functions of $z$. For shorthand we define $\hat{\zeta} \equiv i k \hat{v} - i \ell \hat{u}$ and $\hat{\delta} \equiv i k \hat{u} + i \ell \hat{v}$. \textbf{Importantly, $(\hat{u},\hat{v},\hat{w},\hat{\eta})$ are to be treated a placeholders for function arguments, not the actual dynamical variables.}

    \begin{align*} 
\braopket{\wmode}{\op{H}}{\psi}
     = & \braopket{\wmode}{\op{H}}{\psi-A_0 \gmode} \\\nonumber
       =&   \frac{e^{\mp i \omega t}}{2\omega \kappa} \int_
        {-D}^0\left( \left( k \omega \pm i f \ell \right) F_\kappa \left( \hat{u} - \hat{u}_g \right) + \left( \ell \omega \mp i f k \right) F_\kappa \left( \hat{v} - \hat{v}_g \right) + i \kappa^2 \omega h G_\kappa \hat{w} \mp  N^2 \kappa^2 h G_\kappa(z) \left( \hat{\eta} - \hat{\eta}_g \right) \right) dz \\  
        =&   \frac{e^{\mp i \omega t}}{2\omega \kappa} \int_{-D}^0\left( -i\omega \hat{\delta} F_\kappa  + i \kappa^2 \omega h G_\kappa \hat{w} \mp  f \left(\hat{\zeta}-\hat{\zeta}_g \right) F_\kappa \mp  N^2 \kappa^2 h G_\kappa(z) \left( \hat{\eta} - \hat{\eta}_g \right) \right) dz \\
        =&   \frac{e^{\mp i \omega t}}{2 \kappa} \underbrace{ \int_{-D}^0\left( -i \hat{\delta} F_\kappa  + i \kappa^2 h G_\kappa \hat{w} \right) \, dz }_{\mathcal{I}_1}  \mp \frac{e^{\mp i \omega t}}{2 \kappa} \underbrace{\int_{-D}^0\left( \frac{f}{\omega} \left(\hat{\zeta}-\hat{\zeta}_g \right) F_\kappa +  \frac{N^2}{\omega} \kappa^2 h G_\kappa(z) \left( \hat{\eta} - \hat{\eta}_g \right) \right) dz}_{\mathcal{I}_2}
\end{align*}
Starting with $\mathcal{I}_2$
\begin{align*}
    \mathcal{I}_2 =& \int_{-D}^0\left(   \frac{f}{\omega} \left(\hat{\zeta}-\hat{\zeta}_g \right) F_\kappa +  \frac{1}{g \omega} N^2 \left( \omega^2 - f^2 \right) G_\kappa(z) \left( \hat{\eta} - \hat{\eta}_g \right) \right) dz \\
    = & \int_{-D}^0\left(   \frac{f}{\omega} \left(\hat{\zeta}-\hat{\zeta}_g \right) F_\kappa +\frac{f^2}{g \omega}  \left( \omega^2  - N^2 \right) G_\kappa(z) \left( \hat{\eta} - \hat{\eta}_g \right)  \right) dz + \int_{-D}^0\left(   \frac{\omega}{g} \left(N^2-f^2\right) G_\kappa(z) \left( \hat{\eta} - \hat{\eta}_g \right) \right) dz \\
    =& \omega \mathcal{G}_\kappa \left[\hat{\eta} - \hat{\eta}_g \right] +\frac{1}{\omega}  \int_{-D}^0\left(   f \left(\hat{\zeta}-\hat{\zeta}_g \right) F_\kappa + f^2 \partial_z F_\kappa(z) \left( \hat{\eta} - \hat{\eta}_g \right)  \right) dz \\
    =& \omega \mathcal{G}_\kappa \left[\hat{\eta} - \hat{\eta}_g \right] +\frac{1}{\omega}  \int_{-D}^0\left(    \left(\hat{\zeta}-\hat{\zeta}_g \right) - f \partial_z  \left( \hat{\eta} - \hat{\eta}_g \right)  \right) F_\kappa(z) dz \\
    =& \omega \mathcal{G}_\kappa \left[\hat{\eta} - \hat{\eta}_g \right] 
\end{align*}
where the integral in the final step vanishes because the geostrophic mode contains all of the QGPV in the flow. The first integral, $\mathcal{I}_1$ needs to be rewritten as a projection operator $\mathcal{G}_\kappa$, which means that we will need to use integration-by-parts to flip the $F_\kappa$ to a $G_\kappa$, either as an integral of $\hat{\delta}$,
\begin{equation}
\label{second-integral-integrated-version}
    \mathcal{I}_1 = -i\left[  \mathcal{G}_\kappa \left[ \Delta \right] - g h_\kappa \kappa^2 \mathcal{G}_\kappa \left[ \frac{\Delta  + \hat{w}}{N^2 - f^2}\right]  \right]
\end{equation}
where $\Delta$ as the integral of $\hat{\delta}$ or a derivative of $\hat{\delta}$,
\begin{equation}
\label{second-integral-derivative-version}
    \mathcal{I}_1 = i g h_\kappa \mathcal{G}_\kappa \left[ \frac{\hat{\delta}_z + \kappa^2 \hat{w}}{N^2 - f^2} \right].
\end{equation}
As it will turn out, the integral version will lead to a more computationally efficient version of the projection of the fluid state, while the derivative version will be more computationally efficient for the non-hydrostatic flux. When projecting the fluid state, the continuity condition applies and the second integral in \eqref{second-integral-integrated-version} vanishes, leading to the state projection operator,
\begin{equation}
        A_w^i = \frac{\braopket{\wmode}{\op{H}}{\psi}}{\braopket{\wmode}{\op{H}}{\wmode}} 
        = \frac{e^{\mp i \omega_\kappa^j t}}{2 \kappa h_\kappa^j}  \left( i \mathcal{G}^j_\kappa \left[\hat{w}(z) \right] \mp \omega_\kappa^j \mathcal{G}^j_\kappa \left[\hat{\eta} - \hat{\eta}_g \right].
        \right)
\end{equation}
In practice $\hat{w}$ is always computed from the integral of the horizontal divergence,
\begin{equation}
    \Delta = \mathcal{G}_g^{-1} \left[ h_g^j \mathcal{F}_g \left[ \delta \right] \right].
\end{equation}
When projecting the nonlinear flux, we recommend \eqref{second-integral-derivative-version} so that,
\begin{equation}
        A_w^i = \frac{\braopket{\wmode}{\op{H}}{\psi}}{\braopket{\wmode}{\op{H}}{\wmode}} \\ \nonumber
        = \frac{e^{\mp i \omega_\kappa^j t}}{2 \kappa h_\kappa^j}  \left( i g h_\kappa \mathcal{G}_\kappa \left[ \frac{\hat{\delta}_z + \kappa^2 \hat{w}}{N^2 - f^2} \right] \mp \omega_\kappa^j \mathcal{G}^j_\kappa \left[\hat{\eta} - \hat{\eta}_g \right].
        \right)
\end{equation}
where one would use that,
\begin{equation}
    \hat{\delta}_z = -\frac{N^2}{g} \mathcal{G}_g^{-1} \left[ \mathcal{F}_g \left[ \hat{\delta} \right] \right]
\end{equation}
to compute the derivative spectrally.

%%%%%%%%%%%%%%%%%%%%%%%%%%%%%%%%%%%%
%
\section{Proof of orthogonality}\label{sec:proof-of-orthogonality}
%
%%%%%%%%%%%%%%%%%%%%%%%%%%%%%%%%%%%%

Orthogonality between modes is trivial for modes at different wavenumbers following \eqref{eqn:fourier-orthogonality},
\begin{equation}
\label{eqn:two-mode-general-orthogonality}
    \braopket{\psi^a}{\op{H}}{\psi^b} = \frac{1}{2} \delta_{k_a k_b} \delta_{\ell_a \ell_b} \int_{-D}^0 \hat{u}_a \conj{\hat{u}}_b + \hat{v}_a \conj{\hat{v}}_b + \hat{w}_a \conj{\hat{w}}_b + N^2(z) \hat{\eta}_a \conj{\hat{\eta}}_b \, dz
\end{equation}
and so we need only consider orthogonality of modes at the same wavenumber and thus can drop the $\delta_{k_a k_b} \delta_{\ell_a \ell_b}$ from the proofs that follow.

%%%%%%%%%%%%%%%%%%%%%%%%%%%%%%%%%%%%
\paragraph{Orthogonality between two geostrophic modes}
%%%%%%%%%%%%%%%%%%%%%%%%%%%%%%%%%%%%

Inserting two geostrophic modes from Table~\ref{tab:solutions} into \eqref{eqn:two-mode-general-orthogonality} leads to the condition that,
\begin{equation}
    \braopket{\gmode^a}{\op{H}}{\gmode^b} =\frac{1}{2} \int_{-D}^0 \frac{g^2 \kappa^2}{f^2} F_g^a F_g^b + N^2 (z) G_g^a G_g^b \, dz = 0
\end{equation}
when $a \neq b$. Orthogonality follows immediately from the orthogonality of the $F_g$ modes \eqref{eqn:f-norm-geostrophic} and the $G_g$ modes \eqref{eqn:g-norm-geostrophic}.

%%%%%%%%%%%%%%%%%%%%%%%%%%%%%%%%%%%%
\paragraph{Orthogonality between geostrophic and wave modes}
%%%%%%%%%%%%%%%%%%%%%%%%%%%%%%%%%%%%

First note that using integration by parts we can show that
\begin{equation}
    \int_{-D}^0 g F_\kappa F_g \, dz = \int_{-D}^0 g h_\kappa \partial_z G_\kappa F_g \, dz = - g h_\kappa \int_{-D}^0 G_\kappa \partial_z F_g \, dz, 
\end{equation}
using that $F_\kappa = h_\kappa \partial_z G_\kappa$ and the boundary conditions $G_\kappa(0) = G_\kappa(-D) = 0$. Inserting one wave mode and one geostrophic mode from Table~\ref{tab:solutions} into \eqref{eqn:two-mode-general-orthogonality} leads to the condition that
\begin{subequations}
    \begin{align}
        \braopket{{\wmode}^a}{\op{H}}{\gmode^b} =& \pm \frac{k}{\omega_\kappa^a} e^{i k x \pm i \omega_\kappa^a t} \int_{-D}^0 g F_\kappa^a F_g^b - h_\kappa^a N^2 G_\kappa^a G_g^b \, dz \\
        =& \mp \frac{k}{\omega_\kappa^a} h_\kappa^a e^{i k x \pm i \omega_\kappa^a t} \int_{-D}^0  G_\kappa^a \left( g \partial_z F_g^b +  N^2 G_g^b \right)\, dz \\
        =& 0
    \end{align}
\end{subequations}
when $a \neq b$. The last step follows using that $g \partial_z F_g^b = - N^2 G_g^b$.

%%%%%%%%%%%%%%%%%%%%%%%%%%%%%%%%%%%%
\paragraph{Orthogonality between two wave modes}
%%%%%%%%%%%%%%%%%%%%%%%%%%%%%%%%%%%%

Inserting two wave modes from Table~\ref{tab:solutions} into \eqref{eqn:two-mode-general-orthogonality} leads to the condition that
\begin{subequations}
    \begin{align} \nonumber
        \braopket{{\wmode}^a}{\op{H}}{\gmode^b} =& \frac{1}{2} e^{i(\omega_\kappa^a- \omega_\kappa^b)t} \int_{-D}^0 \left[ 1+ \frac{f^2}{\omega_\kappa^a \omega_\kappa^b}\right] F_\kappa^a F_\kappa^b + \kappa^2 h_\kappa^a h_\kappa^b G_\kappa^a G_\kappa^b + \frac{\kappa^2 h_\kappa^a h_\kappa^b}{\omega_\kappa^a \omega_\kappa^b} N^2 G_\kappa^a G_\kappa^b \, dz \\ \label{eqn:two-wave-proof}
        =& \frac{1}{2} e^{i(\omega_\kappa^a- \omega_\kappa^b)t} \int_{-D}^0 \left[ 1+ \frac{f^2}{\omega_\kappa^a \omega_\kappa^b}\right] \left( F_\kappa^a F_\kappa^b + \kappa^2 h_\kappa^a h_\kappa^b G_\kappa^a G_\kappa^b \right) dz 
    \end{align}
\end{subequations}
where we used that $\int (N^2-f^2) G_\kappa^a G_\kappa^b dz = 0$ for $a \neq b$ from \eqref{eqn:g-norm-igw}. Using integration by parts and $G_\kappa(0) = G_\kappa(-D) = 0$, this condition also leads to
\begin{equation}
    \int_{-D}^0 F_\kappa^a F_\kappa^b + h_\kappa^a h_\kappa^b \kappa^2 G_\kappa^a G_\kappa^b \, dz = 0
\end{equation}
which means the integral in \eqref{eqn:two-wave-proof} vanishes, concluding the proof.

%%%%%%%%%%%%%%%%%%%%%%%%%%%%%%%%%%%%
\paragraph{Orthogonality between the mda and inertial modes}
%%%%%%%%%%%%%%%%%%%%%%%%%%%%%%%%%%%%

Table~\ref{tab:solutions} shows that mda modes have only non-trivial isopycnal deviation, while inertial oscillations have only non-trivial horizontal velocity, and thus \eqref{eqn:two-mode-general-orthogonality} is trivially satisfied.

\end{document}